%% file: main.tex
\documentclass[prd,amsmath,aps,floats,amssymb, floatfix, superscriptaddress,nofootinbib,preprintnumbers,twocolumn]{revtex4-1}  
\usepackage[linktocpage=true]{hyperref}
\usepackage{makecell}
\usepackage{amsmath,amssymb,bm,natbib,latexsym,times}
\usepackage{bm}
\usepackage{graphics}
\usepackage{todonotes,multirow,enumitem}
\usepackage{xspace}
\xspaceaddexceptions{]\}}
\usepackage{ulem}

\makeatletter
\g@addto@macro\bfseries{\boldmath}
\makeatother

\newcommand{\sect}[1]{Sec.~\ref{#1}\xspace}
\newcommand{\sects}[1]{Secs.~\ref{#1}\xspace}
\newcommand{\app}[1]{Appendix~\ref{#1}\xspace}
\newcommand{\apps}[1]{Appendices~\ref{#1}\xspace}
\newcommand{\fig}[1]{Fig.~\ref{#1}\xspace}
\newcommand{\figs}[1]{Figs.~\ref{#1}\xspace}
\newcommand{\eq}[1]{Eq.~(\ref{#1})\xspace}
\newcommand{\eqs}[1]{Eqs.~(\ref{#1})\xspace}
\newcommand{\tab}[1]{Table~\ref{#1}\xspace}

\newcommand{\lcdm}{\ensuremath{\Lambda\text{CDM}}\xspace}
\newcommand{\wcdm}{\ensuremath{w\text{CDM}}\xspace}
\newcommand{\lwcdm}{\ensuremath{\Lambda{\text{/}}w\text{CDM}}\xspace}
\newcommand{\mpp}{3\ensuremath{\times}2{\rm pt}\xspace}
\newcommand{\cosmosis}{{\sc CosmoSIS}\xspace}
\newcommand{\mgcamb}{{\sc mgcamb}\xspace}
\newcommand{\halofit}{{\sc Halofit}\xspace}
\newcommand{\EuclidEmulator}{{\sc Euclid Emulator}\xspace}
\newcommand{\ythree}{Y3-3\ensuremath{\times}2pt\xspace}

\newcommand{\metacal}{{\sc Metacalibration}\xspace}
\newcommand{\redmagic}{{\sc redMaGiC}\xspace}
\newcommand{\maglim}{MagLim\xspace}
\newcommand{\fastpt}{Fast-PT\xspace}
\newcommand{\mnu}{\ensuremath{\sum m_{\nu}}\xspace}

\newcommand{\summnu}{\ensuremath{\sum m_{\nu}}\xspace}
\newcommand{\omnuhh}{\ensuremath{\Omega_{\nu}h^2}\xspace}
\newcommand{\sigeight}{\ensuremath{\sigma_8}\xspace}
\newcommand{\seight}{\ensuremath{S_8}\xspace}
\newcommand{\fsig}{\ensuremath{f\sigma_8}\xspace}

\newcommand{\chisq}{\ensuremath{\chi^2}\xspace}
\newcommand{\dchisq}{\ensuremath{\Delta\chi^2}\xspace}
\newcommand{\As}{\ensuremath{A_{\rm s}}\xspace}
\newcommand{\om}{\ensuremath{\Omega_{\rm m}}\xspace}
\newcommand{\ok}{\ensuremath{\Omega_k}\xspace}
\newcommand{\ob}{\ensuremath{\Omega_{\rm b}}\xspace}
\newcommand{\Neff}{\ensuremath{N_{\mathrm{eff}}}\xspace}

\newcommand{\neff}{\ensuremath{N_{\mathrm{eff}}}\xspace}
\newcommand{\meff}{\ensuremath{m_{\mathrm{eff}}}\xspace}
\newcommand{\neffmeff}{\ensuremath{\neff-\meff}\xspace}
\newcommand{\wo}{\ensuremath{w_0}\xspace}
\newcommand{\wa}{\ensuremath{w_a}\xspace}
\newcommand{\wowa}{\ensuremath{w_0-w_a}\xspace}
\newcommand{\npg}{binned $\sigma_8(z)$\xspace} 
\newcommand{\Anpg}{\ensuremath{A^{P_{\rm lin}}}} 
\newcommand{\npgsig}[1]{\ensuremath{\sigma_8^{[{\rm bin}\,#1]}}\xspace}
\newcommand{\npgsigcmb}{\ensuremath{\sigma_8^{[{\rm CMB}]}}\xspace}

\newcommand{\xlens}{\ensuremath{X_{\rm Lens}}\xspace}
\newcommand{\sigo}{\ensuremath{\Sigma_0}\xspace}
\newcommand{\muo}{\ensuremath{\mu_0}\xspace}
\newcommand{\sigmu}{\ensuremath{\Sigma_0-\mu_0}\xspace}

\newcommand{\aap}{{A\&A}}
\newcommand{\apjs}{{Astrophys.~J.~Supp.}}
\newcommand{\aj}{{Astro.~Journal}}

\newcommand{\mnras}{{Mon.~Not.~R.~Astron.~Soc.}}
\newcommand{\jcap}{{JCAP}}

\newcommand{\fastismore}{{\sc{FastISMoRE}}\xspace}
\newcommand\be{\begin{equation}}
\newcommand\ee{\end{equation}}

\def\bea{\begin{eqnarray}}
\def\eea{\end{eqnarray}}

\newcommand{\magcoeff}{{\bf\mathcal{C}}}
\newcommand{\magcoeffi}{{\bf\mathcal{C}}_{i}}
\newcommand{\magcoeffj}{{\bf\mathcal{C}}_{j}}

\makeatletter
\def\l@subsubsection#1#2{}
\makeatother

\begin{document}

\title{Dark Energy Survey Year 3 Results: Constraints on extensions to \texorpdfstring{$\bm{\Lambda}$CDM}{Lg} with weak lensing and galaxy clustering}
\collaboration{DES Collaboration}
\input{july14_DES-2021-0654_author_list}
\preprint{DES-2021-0654}
\preprint{FERMILAB-PUB-22-470-PPD}

\date{\today}

\label{firstpage}
\begin{abstract}
We constrain six possible extensions to the $\Lambda$CDM model using measurements from the Dark Energy Survey's first three years of observations, alone and in combination with external cosmological probes.  The DES data are the two-point correlation functions of weak gravitational lensing, galaxy clustering, and their cross-correlation.  We use simulated data vectors and blind analyses of real data to validate the robustness of our results to astrophysical and modeling systematic errors.  In many cases, constraining power is limited by the absence of theoretical predictions beyond the linear regime that are reliable at our required precision.  The $\Lambda$CDM extensions are: dark energy with a time-dependent equation of state,  non-zero spatial curvature,  additional relativistic degrees of freedom, sterile neutrinos with eV-scale mass, modifications of gravitational physics, and a binned $\sigma_8(z)$ model which serves as a phenomenological probe of structure growth. For the time-varying dark energy equation of state evaluated at the pivot redshift we find $(w_{\rm p}, w_a)= (-0.99^{+0.28}_{-0.17},-0.9\pm 1.2)$ at 68\% confidence with $z_{\rm p}=0.24$ from the DES measurements alone, and $(w_{\rm p}, w_a)= (-1.03^{+0.04}_{-0.03},-0.4^{+0.4}_{-0.3})$ with $z_{\rm p}=0.21$  for the combination of all data considered.  Curvature constraints  of $\Omega_k=0.0009\pm 0.0017$ and effective relativistic species $N_{\rm eff}=3.10^{+0.15}_{-0.16}$ are dominated by external data, though adding DES information to external low redshift probes tightens the $\Omega_k$ constraints that can be made without CMB observables by 20\%.  For massive sterile neutrinos, DES combined with external data improves the upper bound on the mass $m_{\rm eff}$ by a factor of three compared to previous analyses, giving 95\% limits of $(\Delta N_{\rm eff},m_{\rm eff})\leq (0.28, 0.20\, {\rm eV})$ when using priors matching a comparable {\it Planck} analysis. For modified gravity, we constrain changes to the lensing and Poisson equations controlled by functions $\Sigma(k,z) = \Sigma_0 \Omega_{\Lambda}(z)/\Omega_{\Lambda,0}$ and $\mu(k,z)=\mu_0 \Omega_{\Lambda}(z)/\Omega_{\Lambda,0}$ respectively to $\Sigma_0=0.6^{+0.4}_{-0.5}$ from DES alone and $(\Sigma_0,\mu_0)=(0.04\pm 0.05,0.08^{+0.21}_{-0.19})$ for the combination of all data, both at 68\% confidence.  Overall, we find   no significant evidence for physics beyond $\Lambda$CDM.
\end{abstract}

\maketitle


\tableofcontents 

\section{Introduction}\label{sec:intro}

The discovery of the accelerated expansion of the universe made about two decades ago \cite{SupernovaSearchTeam:1998fmf, SupernovaCosmologyProject:1998vns} established  \lcdm as the standard model in cosmology. This paradigm  relies on three pillars: that general relativity  correctly describes gravitational interactions at cosmological scales; that at those scales the Universe appears homogeneous, isotropic and spatially flat; and that the Universe's content at late times is dominated by non-relativistic, pressureless cold dark matter (CDM), and the cosmological constant term $\Lambda$. The resulting $\Lambda$CDM model is in good agreement with cosmological observations from a wide range of temporal and spatial scales  \cite{SNLS:2005qlf, ESSENCE:2007acn, 2010A&A...523A...7G, 2011ApJS..192....1C, Rest:2013mwz, Scolnic:2017caz, DES:2018paw, Hinshaw:2012aka, Planck:2018vyg, ACT:2020gnv, 2012MNRAS.425..405B, Elvin-Poole:2017xsf, eBOSS:2020yzd, Heymans:2013fya, Troxel:2017xyo, HSC:2018mrq, Heymans:2020gsg, Bonvin:2016crt, Huterer:2016uyq, Ivanov:2019hqk}.

The impressive phenomenological success of the $\Lambda$CDM model has not been matched in our understanding of the physical nature of dark energy \cite{Frieman:2008sn,Weinberg:2012es}, nor in insights as to  why the cosmological constant appears to be so small relative to natural scales in particle physics \cite{RevModPhys.61.1,Carroll:2000fy,Martin:2012bt,Padilla:2015aaa}. Therefore, cosmology 
is in need of new and better data that can help shed light on these cosmological conundrums. The quest to understand dark energy has spawned a worldwide effort to 
measure the growth and evolution of cosmic structures in the Universe.
Ongoing experiments focused on dark energy include wide field photometric surveys such as the Dark Energy Survey
(DES)\footnote{\url{http://www.darkenergysurvey.org/}}
\cite{flaugher15,Abbott:2017wau,Abbott:2016ktf}, the Hyper Suprime-Cam Subaru Strategic Program (HSC-SSP)\footnote{\url{https://www.naoj.org/Projects/HSC/}}
\cite{Aihara:2017paw,HSC:2018mrq}, the Kilo-Degree Survey
(KiDS)\footnote{\url{http://kids.strw.leidenuniv.nl/}}
\cite{Kuijken:2015vca,Heymans:2020gsg}, in addition to 
  ongoing spectroscopic surveys like the Extended Baryon Oscillation Spectroscopic Survey (eBOSS)\footnote{\url{https://www.sdss.org/surveys/eboss/}} \cite{Dawson:2015wdb} and the Dark Energy Spectroscopic Instrument (DESI) \footnote{\url{https://www.desi.lbl.gov/}} \cite{desi_paper}. 
These surveys have demonstrated the feasibility of ambitious large-scale structure analyses, featured development of state-of-the-art systematics calibration, and established new standards in protecting analyses against observer bias before the results are revealed. 
Thus far, these surveys have provided constraints
consistent with the \lcdm\ model, and
contributed to tightening the constraints on several key cosmological parameters.

Using data from these surveys to search for deviations from the predictions of \lcdm is one of the primary goals of modern cosmology.
Such deviations could provide clues as to where that minimal cosmological model needs to be extended, and thus a deeper understanding of the fundamental physics impacting the large-scale properties of the Universe.
One approach to testing the \lcdm model is to compare \lcdm parameter estimates inferred from different sets of observables.
This is the motivation behind the ongoing exploration of the 3--5$\sigma$ tension in measurements of the Hubble constant, $H_0$, between low-redshift distance-ladder measurements and those from the CMB at $z\approx1100$ (see Refs.~\cite{Verde:2019ivm,Knox:2019rjx,DiValentino:2021izs,Shah:2021onj,Riess:2021jrx,Abdalla:2022yfr} for a summary), as well as the scrutiny of 1--3$\sigma$ offsets between large scale structure~\cite{Heymans:2013fya,Troxel:2017xyo,Abbott:2017wau,HSC:2018mrq,KiDS:2020suj, Heymans:2020gsg,Krolewski:2021yqy,y3-3x2ptkp,y3-5x2pt_cosmo,y3-6x2pt_cosmo} and CMB-based~\cite{Planck:2018vyg,ACT:2020gnv} constraints on the amplitude of matter density fluctuations scaled by the square root of the matter density, $S_8\equiv \sigma_8 (\Omega_{\rm m}/0.3)^{0.5}$.

In the present analysis we adopt a complementary approach by constraining cosmological models which add physics beyond that of the standard \lcdm paradigm. 
While future precision measurements and careful characterization of existing data (e.g. as in Refs.~\cite{LSSTDarkEnergyScience:2018bov, DES:2021ljl}) will undoubtedly be required to resolve the origin of any tensions between datasets, it is also valuable to investigate whether (or to what extent) any observed offsets may be alleviated by new physics.
Additionally, constraining the parameters of extended models can offer greater sensitivity to signatures of beyond-\lcdm physics that may not manifest clearly as a tension between different measurements of \lcdm parameters. 

This paper constrains a range of extended models using a combined analysis of weak-lensing and galaxy-clustering observations from the first three years of data\footnote{Publicly available at: \url{https://des.ncsa.illinois.edu/releases/y3a2}} of the Dark Energy Survey (henceforth DES Y3)~\cite{des_dr1}. The models, are as follows:
\begin{itemize}
\item Dynamical dark energy parameterized via the linear expansion of the dark energy equation of state $w(a) = \wo + (1-a)\wa$;
\item Non-zero spatial curvature \ok; 
\item Varying the effective number of relativistic species \neff;
\item Sterile neutrinos varying parameters \neff and \meff to control the particles' temperature and effective mass respectively;
\item Deviations from General Relativity  introduced via the functions $\Sigma(k,z) = \sigo \Omega_{\Lambda}(z)/\Omega_{\Lambda,0}$ and $\mu(k,z)=\muo \Omega_{\Lambda}(z)/\Omega_{\Lambda,0}$ respectively modifying the lensing and Poisson equations; 
\item Variation  of the growth rate of structure parameterized by independent $\sigma_8$ values in different redshift bins. 
\end{itemize}
The selection of these models dictated both by their interest to the cosmology community (this is the primary motivation for \wowa, \ok, \neff), and by the kinds of physics about which DES measurements add qualitatively new information (primarily motivating \neffmeff, \sigmu, \npg).

This work is a successor to the DES Y1 extended-model analysis \cite{Abbott:2018xao}, which for conciseness we will reference as  DES-Y1Ext, and complements the main DES Y3 galaxy clustering and weak lensing  analysis \cite{y3-3x2ptkp} (henceforth  DES-Y3KP) that presented cosmology results for \lcdm and the \wcdm model testing for a constant dark energy equation of state different from $-1$. All of these studies extract cosmological information from DES data via a so-called `\mpp' analysis, in which parameter estimation is based on the combined analysis of three types of projected two-point correlation functions: cosmic shear measurements capturing weak lensing distortions to the shape of background source galaxies, galaxy clustering measurements of the positions of foreground lens galaxies, and the tangential shear of source galaxy shapes around each of the lens positions. Compared to DES-Y1Ext, this work includes a number of updates, the most notable being that the DES Y3 data cover roughly three times the sky area included in the Y1 analysis.  To maximize the constraining power of our cosmological data, we will additionally combine the DES Y3 \mpp constraints with the following external datasets:  baryon acoustic oscillations (BAO) and redshift-space distortion (RSD) measurements from the eBOSS, 6dF, and MGS galaxy surveys \cite{eBOSS:2020yzd}, the Pantheon type Ia supernova (SN) catalogue \cite{Scolnic:2017caz}, and the Planck 2018 cosmic microwave background (CMB) data \cite{Planck:2019nip}.

The paper is organized as follows: In Sec.~\ref{sec:DESY3data} we describe the DES Y3 data used in this analysis, and the baseline modeling of the observables. Sections \ref{sec:extmodels} and \ref{sec:external_data} are devoted to a presentation of the extended models and the main datasets exploited in this work, respectively. In Sec.~\ref{sec:analysis} we discuss the details of our analysis validation. We present our main results in Sec.~\ref{sec:res}, and conclude
in Sec.~\ref{sec:ccl}.

Data supplementing this paper, including chains, scale cuts, and numerical versions of summary plots, will be available online as part of the DES Y3 data release.~\footnote{\url{https://des.ncsa.illinois.edu/releases/y3a2/Y3key-extensions}}


\section{Data and baseline modeling} \label{sec:DESY3data}

In this section we describe the DES data used in this analysis and the likelihood used to perform parameter estimation based on  the angular two-point correlation function (2PCF) summary statistics into which those data are condensed.

\subsection{Source and lens catalogues}

The Dark Energy Survey (DES) is a 5000 deg$^2$ photometric galaxy survey which, over the course of six years, collected data using the Dark Energy Camera (DECam \cite{DECam}), mounted on the Víctor Blanco 4m telescope at the Cerro Tololo Inter-American Observatory (CTIO) in Chile.  In this work we employ data from the first three years of DES observations (DES Y3), which constitute the DES Data Release 1 (DR1 \cite{des_dr1}). 
That data was processed to produce a photometric catalog of 399 million objects with signal-to-noise ratio of $\sim$ 10 in \textit{r,i,z} co-add images.
For cosmological inference, we further refine this catalog to produce a `Gold' sample \cite{y3-gold} containing 319 million objects, extending to a limiting magnitude of 23 in the $i$-band.

From the Gold sample galaxies we select two samples: `source' (background) galaxies, whose shear is used for measurements of gravitational lensing, and `lens' (foreground) galaxies, whose positions are recorded and used for measurements of galaxy clustering.
The source galaxy sample is used to produce the DES Y3 shape catalogue~\cite{y3-shapecatalog}. 
We measure galaxy shapes with the \metacal\ pipeline \cite{Huff:2017qxu, Sheldon:2017szh}, which uses \textit{r,i,z}-band information to infer objects' ellipticity and other photometric properties,  employing updates to PSF solutions~\cite{y3-piff}, astrometric solutions~\cite{y3-gold}, and inverse-variance weighting for the galaxies to improve upon a similar pipeline used for the DES Y1 analysis~\cite{Abbott:2017wau}.
Multiplicative shear bias is calibrated using image simulations~\cite{y3-imagesims} and redshifts are inferred using a self-organizing-map approach~\cite{y3-sompz, y3-sompzbuzzard} which connects wide and deep-field~\cite{y3-deepfields} galaxy measurements using Balrog simulations~\cite{y3-balrog}. The final DES Y3 shape catalog contains 100 million galaxies covering an area of 4143 deg$^2$, with a weighted effective number density $n_{\rm eff}=5.9$ per arcmin$^2$ and corresponding shape noise {$\sigma_e = 0.26$}.

The lens galaxy sample that we use in this paper, \maglim~\cite{y3-2x2maglimforecast}, is one of the two lens samples considered in DES-Y3KP. The \maglim sample contains sources selected to be in the redshift range $ 0.2\leq z \leq 1.05 $, and is divided into six tomographic bins with nominal edges $z = [0.20, 0.40, 0.55, 0.70, 0.85, 0.95, 1.05]$.  Uncertainties in the photometric redshift estimator used to define these bins cause the bins' actual $n(z)$ redshift distributions to extend outside those bounds. 
The inferred lens redshift distributions have been further validated using cross-correlations between galaxies in \maglim with spectroscopic galaxy samples~\cite{y3-lenswz}.
Weights based on the correlation between number density with survey properties mitigate the impact of observing systematics~\cite{y3-galaxyclustering}. 
Refs.~\cite{y3-galaxyclustering,y3-2x2ptaltlensresults} describe the sample's validation and characterization in more detail.

We follow DES-Y3KP in removing the two highest redshift \maglim bins from our analysis, as studies after unblinding the \lwcdm results revealed issues with the sample at $z>0.85$. These issues, which manifest as an inability of the model to consistently match the clustering amplitude of galaxy clustering and galaxy--galaxy lensing in the last two tomographic bins, were initially detected because they contribute to poor \lwcdm model fits. While this might be of particular interest to searches for beyond-\lcdm physics, investigation since suggests that  the issue is most plausibly caused by systematics related to photometric calibration.
Further discussion can be found in Refs.~\cite{y3-2x2ptaltlensresults,y3-3x2ptkp,y3-6x2pt_cosmo}.
Given these indications, we choose to adopt a conservative approach of removing the impacted \maglim bins from our analysis. 

\subsection{2PCF -- measurements}

The cosmological information contained in the lens and source samples described above is then summarized in three two-point correlation functions (2PCF):
\begin{itemize}[noitemsep,topsep=0pt]
\item \emph{Galaxy clustering}: the auto-correlation of lens galaxy positions $w^i(\theta)$ in each redshift bin $i$, i.e.~the fractional excess number of galaxy pairs of separation $\theta$ relative to the number of pairs of randomly distributed points within our survey mask \cite{y3-galaxyclustering},
\item \emph{Cosmic shear}: the auto-correlation of source galaxy shapes within and between source redshift bins $i,j$, of which there are two components $\xi_{\pm}^{i,j}(\theta)$, taking the products of the ellipticity components of pairs of galaxies, either adding ($+$) or subtracting ($-$) the component tangential to the line connecting the galaxies and the component rotated by $\pi/4$ \cite{y3-cosmicshear1,y3-cosmicshear2},
\item \emph{Galaxy--galaxy lensing}: the mean tangential ellipticity of source galaxy shapes around lens galaxy positions $\gamma_t^{i,j}(\theta)$, for each pair of redshift bins $i,j$ \cite{y3-gglensing}.
\end{itemize}
Details of these measurements and the checks for potential systematic effects in them are described in detail in Refs.~\cite{y3-galaxyclustering,y3-gglensing, y3-cosmicshear1,y3-cosmicshear2}, and an overview of the full data vector is given in Ref.~\cite{y3-3x2ptkp}. We follow DES-Y3KP and refer to the combined list of $\{w^i(\theta), \xi^{ij}_{\pm}(\theta), \gamma^{ij}_t(\theta)\}$, for all angles $\theta$ and redshift bins $i$ and $j$,  as the `data vector'. Section \ref{sec:2pcfmodel} below has more details about the component pieces of the data vector.

Each of these measurements is performed in a set of 20 logarithmic bins of angular separation between 2.5' and 250' using the software \textsc{treecorr} \cite{Jarvis:2003wq}. We only use a subset of these bins, removing angular scales where our model is not sufficiently accurate, as discussed in Sec.~\ref{sec:fidscuts}.

\subsection{2PCF -- baseline  modeling}\label{sec:2pcfmodel}

Our baseline modeling methodology generally follows that used in DES-Y3KP \cite{y3-3x2ptkp}, and described in detail in the methodology Y3 paper \cite{y3-generalmethods}. 
Notable differences from the DES-Y3KP analysis include the use of a simpler non-linear alignment (NLA) intrinsic alignment model as opposed to the tidal alignment and tidal torquing (TATT) model  as our fiducial intrinsic alignment model, and not using the shear-ratio likelihood \cite{y3-shearratio} in most of our analysis.
Below we summarize the modeling used to compute the \mpp likelihood, as well as these differences.

As noted above, the \ythree analysis consists of a set of 2PCF measurements describing the angular correlation of lens galaxy positions and source galaxy shapes for several redshift bins.  We model the likelihood as Gaussian in the data vector $\mathbf D$, 
\begin{equation}
\ln \mathcal{L}(\mathbf D | \mathbf \Theta) = -\frac{1}{2}\left[\left({\mathbf D}-\mathbf M(\mathbf \Theta)\right)^{\rm{T}} \mathbf C^{-1}\left({\mathbf D}-\mathbf M(\mathbf \Theta)\right)\right] + L_0,
\label{eq:like}
\end{equation}
where $\mathbf \Theta$ is the vector of cosmological and nuisance parameters,  $\mathbf C$ is the covariance, and $L_0$ is a normalization constant.  The covariance is computed analytically using 
\textsc{CosmoLike} \cite{y3-generalmethods} including \textsc{CosmoCov} \cite{Fang:2020vhc}. The likelihood and the covariance were validated in Ref.~\cite{y3-covariances}, where it has been shown that, for the precision level attained by the DES Y3 analysis, assuming a Gaussian likelihood with the various assumptions involved in the computation of $\mathbf C$ are all excellent approximations (see in particular Fig.~1 of Ref.~\cite{y3-covariances}).
We sample the above likelihood to obtain posterior and evidence estimates using the {\sc Polychord} nested sampler~\cite{Handley:2015a,Handley:2015b}, following guidelines for settings described in Ref.~\cite{y3-samplers}.
The length of the fiducial data vector $\mathbf D$ is 462 though for some models data points will be removed to account for modeling uncertainties (see Table \ref{tab:scuts}). The length of the parameter vector $\mathbf \Theta$ for \lcdm is 28 when fitting DES data alone (additional parameters are introduced when combining with external data).

Full details of how the data vector $\mathbf D$ is theoretically predicted can again be found in Refs.~\cite{y3-generalmethods,y3-3x2ptkp}, but here we give a brief overview.
The 2PCF are computed from the observed projected galaxy density contrast $\delta_{g}$ and the shear field  decomposed into E- and B-modes. The 2PCF forming the data vector $\mathbf D$ can be expressed in terms of the angular power spectra as:
\begin{equation}
    \begin{aligned}
\label{eq:xi_in_terms_of_Cell}
w^i(\theta) &= \sum_\ell \frac{2 \ell +1}{4\pi} P_\ell(\cos \theta) C^{ii}_{\delta_g \delta_g}(\ell)\\[0.2cm]
\gamma^{ij}_t(\theta) &= \sum_\ell \frac{2\ell + 1}{4\pi} \frac{P_\ell^2\left( \cos \theta \right)}{\ell(\ell + 1)} C_{\delta_g \mathrm{E}}^{ij}(\ell)\\[0.2cm]
\xi_{\pm}^{ij}(\theta) &= \sum_{\ell \geqslant 2} \frac{2\ell + 1}{4\pi}\ \frac{2(G_{\ell, 2}^+(x) \pm G_{\ell, 2}^-(x))}{\ell^2(\ell + 1)^2}\ \\ 
 & \times [ C^{ij}_{\mathrm{EE}}(\ell) \pm C^{ij}_{\mathrm{BB}}(\ell) ],
\end{aligned}
\end{equation}
where $i,j$ denote redshift bins.
Here $P_\ell$ are the Legendre polynomials of order $\ell$, $P_\ell^m$ are the associated Legendre polynomials, $x = \cos \theta$ and the functions $G_{\ell,m}^{+/-}(x)$ are a combination of the associated Legendre polynomials $P_{\ell}^m(x)$ and $P_{\ell-1}^m(x)$ and are given explicitly in Eq.~(4.19) of Ref.~\cite{Stebbins:1996wx}.
Following DES-Y3KP, we only consider the auto-correlations $w^i(\theta)$ for each tomographic bin since  in the DES Y1 and Y3 analyses it was shown that the cross correlations do not add much constraining power and  would make our analysis much more susceptible to systematic errors related to the modeling of magnification and redshift distributions~\citep{Elvin-Poole:2017xsf,y3-2x2maglimforecast}.

The angular power spectra $C(\ell)$ that enter Eqs.~(\ref{eq:xi_in_terms_of_Cell})   combine integrals over tracer distributions with astrophysical contributions from intrinsic alignments (IA), magnification, and redshift space distortions (RSD). The shear-shear (EE, BB) and galaxy-shear ($\delta_g$E)  spectra are computed using the Limber approximation and include magnification and IA, but neglect RSD. Galaxy-galaxy clustering ($\delta_g \delta_g$) is computed with via non-Limber integrals with contributions from both RSD and magnification. These calculations are described below.

Using the Limber approximation~\cite{LimberApprox} and assuming spatial flatness, the angular power spectra for for cosmic shear and galaxy-galaxy lensing can be written in a general form
\begin{equation}\label{eq:generalLimber}
     C_{\mathcal{AB}}^{ij}(\ell) = \int d\chi \frac{q_{\mathcal A}^i(\chi)q_{\mathcal B}^j(\chi)}{\chi^2}P_{\delta}\left(k = \frac{\ell+0.5}{\chi},z\right),
\end{equation}
where $\{\mathcal{A, B}\}\in \{\delta_g, \kappa\}$ and $\kappa$ is the weak lensing convergence whose contributions to shear correlations will be detailed below. Here, $P_{\delta}(k, z)$ is the three-dimensional matter power spectrum evaluated at wavenumber $k$ and redshift $z$. We use {\sc CAMB} to compute the linear $P_{\delta}(k,z)$ and \halofit~\cite{Smith:2002dz,Takahashi:2012em,Bird:2011rb} to do nonlinear modeling. The radial weight functions $q^{i}_{\mathcal{A, B}}$ are given by
\begin{equation}
\begin{aligned}
\label{eq:los_weights}
 q^i_\kappa (\chi) & = \frac{3H_0^2 \Omega_{\rm m}\chi}{2a(\chi)} \int_{\chi}^{\chi_h} d\chi^\prime \left(\frac{\chi^\prime - \chi}{\chi'}\right) n^i_\kappa(z(\chi^\prime)) \frac{d z}{d\chi^\prime} \\[0.2cm]
 q^i_{\delta_g} (\chi) & = b^i\  n^i_\delta(z(\chi)) \frac{d z}{d\chi}.
\end{aligned}
\end{equation}
Here $H_0$ is the Hubble parameter today, $\Omega_{\rm m}$ is the ratio of today's matter density to today's critical density, and $a(\chi)$ is the Universe's scale factor at comoving distance $\chi$. For conciseness, we refer Refs.~\cite{Fang:2019xat,Krause:2017ekm} for the full non-Limber expressions used for $C_{\delta_g\delta_g}^{ij}(\ell)$.

In the expression above, we adopt a linear galaxy bias model to relate the galaxy density $\delta_g$ to the matter density: $\delta_g = b^i \delta$, with $b^i$ the galaxy bias in lens redshift bin $i$, which we vary in the analysis.  
Furthermore, $n^i_{\kappa}(z)$ and $n^i_{\delta}(z)$ denote the redshift distributions of the different DES Y3 redshift bins of source and lens galaxies respectively, normalized so that $\int dz \; n^i_{\kappa, \delta}(z) = 1$. 

Contributions to observed spectra from intrinsic alignments and magnification are included as follows:
\begin{equation}\label{eq:cell}
    \begin{aligned}
C^{ij}_{\mathrm{EE}}(\ell) & = 
		C^{ij }_{\kappa \kappa}(\ell) + 
		C^{ij}_{\kappa I_{\mathrm{E}}}(\ell) +  C^{ji}_{\kappa I_{\mathrm{E}}}(\ell) + C^{ij}_{I_{\mathrm{E}} I_{\mathrm{E}}}(\ell)  \\
C^{ij}_{\mathrm{BB}}(\ell) & = C^{ij}_{I_{\mathrm{B}} I_{\mathrm{B}}}(\ell)\\
C^{ij}_{\delta_g \mathrm{E}}(\ell) & =
		C^{ij \prime}_{\delta_g \kappa}(\ell) + 
		C^{ij \prime}_{\delta_g I_{\mathrm{E}}}(\ell) + 
		{\magcoeffi}{\magcoeffj} C^{ij}_{\kappa_g \kappa_g}(\ell) +
		{\magcoeffi} C^{ij}_{\kappa_g I_{\mathrm{E}}}(\ell)
	    \\
C^{ii}_{\delta_g \delta_g}(\ell) & =
		C^{ii \prime}_{\delta_g \delta_g}(\ell) + 
		2 { \magcoeffi} C^{ii}_{\delta_g \kappa_g}(\ell) + 
		{\magcoeffi}^2 C^{ii}_{\kappa_g \kappa_g}(\ell).
\end{aligned}
\end{equation}
Here $I_{\mathrm{E/B}}$ refers to the E/B-modes of the intrinsic alignment (IA), the prime denotes non-magnified power spectra, $\kappa_g$ refers to the convergence of lens galaxies (the corresponding power spectrum is computed using the first line of equation \ref{eq:los_weights} replacing $n_{\kappa}^i(z)$ by $n_{\delta}^i(z)$) and ${\bf \magcoeff_{i}}$ are  magnification constants. We emphasize that compared to the calculations that we employ in practice, the expression for galaxy clustering in \eq{eq:cell} neglects contributions from RSD. These RSD contributions, which depend on  the linear growth factor $f(\chi)$, are incorporated --- along with magnification --- in the non-Limber calculation of  $C^{ii}_{\delta_g \delta_g}(\ell)$.

We now describe the IA and magnification effects, as well as their modeling, in more detail. IA refers to the fact that galaxies tend to align because of their gravitational environment, thus contributing to the cosmic shear signal. 
Here, we adopt the non-linear tidal alignment model as our fiducial IA model \cite{hirata04,bridle07}. The NLA model assumes that intrinsic galaxy shapes are linearly proportional to the fully nonlinear tidal field, calculated using the nonlinear power spectrum. While this ansatz is not a fully consistent nonlinear model, it is straightforward to calculate and has been shown to more accurately describe observed IA than linear theory (see, e.g., Ref.~\cite{blazek15}).
Our fiducial NLA model has two free parameters, $a$ and $\eta$, which control the amplitude and redshift dependence of IA, respectively.
IA includes both gravitational lensing--intrinsic (GI) and intrinsic--intrinsic (II) contributions, whose power spectra are then given by
\begin{equation} \label{eq:theory:nla}
    P_{\rm GI}(k,z) = A_1(z) P_{\delta}(k,z),\quad
    P_{\rm II}(k,z) = A_1^2(z) P_{\delta}(k,z).
\end{equation}
The pre-factor $A_1(z)$ is
\begin{equation} 
\label{eq:AIA}    
A_1(z) = -a \bar{C}_{1} \frac{\rho_{\rm crit}\Omega_{\rm m}}{D(z)} \left(\frac{1+z}{1+z_{0}}\right)^{\eta} \,,
\end{equation}
where $D(z)$ is the linear growth factor normalized to be equal to $(1+z)^{-1}$ at high redshifts, $\rho_{\rm crit}$ is the critical density and $\bar{C}_1$ is a normalisation constant, by convention fixed at
$\bar{C}_1=5\times10^{-14}M_\odot^{-1} h^{-2} \mathrm{Mpc}^3$.
The IA angular power spectra are then computed using \eq{eq:generalLimber} with the kernel $q^i_{\rm I} (\chi) =  n^i_\kappa(z(\chi)) \, dz/d\chi$ (see Ref.~\cite{y3-cosmicshear2} for more detailed discussion of the IA modeling and implementation for DES Y3 cosmic shear).

Our decision to adopt the NLA model contrasts with the DES-Y3KP \lwcdm analysis, 
which adopted a more complicated tidal alignment and tidal torquing (TATT) IA model. The systematic tests carried out prior to the DES-Y3KP analysis motivated the use of the TATT model because when analyzing  synthetic data containing tidal torquing effects of a size allowed by DES Y1 constraints, cosmological constraints using the simpler NLA model were found to be  biased. However, the \lcdm analysis of the \ythree data has subsequently shown preference for a generally lower amplitude of intrinsic alignments, finding that the NLA model is sufficient for unbiased modeling at the Y3 precision level.  With the benefit of these \lcdm\ results, and the desire to limit the number of nuisance parameters in our extended-model analysis, we thus opt for the NLA model. We do however run additional chains that use TATT, and prior to unblinding we check  if there is a preference for the TATT model over NLA in any of the beyond-\lcdm models.
We use the \fastpt code \cite{McEwen:2016fjn} implemented in \cosmosis in order to compute both the NLA and TATT contributions, unless specified otherwise.

As noted above, we include the contribution to galaxy clustering from the magnification of the lens sample density in Eq.~(\ref{eq:cell}) using magnification constants $\magcoeffi$. These constants are determined by the selection function of the lens sample tomographic bin such that the magnified number density is related to the convergence experienced by lens galaxies through: $\delta^i_{g,\text{mag}} =\magcoeffi \kappa^i $. 
The prime in Eq.~(\ref{eq:cell}) indicates the power spectrum unmodified by magnification.  We fix the coefficients $\magcoeffi$ to values indicated in Table~\ref{tab:params}, which were determined in Ref. \cite{y3-2x2ptmagnification} using the Balrog image simulations \cite{y3-balrog}. Note  that in \ref{sec:analysis} we test the sensitivity of our cosmology results to inaccuracies in these assumed values.

The baseline analysis of DES-Y3KP includes a shear ratio likelihood. This quantity incorporates the ratio between $\gamma_t$ measurements with the same lens bin and different shear bins \cite{y3-shearratio}. While previously studied in the context of constraining dark energy models \cite{Jain:2003tba}, it has more recently been found that shear ratio's particular strength is its sensitivity to the redshift distribution of source galaxies \cite{Prat:2017goa}.
In all model extensions other than \npg, we do not include this shear ratio likelihood. Recall that the motivation for including shear ratio is to add additional geometric constraining power which for instance helps reduce photometric redshift uncertainties. 
However, simulated analyses for our extended model showed that the inclusion or not of the shear ratio likelihood had a minimal impact on \mpp constraints. Given this, and the lack of extended modeling validation of that likelihood, we have opted to not include it as part of our baseline analysis.

\subsection{Scale cuts}\label{sec:fidscuts}

As in the DES-Y3KP analysis, we define scales below which we remove measurements from our analysis to mitigate the limits of the \mpp modeling. Modeling uncertainties of measurements at small angular scales may otherwise lead to systematic biases in cosmological parameter estimates. 
We refer to this approach as scale cuts. In the end the likelihood calculation in \eq{eq:like} only uses 2PCF measurements that remain after such cuts.
Our baseline scale cuts are the same as those used for the \lcdm analysis of DES-Y3KP.
As described in detail in the DES Y3 methods paper \cite{y3-generalmethods}, these cuts were defined based on the iterative analysis of synthetic data. Specifically, that data was  a theoretical prediction of the 2PCF observables that included two significant systematic effects not included in our model:  baryonic feedback effects extracted from the OWLS AGN hydrodynamic simulations and non-linear galaxy bias. By repeatedly analyzing that synthetic data while removing successively more  small-angle data points, we determined scale cuts at which the biases on \om and \seight due to each of the unmodeled systematics were below $0.3\sigma$. 
This determines the fiducial scales used for the DES Y3 \mpp analysis, where the number of data points for each of the 2PCF is summarized in the first line of \tab{tab:scuts}.\footnote{Minimum and maximum scales used after the scale cuts procedure are indicated in the \cosmosis files shared as part of the data release.} These same cuts are used for the \wcdm,  $\wowa$, $\neff$ and \npg models.

\begin{table}
\begin{tabular}{lcccccl}
    & \multicolumn{5}{c}{Data points} \\
    Scale cuts & $\xi_+$ & $\xi_-$ & $\gamma_t$ & $w$ & Total & Used for extended models  \\\hline
    Fiducial  & 166 & 61 & 192 & 43  &  462 & $\wowa$, \neff, \npg  \\
    Linear & 105 & 3 & 105 &43 & 256  & $\neffmeff$, $\sigmu$ \\
    Linear+Limber &100 & 2 & 100 & 19 & 221 & \ok \\
\end{tabular}
\caption{Number of \mpp data points remaining after the different sets of scale cuts used in this analysis. The fiducial cuts are the same as those used in DES-Y3KP, linear cuts remove additional points at small scales affected by nonlinear structure growth, and linear+Limber cuts remove data points both at nonlinear scales and where non-Limber calculations are needed to accurately model large-angle galaxy clustering. Unless otherwise noted, whenever a comparison is shown between an extended model and \lcdm, the \lcdm results will use scale cuts matching those of the extended model.} \label{tab:scuts}
\end{table}

For several of the models studied in this paper, we have chosen a stricter set of scale cuts than the fiducial case. Specifically, for models with nonzero \ok, at the time of this analysis \halofit had not been sufficiently validated on non-linear scales\footnote{Ref.~\cite{PhysRevD.106.083504}, which was released when this paper was in final stages of preparation, represents a promising approach to improve this.}; for $\neffmeff$ models, \halofit is known to be not sufficiently well calibrated; and finally the $\sigmu$ tests of gravity are only well-defined on linear scales. For these classes of models, we restrict our analysis to purely linear scales (for nonzero curvature there will be an additional scale cut, discussed below). 
To determine those scales, we follow the procedure first applied in the \textit{Planck} 2015 analysis~\cite{Ade:2015rim} and followed later in DES-Y1Ext. 
We compute the difference between the nonlinear and linear theory predictions of the 2PCF in the standard $\Lambda$CDM model at a fiducial cosmology on scales left after fiducial scale cuts.
Using the respective data vector theory predictions, $\mathbf{D}_{\rm NL}$ and $\mathbf{D}_{\rm lin}$, and full error covariance of DES Y3, $ \mathbf{C}$, we
calculate the quantity
\begin{equation}
\Delta\chi^2 \equiv (\mathbf{D}_{\rm NL}-\mathbf{D}_{\rm lin})^{\rm{T}}\,
\mathbf{C}^{-1}\,(\mathbf{D}_{\rm NL}-\mathbf{D}_{\rm lin})
\end{equation}
and identify the single data point that contributes most to this quantity. We
remove that data point, and repeat the process until $\Delta\chi^2<1$. This constitutes our set of linear scales, used for \neffmeff and \sigmu. The resulting linear scale cuts lead to a \mpp data vector of 256 elements (see second line of Table~\ref{tab:scuts}).

The third and final choice of scale cuts removes both scales which are impacted by nonlinear structure growth, and those requiring non-Limber projection calculations to accurately model galaxy clustering.
This is relevant for the \ok analysis, because the accuracy of the fast non-Limber method~\cite{Fang:2019xat} used to model large-angle galaxy clustering has not been tested for $\ok\neq0$. 
The procedure to identify these scales is the same as that described for linear scale cuts  above, except that the synthetic data vector $\mathbf{D}_{\rm lin}$ is replaced with $\mathbf{D}_{\rm lin+Limb}$, which, in addition to having only linear modeling for the matter power spectrum, is computed using the  Limber approximation at all angular scales. The linear+Limber cuts are thus slightly more stringent than the linear-only cuts.  \tab{tab:scuts} shows that the resulting cuts lead to a \mpp data vector of 221 elements that are then used in the \ok analysis.  Note that the iterative nature of our scale-cut definition causes the Linear+Limber cuts to remove additional points from   $\xi_{\pm}(\theta)$ and $\gamma_t(\theta)$ compared to the Linear cuts, even though non-Limber calculations are only used for galaxy clustering $w(\theta)$ calculations.

\begin{table}
\begin{center}
\begin{tabular*}{\columnwidth}{ l  @{\extracolsep{\fill}} c  c}
\hline
\hline
Parameter & \multicolumn{2}{c}{Prior}  \\  
\hline 
\multicolumn{2}{l}{{\bf Base Cosmology}} \\
$\Omega_{\mathrm{m}}$  &  Flat  & (0.1, 0.9)  \\ 
$10^{9}A_{\mathrm{s}}$ &  Flat  & ($0.5,5.0$)  \\ 
$n_{\mathrm{s}}$ &  Flat  & (0.87, 1.07)  \\
$\Omega_{\mathrm{b}}$ &  Flat  & (0.03, 0.07)  \\
$h$  &  Flat  & (0.55, 0.91)   \\
$10^{3}\Omega_\nu h^2$  & Flat  & ($0.60$, $6.44$) \\
  &    & $0.005<\Omega_{\mathrm{b}}h^2< 0.040$  \\
\hline 
\multicolumn{2}{l}{{\bf Extended Cosmology}} \\
$w_0, w_a$ &   Flat  & $\wo\in(-3.0, -0.33)$\\
& &  $\wa\in(-3.0, 3.0)$\\
& & $\wo+\wa<0$ \\
$\ok$ &   Flat  & (-0.25, 0.25)   \\
$\neff$ &   Flat  & (1.0, 10.0)   \\
$\neff,\meff$ &   Flat  & $\neff\in(3.044,10)$\\
 & & $\meff\in(0.0,3.0)$~eV\\
$\sigo,\muo$ &   Flat  & $\sigo,\muo\in(-1.5,1.5)$   \\
 & &$ \muo > 2\sigo + 1$ \\
$\Anpg_i (i \in [2,4])$ & Flat & ($0.1,3$) \\ 
\hline
\multicolumn{2}{l}{{\bf Lens Galaxy Bias} } \\
$b_{i} (i\in[1,4])$   & Flat  & (0.8, 3.0) \\
\hline
\multicolumn{2}{l}{{\bf Lens magnification} } \\
$\magcoeff_{1}$ & Fixed &  $0.42$ \\
$\magcoeff_{2}$ & Fixed &  $0.30$ \\
$\magcoeff_{3}$ & Fixed &  $1.76$ \\
$\magcoeff_{4}$ & Fixed &  $1.94$ \\
\hline
\multicolumn{2}{l}{{\bf Lens photo-z } } \\
$\Delta z^1_{\rm l} \times 10^{2}$  & Gaussian  & ($-0.9, 0.7$) \\
$\Delta z^2_{\rm l} \times 10^{2}$  & Gaussian  & ($-3.5, 1.1$) \\
$\Delta z^3_{\rm l} \times 10^{2}$  & Gaussian  & ($-0.5, 0.6$) \\
$\Delta z^4_{\rm l} \times 10^{2}$  & Gaussian  & ($-0.7, 0.6$) \\
$\sigma^1_{z,\rm l}$  & Gaussian  & ($0.98, 0.06$) \\
$\sigma^2_{z,\rm l}$  & Gaussian  & ($1.31, 0.09$) \\
$\sigma^3_{z,\rm l}$  & Gaussian  & ($0.87, 0.05$) \\
$\sigma^4_{z,\rm l}$  & Gaussian  & ($0.92, 0.05$) \\
\hline
\multicolumn{2}{l}{{\bf Intrinsic Alignment}} \\
$a$   & Flat &  ($-5,5$) \\
$\eta$ & Flat  & ($-5,5$) \\
\hline
\multicolumn{2}{l}{{\bf Source photo-z}} \\
$\Delta z^1_{\rm s} \times 10^{2}$  & Gaussian  & ($0.0, 1.8$) \\
$\Delta z^2_{\rm s} \times 10^{2}$  & Gaussian  & ($0.0, 1.5$) \\
$\Delta z^3_{\rm s} \times 10^{2}$  & Gaussian  & ($0.0, 1.1$) \\
$\Delta z^4_{\rm s} \times 10^{2}$  & Gaussian  & ($0.0, 1.7$) \\
\hline
\multicolumn{2}{l}{{\bf Shear calibration}} \\
$m^1 \times 10^{2}$ & Gaussian  & ($-0.6, 0.9$)\\
$m^2 \times 10^{2}$ & Gaussian  & ($-2.0, 0.8$)\\
$m^3 \times 10^{2}$ & Gaussian  & ($-2.4, 0.8$)\\
$m^4 \times 10^{2}$ & Gaussian  & ($-3.7, 0.8$)\\
\hline
\multicolumn{2}{l}{{\bf External data}} \\
$\tau$ (\textit{Planck}) & Flat & ($0.01,0.8$) \\
$A_{\rm{P}}$ (\textit{Planck}) & Gaussian & (1.0,0.0025) \\
$M$ (SN) & Flat &  ($-20,-18$)\\
\hline
\end{tabular*}
\end{center}
\caption{Parameters and priors describing the baseline cosmology, extended models and nuisance parameters used in this analysis. 
We quote the lower and upper limits of flat priors and the mean and standard deviation of Gaussian priors.
The parameter $w$ is fixed to $-1$ for all models other than \wcdm and  $\wowa$, and for \wcdm it uses the same prior as for \wo.  
For the $\neffmeff$ model we fix the sum of active neutrino masses to $0.06$~eV.
}
\label{tab:params}
\vspace{-1.2cm}
\end{table}

\subsection{Parameter space}\label{sec:baseparams}

The parameters and their priors used in our baseline analysis match those of DES-Y3KP. The cosmological parameters are
\begin{equation}
    \mathbf{\Theta}_{\rm base}=\{\om, \ob, \As, n_{\rm s}, h,  \omnuhh\}
\end{equation}
(equivalently, $\omnuhh$ can be replaced by $\summnu$), with the neutrinos modeled as three degenerate species with equal masses. The priors on these parameters are listed in the top section of   \tab{tab:params}. 
We adopt an additional 
prior requiring $0.005<\Omega_{\rm b} h^2<0.040$. This baryon-density prior is introduced because we include a Big Bang Nucleosynthesis (BBN) consistency condition which imposes a relation between the physical baryon density $\Omega_{\rm b}h^2$, the relativistic degrees of freedom $\neff$, and the Helium abundance $Y_{\rm He}$ \cite{Pisanti2008}.\footnote{\url{http://parthenope.na.infn.it/}} This consistency relation only alters calculations when \neff is varied, but introduces the $\Omega_b h^2$ prior for all models because it relies on a table defined for a finite range of physical baryon density and so rejects samples outside that range.  

We also vary a number of nuisance parameters to describe systematic effects.
The intrinsic alignment is described by two parameters, $a$ and $\eta$, and the linear galaxy bias by one parameter $b_i$ for each of the lens bins; these parameters are assigned flat priors.
Additionally, each lens bin has two nuisance parameters: one that controls the mean of the photometric redshift distribution in redshift bin $i$, $\Delta z^i_{\rm l}$,  and another which stretches or compresses the $n(z)$ distribution in $z$, $\sigma^i_{z,\rm l}$. Each of the four source bins has a photometric redshift uncertainty parameter $\Delta z^i_{\rm s}$, as well as a shear calibration parameter $m^i$. 

As indicated at the bottom of \tab{tab:params}, we also vary the optical depth $\tau$ along with a number of nuisance parameters associated with the \textit{Planck} likelihood, as described in \sect{sec:Planck}, when using \textit{Planck} data. 
When the Pantheon supernovae likelihood is included, we additionally sample over the absolute magnitude of the supernovae $M$, as described in \sect{sec:pantheon}.

\section{Beyond-\texorpdfstring{\lcdm}{Lg} models}
\label{sec:extmodels}
We now introduce the beyond-\lcdm models constrained in this paper.
For each, we introduce its physics and parameterization, and describe any alterations required to the baseline approach for modeling \mpp observables and performing parameter estimation. 


\subsection{Dark energy: \texorpdfstring{\wowa}{Lg}}\label{sec:w0wa}

We use the phenomenological model proposed in~\cite{Linder_wa,Chevallier:2000qy} , often referred to as the CPL model, for a time-varying dark energy equation of state:
\begin{equation}
    w(a) = \wo + (1-a)\wa,\label{eq:wowa}
\end{equation}
where $a$ is the scale factor, and $w_0$ and $w_a$ two new parameters. This is the most commonly considered parameterization of dark energy equation of state with more than one parameter, and has been shown to provide a good fit to a number of dynamical dark energy models that have a more complete physical description \cite{Linder_wa}.
    
We will also report constraints on the value of $w(a)$ at the so-called pivot redshift \cite{Huterer:2000mj} $z_{\rm p}= a_{\rm p}^{-1} -1$, $w_{\rm p}\equiv w(a_{\rm p})$. Here $a_{\rm p}$ is the scale factor where we have the strongest constraints on $w(a)$, and therefore where the value of the equation of state and its derivative with respect to the scale factor are uncorrelated. Using this parameter, \eq{eq:wowa} can be rewritten as 
\begin{equation}
    w(a) = w_{\rm p} + (a_{\rm p}-a)\wa.
\end{equation}
The value of the pivot scale factor is determined using the marginalized parameter covariance as 
\begin{equation}
    a_{\rm p} = 1+\frac{\mathbf{C}_{\wo\wa}}{\mathbf{C}_{\wa\wa}}.
\end{equation}

In the flat $\wowa$ model, the expansion rate becomes
\begin{equation}
\frac{H^2(a)}{H^2_0} = \om a^{-3} + (1-\om) a^{-3(1+\wo +\wa)}e^{-3\wa(1-a)}
\end{equation}

As described in \app{sec:w0wa_casarini_check}, we performed additional validation tests to ensure that the use of \halofit in our calculation of the nonlinear matter power spectrum is valid for the $\wowa$ model. We use the fiducial scale cuts for this model (see \tab{tab:scuts}).


\subsection{Curvature: \texorpdfstring{\ok}{Lg}}\label{sec:ok}

To define the curvature density \ok, it is most convenient to start from the Friedmann--Lema\^{i}tre--Robertson--Walker (FLRW) metric in the form 
\begin{equation}\label{eq:curvedflrw}
    ds^2 = -dt^2 + a^2\left[d\chi^2 +r^2\left(\chi\right)\left(d\theta^2 + \sin^2\theta d\phi^2\right)\right],
\end{equation}
where $\chi$ is the comoving  (radial) distance and $a$ is the scale factor. Then  the  angular diameter  distance $r\left(\chi\right)$ is defined as:
\begin{equation}
r(\chi) =  \left\{  \begin{array}{cl}
  K^{-1/2}\sin{\left (K^{1/2}\chi\right )}& \mbox{for}\quad K>0  \\[0.2cm]
\chi  & \mbox{for}\quad K=0 \\[0.2cm]
 |K|^{-1/2}\sinh{\left(|K|^{1/2}\chi\right)}   & \mbox{for}\quad K<0,
\end{array} \right .    
\end{equation}
where $K$ is the curvature term. A positively curved space has $K>0$, negatively curved corresponds to $K<0$, and flat space has $K=0$.
With the curvature of arbitrary sign, the expansion rate can be written as
\begin{equation}
\frac{H^2(a)}{H^2_0} = \om a^{-3} + (1-\om-\ok) + \ok a^{-2}\;, 
\end{equation}
where $\Omega_k=-K/H_0^2$. It then follows that  $\ok<0$ corresponds to positive spatial curvature, and $\ok>0$ to negative. 

As noted in Sec.~\ref{sec:fidscuts}, for the curved-universe analysis, due to lack of validated modeling we use a conservative set of scale cuts which avoid both  nonlinear scales and the large angular scales where the non-Limber calculation is adopted to model galaxy clustering.\footnote{This is a more conservative choice than the approach taken in DES-Y1Ext, where \ok constraints used the fiducial scale cuts as the Limber approximation was used at all scales.} For the angular scales where the Limber approximation is used, we apply the commonly-used angular-diameter rescaling approximation for the impact of curvature on line-of-sight projection, which replaces $\chi$ in \eq{eq:generalLimber} with the angular diameter distance $r(\chi)$.


\subsection{Extra relativistic degrees of freedom: \texorpdfstring{\Neff}{Lg} } \label{sec:neff}

We next consider a model that allows for new radiative degrees of freedom in the early Universe, described by the parameter \neff. 
This parameter relates contributions to the energy density in radiation in the early Universe from relativistic species to that of photons via
\begin{equation}
    \rho_{\rm rad} = \left[1+ \neff\frac{7}{8}\left(\frac{T_{{\rm \nu},0}}{T_{\gamma,0}}\right)^{4}\right]\rho_{\gamma}.\label{eq:rad_energy_density}
\end{equation} 
Here $\rho_{\rm rad}$ and $\rho_{\gamma}$ are the co-moving energy densities of radiation and  photons after electron--positron annihilation. In the standard cosmological model, all contributions to \neff  come from neutrinos and its value is $\neff=3.044$,  corresponding to three neutrino species plus small corrections due to their non-instantaneous decoupling from photons~\cite{Bennett:2020zkv,Akita:2020szl}. 

We capture the effects of \Neff by using {\sc CAMB}'s predictions for its impact on the expansion history and power spectra, using a modified version of the \cosmosis~{\sc CAMB} interface.  We set the {\sc CAMB} parameters so that each of the three massive neutrino species is assigned a degeneracy~\cite{cambnotes:2014} of $\tfrac{1}{3}\neff$. This means that varying \neff has a continuous effect on the neutrino temperature $T_{\nu}$, with $\Delta\neff=0$ corresponding to the temperature if there are no additional relativistic species beyond the standard model.
We do not apply any other modifications to the fiducial model. 

\subsection{Massive sterile neutrinos: \texorpdfstring{$\neffmeff$}{Lg}}\label{sec:neffmeff}

We additionally constrain the properties of a light relic particle with non-zero mass, modeled as single species of thermal sterile neutrino. The properties of the sterile neutrino are controlled by the parameters \neff and \meff. The impact of sterile neutrinos on CMB observables is fairly similar to that of varying \neff alone, while its impact on large scale structure has  a richer phenomenology. Like active neutrinos, sterile neutrinos suppress large scale structure formation at scales smaller than a free-streaming length scale ($k>k_{\rm fs})$, with the magnitude of that suppression at high $k$ controlled by their contribution to cosmological energy density $\Omega_{{\nu_{\rm s}}}$. The free-streaming scale is set by both the particle's physical mass and temperature, and the relationship between those properties and the parameters \neff and \meff depend on the specifics of the model considered. In this analysis, we choose to model the sterile neutrino as a thermal relic, that is, a stable particle species which was once in thermal equilibrium with standard model particles but decoupled at an early time. With this assumption, the particle's physical mass is 
$m_{\rm th} = \meff (\Delta\neff)^{-3/4}$ and in linear theory the
free-streaming scale is~\cite{Xu:2021rwg}
\begin{equation}
    k_{\rm fs} = \frac{0.8 h{\rm Mpc}^{-1}}{\sqrt{1+z}}\left(\frac{\meff}{(1 {\rm eV})\Delta\neff}\right).
\end{equation}
While this thermal model is just one of several possible choices one could make for describing sterile neutrinos, our constraints will represent a more general search for new physics.
As is discussed in  Refs.~\cite{Xu:2021rwg,Munoz:2018ajr,Colombi:1995ze,Lesgourgues:2013sjj}, this  kind of two-parameter $\neffmeff$ model is sufficient to perform a generic search for a population of stable, non-interacting massive relic particles. 
\cite{Colombi:1995ze,Lesgourgues:2013sjj}.

Here the parameter $\Delta\neff\equiv\neff-3.044$  determines the temperature of the sterile neutrino, which is related to the standard model temperature of active neutrinos via $T_{\nu_{\rm s}} = (\Delta\neff)^{1/4} T_{\nu_{\rm a}}$.  Thus, $\Delta\neff=1$ corresponds to a sterile neutrino that thermalizes at the same temperature as the active neutrinos, while lower \neff  means the sterile neutrinos are colder.  
The parameter \meff is an effective mass which captures how the sterile neutrino contributes to the cosmological energy densities, defined so that
\begin{equation}
    \Omega_{{\nu_{\rm s}}}h^2 = \frac{\meff}{ 94.1{\rm eV}}. 
\end{equation}
When we consider sterile neutrinos the conversion factor between the particle mass and $\Omega_{\nu}h^2$ is slightly different from the 93.14eV value used for active neutrinos. This is because sterile neutrinos are assumed  
not to be affected by electron--positron annihilation in the same way as active neutrinos. Note that both versions of the $\Omega_{\nu}$-to-mass conversion factor encode a number of standard model assumptions which cannot be disentangled from cosmological constraints on neutrino mass. Our measurements thus serve as both a test of the mass of neutrinos and of those assumptions. 
As with the \neff model described above, we  use {\sc CAMB} along with a modified version of the  \cosmosis~{\sc CAMB} interface to compute the impact of the sterile neutrinos on expansion history and the linear matter power spectrum. 

We assume that active neutrino temperatures are at their standard model  value in the instantaneous decoupling approximation, $T_{\nu_{\rm a}}=(4/11)^{1/3}T_\gamma$, and following the \textit{Planck} 2018 cosmology analysis~\cite{Planck:2018vyg}, we fix the active neutrino mass to the minimum allowed by neutrino oscillation experiments, $\summnu=0.06$ eV. 
Additionally, because the presence of massive light relics like sterile neutrinos complicates the modeling of the nonlinear matter power spectrum as well as galaxy bias \cite{Xu:2021rwg,Munoz:2018ajr,LoVerde:2014pxa,Aviles_2021} and there are not readily available tools to account for the impact of sterile neutrinos on nonlinear power spectrum modeling (see e.g.\ Refs.~\cite{Brandbyge:2017tdc,Banerjee:2022}),
when constraining $\neffmeff$ we restrict our analysis to linear scales using the scale-cut procedure described in Sec.~\ref{sec:fidscuts}. 

Note that our fiducial prior has a lower bound of $\Delta\neff=0$, which means our parameter space will include the small-$\Delta\neff$ regime where the sterile neutrino will be indistinguishable from cold dark matter.  
As we will find in \sect{sec:changepipe}, including this unconstrained region makes parameter estimation  more susceptible to  projection effects and thus less robust to the details of nuisance parameter marginalization and the data's noise realization. Given this, in order to obtain a more robust set of constraints and to allow more direct comparison with other studies,  we report $\neffmeff$ constraints using two alternative priors: one where the lower bound of the prior is raised to require $\Delta\neff>0.047$, corresponding to the minimum temperature for a fermion relic particle that was ever in thermal equilibrium with standard model particles~\cite{Xu:2021rwg}, as well as the same model-specific prior used in {\it Planck} analyses, requiring  $m_{\rm th} \leq 10$ eV. 

\subsection{Test of gravity on cosmological scales: \texorpdfstring{$\sigmu$}{Lg}}\label{sec:sigmu}

We test gravity on cosmological scales by adopting the common $\Sigma$, $\mu$ phenomenological parameterization proposed and developed in Refs.~\cite{Caldwell_2007,amendola:2008,hu:2007,Jain:2007yk,daniel:2008,bertschinger:2008,Zhao_2009,pogosian:2010,Zhao:2010dz,Silvestri:2013ne}. This model has recently been tested using CMB measurements by the {\it Planck} satellite and weak lensing data from surveys such as CFHTLens, KiDS and DES in Refs.~\cite{Simpson:2012ra,Ferte:2017bpf,joudaki_bsm_2017,Ade:2015rim,Planck:2018vyg,Abbott:2018xao}. 
In this approach, deviations from the gravitational physics described by General Relativity (GR) are introduced through modifications to the Poisson and lensing equations which then take the following form in Fourier space: 
\begin{equation}
    \begin{aligned}
\label{eq:sigmu_poisson}
    k^2\Psi &= -4\pi G a^2 \left[1+\mu(a,k) \right] (\rho \delta + 3(\rho + P) \sigma),\\
    k^2 \Phi &= - 4\pi G a^2 
    \left[1+\Sigma(a,k)\right] (2\rho \delta + 3(\rho + P) \sigma).
\end{aligned}
\end{equation}
Here $\Psi$ is the Newtonian gravitational potential, which determines the gravitational interactions of massive particles, $\Phi$ is the Weyl potential with which massless particles interact gravitationally, $\delta$ corresponds to density fluctuations in the comoving gauge, $\rho$ to matter density, and $(P+\rho)\sigma$ to the fluid anisotropic stress potential.  
The functions $\Sigma(a,k)$ and $\mu(a,k)$ represent deviations from GR, with $\Sigma = \mu = 0$ recovering the predictions of GR. This parameterization is equivalent to modifications to the gravitational constant $G$, and $\Sigma$/$\mu$ are sometimes denoted as $G_{\Phi/\Psi}$ respectively.

We assume a time dependence following the energy density of the effective dark energy in units of the  critical density  $\Omega_{\Lambda}(a)$ normalized by its value today $\Omega_{\Lambda,0}$, as done previously in Refs.~\cite{Caldwell_2007,Simpson:2012ra,Ferte:2017bpf,Abbott:2018xao,Planck:2018vyg}: 
\begin{equation}
    \begin{aligned}
\label{eq:sigmudef}
    \Sigma(a,k) &= \sigo \frac{\Omega_{\Lambda}(a)}{\Omega_{\Lambda,0}}, \\
    \mu(a,k)    &= \muo \frac{\Omega_{\Lambda}(a)}{\Omega_{\Lambda,0}}.
    \end{aligned}
\end{equation} 
This parameterization is designed to be sensitive to deviations from GR that are associated with cosmic acceleration. As is pointed out in e.g.~Ref.~\cite{Garcia-Quintero:2020bac}, these assumptions may cause our \sigmu parameterization to lack sensitivity to some modified gravity signals that could be captured by searches with less restrictive assumptions. However, the parameterization of \eq{eq:sigmudef} has the benefit of adding few new parameters, which makes it easier to constrain them robustly. Variations of the ($\Sigma$,$\mu$) model with alternative assumptions about the time and scale dependence of deviations from GR will be explored in a follow-up paper \cite{y3-mg}.

This  phenomenological approach is defined only in linear theory, while possible approaches to define similar functional forms of deviations from GR on all scales have been proposed e.g. in Ref.~\cite{nl_nosim}, allowing the use of halo-model based approaches as proposed in Refs.~\cite{demg_halomodel,fr_halomodel} for ($\Sigma$, $\mu$) models.
However these methods have not yet been tested for the parameterization of ($\Sigma$, $\mu$) considered here, so we restrict our analysis of DES Y3 \mpp measurements to linear scales by imposing scale cuts as described in \sect{sec:fidscuts}.
 
In order to model the impact of \sigmu on the 2PCF, we modify the \cosmosis baseline pipeline to use the Weyl potential power spectrum $P_{\Phi \Phi}(k)$ when computing weak lensing observables. 
This is in contrast to the fiducial analysis, which assumes the Poisson equation: 
\begin{equation}
    k^2\Phi = \frac{3}{2} \Omega_{\rm m} H_0^2 \delta /a. 
\end{equation} 
Although the impact of $\sigo$ can be computed simply modifying the lensing kernel used for 2PCF computations in \eq{eq:generalLimber} (as was done in DES-Y1Ext), we choose to use the Weyl potential directly as it facilitates more flexible applications to other parameterizations of modified gravity and new physics affecting growth, as used in e.g.~Refs.~\cite{dm2dr,y3-mg}.

To model \mpp observables, we need both the Weyl potential auto-correlation $P_{\Phi \Phi}(k,z)$ and its correlation with the matter density $P_{\Phi \delta}(k,z)$. 
We compute their linear predictions using \mgcamb v3.0\footnote{\url{https://github.com/sfu-cosmo/MGCAMB}.} \cite{Zucca:2019}, modifying its interface with \cosmosis.
The corresponding non-linear spectra are then obtained using a non-linear scaling factor:  
\begin{equation}
        P_{\Phi \Phi}^{\rm NL}(k,z) = \frac{P_{\delta \delta}^{\rm NL}(k,z)}{P_{\delta \delta}^{\rm L}(k,z)} P_{\Phi \Phi}^{\rm L}(k,z),
\end{equation}
where the NL and L superscripts refer respectively to the \halofit non-linear and linear predictions of $P(k,z)$ and we use the same non-linear boost to get the cross-power spectrum $P_{\Phi \delta}^{\rm NL}(k,z)$. 

We  modify the \lcdm modeling pipeline so that power spectra of fields derived from the Weyl gravitational potential, namely the convergence $\kappa$ and  magnification, are computed directly using the projected Weyl potential auto- and cross-power spectra. 
The angular power spectra $C(\ell)$ are computed using a version of \eq{eq:generalLimber} with $k^4P_{\Phi\Phi}(k,z)$ replacing $P_{\delta \delta}(k,z)$ for $C_{\kappa \kappa}(\ell)$.  
Similarly $k^2P_{\Phi\delta}(k,z)$ replaces $P_{\delta\delta}(k,z)$ for $C_{\kappa \delta_g}(\ell)$. 
In this formulation, the lensing kernel from \eq{eq:los_weights}  instead reads:  
\begin{eqnarray}
\label{eq:los_weights_weyl}
 q^i_\kappa (\chi) &=& \chi \int_{\chi}^{\chi_h} d\chi^\prime \left(\frac{\chi^\prime - \chi}{\chi'}\right) n^i_\kappa(z(\chi^\prime)) \frac{d z}{d\chi^\prime}.
\end{eqnarray}
Appropriate adjustments must also be made for the modeling of galaxy clustering to account for contributions from magnification as shown in \eq{eq:cell}. Additionally, we compute the GI NLA intrinsic alignment contributions by modifying \eq{eq:theory:nla} such that:
\begin{equation} \label{eq:theory:nla_w}
    P_{\rm GI}(k,z) = A_1(z) k^2 P_{\Phi \delta}(k,z),
\end{equation}
used to compute $C_{\kappa I}(\ell)$ using the lensing kernel in \eq{eq:los_weights_weyl}.  
In a fully rigorous treatment, the modified Newtonian potential $\Psi$ should determine the alignments of galaxies' intrinsic shapes.
However, we choose to model the tidal alignment contributions (corresponding to the I term in the GI and II power spectra of Eqs.~(\ref{eq:theory:nla}) and (\ref{eq:theory:nla_w})) using the matter power spectrum $P_{\delta \delta}(k,z)$ modified by $\mu$, by neglecting the impact of anisotropic stress.  
The angular power spectra $C(\ell)$ of \eq{eq:cell} computed with the Weyl gravitational potential are then converted into real-space 2PCFs $\xi_{\pm}(\theta)$, $\gamma_t(\theta)$, and $w(\theta)$ following the same procedure as in \lcdm.
 
We checked that this modified \cosmosis pipeline reproduces \lcdm results, with negligible shifts in parameter estimation, as shown in Appendix~\ref{app:Weyl-validation}.
We note that the matter power spectrum $P_{\delta \delta}(k,z)$ computed by \mgcamb shows an unexpected dependence on $\Sigma_0$ at large scales, for $k < 10^{-2} \,{\rm Mpc}/h^{-1}$. 
This dependence leads in turn to a slight dependence of the clustering 2PCF $w(\theta)$ on $\Sigma_0$ for $\theta$ above 100 arcmin, more significantly for the highest redshift bins. 
Its impact on $\Sigma_0$ constraints is however negligible at DES Y3 \mpp sensitivity, with a change in the posterior value computed with simulated clustering measurements alone of 0.3$\%$ for $\Sigma_0$ = 1.5 compared to GR.

The \textsc{camb} \texttt{dverk} routine fails due to \mgcamb implementation of the evolution of perturbations, for a large set of \sigmu values satisfying
\begin{equation}\label{eq:sigmu_prior}
\mu_0 > 2\Sigma_0 + 1.
\end{equation} 
We thus impose a prior excluding this region of parameter space.  

As opposed to other cosmological models, we will not test for consistency of \sigmu results against an alternative IA model such as TATT. 
Although its use has not been fully validated against simulations for instance in non-GR theories, the NLA model allows for amplitude and redshift dependences and propagation of \sigmu modification of gravity to IA as described above.
We therefore adopt the NLA model as in other models and as in previous \sigmu studies such as DES-Y1Ext,\cite{joudaki_bsm_2017,Ferte:2017bpf}.
However, to get the next-order terms, the perturbative derivation of the TATT model  
in \cite{PhysRevD.100.103506} assumes GR and would need to be re-derived to capture the tidal physics in modified gravity theories with a similar level of generality.
Adopting the TATT model for the \sigmu analysis would amount to using a GR IA model which would accurately  
describe IA in the case of $\sigmu$ consistent with 0 but could potentially bias results if not.
Additionally, we do not make use of the \fastpt algorithm to predict the IA and galaxy bias models, and instead use the simple linear galaxy bias (as validated in \cite{y3-generalmethods}) and the IA NLA model using the matter and Weyl power spectra directly as computed by \mgcamb. 
We note that we do not use the linear alignment (LA) model, in which case the IA signal is sourced by the linear Weyl and matter power spectra: indeed, although non-linear scales are removed through the scale cuts procedure described in Section~\ref{sec:fidscuts}, we model non-linearities using \halofit so that non-linear information left over after scale cuts is modeled.

\subsection{Binned \texorpdfstring{$\sigma_8(z)$}{Lg}}\label{sec:npg}
Finally, we test \lcdm predictions of the evolution of structure growth without assuming a particular physical mechanism by using what we will refer to as the \npg model. This model introduces amplitudes $\Anpg_i$ which scale the linear matter power spectrum in redshift bins $i$ associated with our lens galaxy sample. For our fiducial \maglim lens sample, the edges of the redshift ranges used to define tomographic bins are [0, 0.4, 0.55, 0.7, 1.5].  
In other words, when we perform our model calculations, in the range $z\in[0,0.4)$ we multiply the linear matter power spectrum  by $\Anpg_1$,  in the range $z\in[0.4,0.55)$ we multiply $P_{\rm lin}(k)$ by $\Anpg_2$, and so on.  As a model-agnostic test of \lcdm growth history, this is in a sense a successor to the growth-geometry split analysis of Ref.~\cite{Muir:2020puy}. 

Because it introduces step functions in the linear growth factor, this parameterization implies delta function features in the linear growth rate $f(z)$. These spikes have no impact on the external RSD modeling because none of the \fsig measurements in that likelihood fall on our $z$-bin boundaries. We neglect their effect on RSD contributions DES galaxy clustering.\footnote{A fully consistent calculation here would account for the  enhanced RSD contributions to $w(\theta)$. This effect could in principle be used to empirically constrain the smoothness of the linear growth factor's redshift evolution, but we neglect it for simplicity because it's impact on overall constraining power is likely to be small, and implementing it would  require significant updates to our modeling pipeline.}

In practice we fix $\Anpg_1=1$ and sample over $\Anpg_{2{\rm -}4}$. Because our measurements are sensitive to the products $\As \Anpg_i$, where $\As$ is the primordial power spectrum amplitude, if we varied all four $\Anpg_i$ amplitudes the parameters would be completely degenerate with $\As$. By fixing\footnote{The choice to fix the amplitude for the lowest redshift  $i=1$ as opposed to some other bin was arbitrary. While a different choice would affect the inferred values of the $\Anpg_i$ parameters,  the resulting model would  have the same degrees of freedom, so  would  not affect the physical interpretation of the results --- i.e.~the inferred \sigeight values.} $\Anpg_1$, we thus avoid those degeneracies and our parameterization of \npg reads: 
\begin{equation}
\Theta_{\sigma_8(z)}\in \{
\Anpg_1\equiv 1,  \Anpg_2,  \Anpg_3,  \Anpg_4\}. \label{eq:npg_params}  
\end{equation} 
This parameterization lends itself to a physical interpretation: $\As$ controls the amplitude of structure observed in the lowest redshift lens bin, while the \npg amplitudes provide a consistency test of whether the time-evolution of the growth of structure is consistent with the \lcdm prediction 
\begin{equation}
    \Anpg_2=\Anpg_3=\Anpg_4=1
    \quad\mbox{(\lcdm)}.
\end{equation} 
We will additionally report constraints on a set of derived parameters,\footnote{Note that the decision to include constraints on the derived \npgsig{i} parameters as part of the presentation of our \npg model results was made after unblinding.}
\begin{equation}\label{eq:news8}
\npgsig{i} \equiv \sigeight \sqrt{\Anpg_i},
\end{equation}
which correspond to the value of \sigeight expected at redshift $z=0$ based on the amplitude of structure in redshift bin $i$.

When we include both \textit{Planck} and low-redshift measurements of structures (from DES or external RSD data), we treat the CMB measurements as an additional high-$z$ bin, and introduce an additional amplitude $\Anpg_{\rm CMB}$. In practice, we implement this by passing the product $\Anpg_{\rm CMB}\As$ as the \As input to {\sc CAMB} when computing CMB observables.
To be fully self-consistent, the amount of that lensing smoothing of the CMB power spectra should account for the modulation of the line-of-sight matter power spectrum by the $\Anpg_i$ parameters. For simplicity, we have chosen not to model this. Instead, when we include CMB constraints for the \npg model, we marginalize over the lensing smoothing amplitude $A_{\mathrm{L}}$~\cite{Calabrese:2008rt} (matching the  parameter used in Planck analyses) in order to remove late-time LSS information from the CMB likelihood. 
We neglect the dependence of the Integrated Sachs-Wolfe effect on late-time modified growth, as sensitivity to the effect is limited by cosmic variance.

Note that the phenomenology probed by this $\Anpg_i$ parameterization is to  similar to  a $\sigmu$ description of modified gravity with fixed $\Sigma=0$ and $\mu(z)$ defined as a piecewise function of $z$. It is therefore comparable to models studied in e.g.\ Refs.~\cite{Garcia-Quintero:2020,Linder:2020xza,Garcia-Quintero:2020bac}.  The distinction between our \npg model and modified gravity parameterizations is largely one of interpretation rather than modeling specifics. Here we are framing \npg as  a consistency test of \lcdm rather than a physical model, so we use the same modeling choices as in \lwcdm, including fiducial scale cuts and  \halofit as our model for the nonlinear matter power spectrum, in contrast to the more conservative approach adopted in the $\sigmu$ model above.

In our \npg analysis  we include the shear ratio likelihood when analyzing  DES \mpp. The shear ratio was forecasted to strengthen the constraints on the \npg amplitudes relative to \mpp alone.  Because shear ratio measurements probe the relative distances between a given lens bin and different source bins, they isolate  geometric information and will be  insensitive to the \Anpg\ parameters.  This means shear ratio measurements help break degeneracies between the \npg amplitudes, nuisance parameters related to photometric redshift uncertainties, intrinsic alignments, and cosmological parameters which affect both geometry and growth (namely \om). As we are treating \npg as a consistency test of \lcdm rather than a physical model, we argue that we can include it without additional validation of the small-scale modeling. 

\section{External data} \label{sec:external_data}

We consider the DES \mpp likelihood described above in comparison to and in combination with measurements from other cosmological experiments. We use the same external measurements as in DES-Y3KP, using public likelihoods from most constraining datasets available at the time of this analysis. These include, as summarized in Table \ref{tab:abb_extdata}:
\begin{itemize}[noitemsep,topsep=0pt]
    \item cosmic microwave background (CMB) temperature and polarization anisotropies measurements by the {\it Planck} satellite as described in \sect{sec:Planck},
    \item distances and growth from 6dFGS, MGS, eBOSS DR16 baryon acoustic oscillations (BAO) and redshift space distortions (RSD) data as described in \sect{sec:eboss}, 
    \item supernovae (SN) distance modulus from Pantheon as described in \sect{sec:pantheon}. \end{itemize}

\begin{table}[]
    \centering
    \begin{tabular}{l l }
        Observables & Data \\
        \hline\\
        CMB (\textit{Planck} in text) & \textit{Planck} 2018 TTTEEE-lowE (no lensing) \\
        BAO & \begin{tabular}{l}eBOSS DR16: LRGs, ELGs, QSOs, \\ Lyman-$\alpha$ QSOs  + 6dFGS + MGS\end{tabular}\\
        RSD & eBOSS DR16:  LRGs, ELGs, QSOs + MGS\\
        SN & Pantheon sample (2018) \\
    \end{tabular}
    \caption{External data used as measurements of additional observables: cosmic microwave background (CMB), baryon acoustic oscillations (BAO), redshift-space distortions (RSD), supernovae (SN).} 
    \label{tab:abb_extdata}
\end{table}
When performing combined analyses of these probes, we assume they are uncorrelated (except for BAO and RSD, whose correlations are taken into account in published likelihoods) so we simply multiply their likelihoods. 

\subsection{Cosmic microwave background}
\label{sec:Planck}

The cosmic microwave background temperature and polarisation primary anisotropies carry information about density and tensor perturbations at the time of the last scattering surface.
In addition, effects such as reionization and gravitational lensing caused by large scale structures produce secondary anisotropies carrying information about the evolution of the Universe since the CMB emission.
In recent decades, measurements of CMB anisotropies have led to the most powerful existing constraints on $\Lambda$CDM cosmological parameters. 

In this paper we therefore use the \textit{Planck} 2018 {TTTEEE-lowE} likelihood presented in Ref.~\cite{Planck:2018vyg}. 
This likelihood combines three components: a high-$\ell$ likelihood based on measurements of multipoles $30 \leq \ell \leq 2508$ for the temperature (TT) angular power spectrum and $30 \leq \ell \leq 1996$ for the TE and EE spectra ({\it plik}), and two low-$\ell$ likelihoods of the temperature TT ({\it commander}) and the polarization EE ({\it simall}) spectra on multipoles $2\leq \ell \leq 29$. To facilitate the study of how cosmological model extensions impact the offset between \seight constraints from \mpp and CMB temperature and polarization, we do not include the CMB lensing likelihood.

When analyzing data we use the full {\it Planck} likelihood provided as part of the \cosmosis standard library. This full likelihood includes 47 nuisance parameters where 21 parameters are marginalized over,  13 of which have Gaussian priors provided with the public {\it Planck} likelihood.
In the case of the \sigmu model, in order to limit computing time we instead use the {\it Planck plik-lite} likelihood, which includes the effects of {\it Planck} nuisance parameter marginalization and only requires us to  sample the absolute calibration parameter $A_{\rm P}$.
We have checked that it gives equivalent results to using the full likelihood in this extended model. 

For simulated analyses of DES \mpp combined with external data,  we use a simplified  {\it Planck} likelihood based on Ref.~\cite{Prince_2019} using two Gaussian likelihoods for $\ell < 30$ and using the TTTEEE {\it Planck plik-lite} likelihood for $\ell \geqslant 30$. 
We also replace the power spectra measurements with model predictions at our fiducial cosmological parameters. 
For both simulated and real analyses, when we include CMB measurements we marginalize over the optical depth to recombination $\tau$ with a flat prior in the range [0.01, 0.8].

\subsection{Baryon acoustic oscillations and redshift space distortions }
\label{sec:eboss}

Baryon acoustic oscillations in the early Universe  imprinted features in the matter distribution at a characteristic scale which can be detected as an excess of galaxy pairs separated by a certain distance in the local universe. 
By measuring the relationship between redshift and the angular diameter distance associated with that excess, we can use BAO as a standard ruler to constrain the expansion history of the Universe.

We use the combinations of BAO likelihoods from eBOSS Data Release (DR) 16 \citep{ahumada202016th} to provide measurements of the Hubble parameter $H(z_i)$ and the evolution of the comoving angular distance $d_{\rm A}(z_i)$. More specifically, we use likelihoods from the reanalysis of BOSS DR 12 Luminous Red Galaxies (LRGs) (dropping its highest redshift measurements), eBOSS LRGs, Emission Line Galaxies (ELGs), quasars (QSOs) and Lyman-$\alpha$ QSOs.
These BAO measurements are made at effective redshifts of $z_{\rm eff} = 0.38, 0.51, 0.70, 0.84, 1.48, 2.33$, respectively.
We additionally include BAO measurements from two lower signal-to-noise galaxy samples, 6dFGS \citep{Beutler:2011hx} and MGS  \citep{Ross:2014qpa}. These likelihoods are based on measurements of the quantity  $d_V(z) \equiv [cz(1+z)^2d_A^2(z)/H(z)]^{1/3}$ evaluated at effective redshifts of  
$z_{\rm eff} = 0.11$ for 6dFGS and $z_{\rm eff} = 0.15$ for MGS.

As an external constraint on the growth rate of structure, we include the eBOSS DR16 redshift-space distortions measurements. 
RSD likelihoods include constraints on the growth of cosmic structure via constraints on the quantity $f(z_i) \sigma_8(z_i)$, where $f$ is the linear growth rate. 
We use the eBOSS DR16 RSD measurements including a reanalysis of BOSS DR12 RSD measurements, LRGs, ELGs and QSOs, at the same redshifts as BAO measurements. 
We also use the MGS RSD measurement from \citep{Tamone:2020rsd} at $z_{\rm eff} = 0.15$. 
When both BAO and RSD measurements from a given sample are included, we account for their covariance using the public eBOSS DR16 likelihoods.

It is worth noting that the RSD likelihoods we use are in the form of marginalized constraints on the quantity $\fsig$ at sample redshifts, and that those constraints are derived quantities from analyses which assume a \lcdm template for RSD features in the galaxy distribution. When studying models beyond-\lcdm, care must be taken in using these likelihoods, as it is possible that inaccuracies in that template could lead to biases in beyond-\lcdm cosmological parameter inferences.  Studies of this in e.g.\   Ref.~\cite{Barreira:2016ovx} demonstrated that using GR-based templates they were able to obtain unbiased $\fsig$ constraints for modified gravity simulations, as long as the modified gravity model did not induce scale-dependent structure growth modifications. Given this, we follow the final eBOSS cosmology analysis~\cite{eBOSS:2020yzd}, which uses these same RSD measurements to constrain \wcdm, \ok, $\wowa$, $\sigmu$, and massive neutrino  cosmologies, and proceed with including RSD measurements among our external likelihoods.
Given the use of these measurements to constrain neutrino mass (e.g. in~\cite{eBOSS:2020yzd}), we assume that they are likely also safe to use for our $\neffmeff$ model, but we highlight that this assumption may be worth investigating for future, more precise, analyses. 

\subsection{Supernovae}\label{sec:pantheon}

Type Ia supernovae are a key cosmological probe that was originally used to discover the accelerated expansion of the universe. Here we adopt the Pantheon SN Ia sample \cite{Scolnic:2017caz}, which combines objects detected and followed up by several different surveys (Pan-STARRS, Sloan Digital Sky Survey (SDSS), Supernova Legacy Survey (SNLS)).  
The resulting sample contains 1048 SN Ia spanning the redshift range $0.01 < z < 2.26$. 

The Pantheon likelihood assigns a Gaussian likelihood to the measured SN distance moduli, $\mu = 5\,\ln\left[d_L/10{\text pc}\right]$. It provides a full  covariance of these measurements, accounting for cosmic variance and the impact of measurement systematics. We model the distance modulus as
\begin{equation}\label{eq:snmu}
\mu = m_B -M + \alpha x_1 - \beta\mathcal{C} +\Delta_M + \Delta_B.
\end{equation}
Here, $x_1$ and $\mathcal{C}$ are the light curve width and color respectively, $\Delta_M$ is a distance correction based on the host-galaxy mass of the SN, and $\Delta_B$
is a distance correction based on predicted biases from
simulations. The calibration parameters $\alpha$, $\beta$, $\Delta_M$ are fit to data as described in Ref.~\cite{Scolnic:2017caz}, while $\Delta_B$ is calibrated using simulations. The absolute magnitude $M$ is a nuisance parameter that we marginalize over in our analysis with a flat prior $-20<M<-18$.

\section{Analysis procedure and validation}\label{sec:analysis}

Our analysis procedure can be divided into five stages. These steps, described in more detail below, proceed as follows: 
\begin{enumerate}
\item  We analyze a fiducial synthetic data vector --- that is, we analyze a noiseless model prediction at a fiducial set of \lcdm parameters as if it were data. In this paper, the term `simulated analysis' will refer to this kind of analysis of synthetic data.  (\sect{subsec:simfid})
\item We validate scale cuts and modeling choices by analyzing a series of alternative simulated data vectors that have been `contaminated' by systematics or by modeling choices which are more complex that those used in our baseline model. (\sect{sec:simdat_changeDV})
\item We perform a set of analysis tests using the fiducial synthetic data vector to study the robustness of our results against changes in the model used for parameter estimation. (\sect{sec:changepipe})
\item We repeat the previous step's robustness tests against variations of the model for real data, without unblinding the cosmology results. Once we completed these robustness tests, a draft of this paper and the analysis plan documented in it underwent a stage of DES internal review (\sect{sec:changepipe}).
\item Finally, we reveal our cosmology results, assess tension metrics, compute model comparison metrics, and describe the results in \sect{sec:res}. 
\end{enumerate}

This procedure is designed to ensure as far as possible that decisions on how to structure the analysis are not influenced by  knowledge of how they affect the main results. Since the data we are working with have already been unblinded for the $\Lambda/w$CDM models in DES-Y3KP, we opted not to use the two-point-function transformation blinding method~\cite{y3-blinding}. 
Instead, we simply blinded at the parameter level, post-processing chains to shift marginalized posterior means onto fiducial values via unknown offsets. Additionally, until pre-unblinding internal review was completed, we avoided looking at comparisons between theory and model predictions for observables, tension metrics between datasets, and any model comparison statistics between extended cosmological models and \lcdm.

Several analysis choices used in this work were adjusted after \lcdm unblinding, in line with changes made to the analysis in DES-Y3KP. These changes are: the choice of the
\maglim rather than the \redmagic~\cite{redmagicSV,y3-galaxyclustering,y3-2x2ptbiasmodelling} lens sample, and the removal of the two highest redshift \maglim lens bins (which would have been bins 5 and 6) from the fiducial analysis. As noted in \sect{sec:2pcfmodel}, we have also adopted the simpler two-parameter NLA description of intrinsic alignments instead of the five-parameter TATT model used in DES-Y3KP. We made the choice to use NLA largely because the DES-Y3KP \lcdm analysis did not find that the  TATT model is favored over NLA. We  emphasize that while these aspects of our analysis have been shaped by findings for unblinded \lcdm results, these changes were frozen in before analyzing any data for the beyond-\lcdm models.

\subsection{Fiducial synthetic data}
\label{subsec:simfid}

We initially validate our analysis by performing a series of analyses of synthetic  measurements --- that is, analyses of the data vector predicted from our fiducial model, with no noise. With the exception of the change in data vector, this analysis is identical to our eventual analysis of real data: the synthetic DES \mpp data are generated using the same redshift distributions used for the final analysis, and the  likelihood is evaluated using a covariance produced at our fiducial cosmology using the same analytical calculations described in Ref.~\cite{y3-covariances}.  In addition to synthetic DES \mpp measurements, we additionally produce synthetic versions of the external likelihoods  for simulated combined analyses. 

We begin with a baseline simulated analysis: using our fiducial model, we analyze synthetic data vectors produced using those same calculations. This can be thought of as a `best case scenario' where our model calculations are exactly correct so that we can estimate the expected constraining power and the relationship between marginalized posteriors and parameters' input values. 

In some cases when constraints are weak, prior volume effects cause marginalized confidence intervals for parameters to be offset from their input (\lcdm) values. This occurs because the prior in our full parameter space can be highly non-uniform when projected onto certain cosmological variables. This occurs notably for the synthetic {\it Planck}-only results, which prefer $\ok<0$ at $1.4\sigma$. This offset from flatness, which can be understood in terms of the CMB's well-known geometric degeneracy~\cite{Efstathiou:1998xx},  is in the same direction as what has been previously reported for the analysis of real CMB data 
but at a lower significance. The preference for $\ok<0$ goes away when the {\it Planck} likelihood is combined with low-redshift geometric likelihoods.

We also see offsets in the DES \mpp and \mpp{}+BAO+RSD+SN $\neffmeff$ constraints, which is due to a positive degeneracy between $\neff$ and $\meff$ constraints for the lower redshift probes, as well as degeneracies between both of those parameters and $H_0$. Adding CMB information introduces a powerful constraint on \neff, breaking those degeneracies and causing the marginalized posterior distribution for the All-data constraints to be more reflective of the input parameter values. Given this concern about the projection effects, for $\neffmeff$ we will focus primarily on constraints from DES Y3 \mpp and all external data (\textit{i.e.} BAO+RSD+SN+\textit{Planck}) rather than DES alone. 

In the \sigmu model,  \muo measurements by DES Y3 \mpp alone are  prior-dominated so will not be reported. We additionally note that a \sigo-\seight degeneracy causes a slight offset for the DES-only \sigo posterior, though the resulting constraints are consistent with the input value. 
The addition of external RSD or {\it Planck} data enable precise measurements of \muo, in turn leading to more precise constraints on \sigo.

\subsection{`Contaminated' synthetic data}
\label{sec:simdat_changeDV}

Next, we analyze synthetic \mpp data that have been `contaminated' by various effects. The goal here is to test robustness of our results to modeling complexities and systematic effects which are not included in our fiducial model. 
To verify this, we  compare the results of the analysis of contaminated synthetic data to those from our baseline simulated analysis, both computed in \lcdm. This allows us to quantify the impact these  modeling uncertainties have on parameter estimates and model comparison statistics used to evaluate tensions with \lcdm.  The priority here --- and what we can evaluate most accurately, given the lack of in-depth study of and available modeling tools for describing systematics in extended cosmological models --- is to assess whether unmodeled systematics are can produce a false detection of tension with \lcdm. 
Specifically, we study three alterations to the synthetic data: 
\begin{itemize}
    \item \textbf{Nonlinear bias + baryons}: A realization of baryonic feedback and non-linear galaxy bias is added to the synthetic \mpp observables. Baryonic feedback effects are added using the method described in Refs.~\cite{DES:2020rmk,Huang:2018wpy}, and are based on the OWLS  hydrodynamic simulations~\cite{OWLS-1,Daalen2011} with large AGN feedback according to the prescription of Ref.~\cite{Booth-2009}. Nonlinear galaxy bias is modeled using an effective 1-loop description with renormalized nonlinear bias parameters as in Refs.~\cite{McDonald-2009,Baldauf-2012,Chan:2012jj,Saito:2014qha}. 
    This synthetic data is produced with the same contaminations used in Ref.~\cite{y3-generalmethods} to define scale cuts for DES-Y3KP. Analyzing it allows us to verify  that the scale cuts defined for \lcdm continue to protect against these small-angle systematics in the beyond-\lcdm models we consider.
    
    \item \textbf{Euclid emulator}: The \halofit computation for the nonlinear, gravity-only matter power spectrum is replaced with that of \EuclidEmulator \cite{Euclid:2018mlb}. Checking that this alternate nonlinear prescription does not shift our results is a test of robustness  against inaccuracies of the small-scale power spectrum modeling.
    
    \item \textbf{Magnification $3\sigma$ offset}: The magnification coefficients $\magcoeffi$ (see \eq{eq:cell}) are offset from their fiducial values by three times their uncertainty, where the latter is determined in Ref.~\cite{y3-2x2ptmagnification} using the Balrog simulations. The coefficients used are $\magcoeff=[1.97, 1.74, 2.93, 2.97]$ compared to the fiducial coefficients $\magcoeff=[0.42, 0.30, 1.76, 1.94]$. 
    This $3\sigma$ offset is designed to demonstrate the robustness of results to the amplitude of the magnification signal and validate the decision to fix the $\magcoeff$ values at their fiducial values.  
\end{itemize}

To facilitate these tests we adopt the newly-developed \fastismore{} (Fast Importance Sampling for MOdel Robustness Evaluation) scheme, which is presented in more detail in \app{app:fastismore} and Ref.~\cite{fastismore}.  Briefly, the approach exploits the fact that if our analysis is robust against a given systematic, the shift in posteriors should be small when the data is `contaminated' with that effect. This allows us to  use the posterior from a baseline chain (run using uncontaminated synthetic \mpp observables) as a proposal distribution for estimating the posterior for a contaminated datavector via  importance sampling (IS). Doing this allows us to quickly  verify whether the change in posterior is indeed negligible. If quality statistics indicate that the IS posterior estimate is of good quality, it can be used to quantify the effect of the systematic on parameter estimates and model comparison statistics. If the estimate is poor, this indicates we need to run a new chain to estimate the contaminated posterior. 

Once we have obtained reliable posterior estimates, we assess shifts in the marginalized constraints on the beyond-\lcdm parameters. We quantify this following  Ref.~\cite{Muir:2020puy}, defining the marginalized posterior shift $\Delta_{\theta}$ for parameter $\Theta$  as
\begin{equation}
\label{eq:peakshift}
\Delta_{\Theta}= \frac{\bar\Theta_A - \bar\Theta_B}{\sqrt{(\bar\Theta_A - \Theta_A^{\rm \tt low68})^2 + (\Theta_B^{\rm \tt up68} - \bar\Theta_B)^2}}.
\end{equation}
Here, $\bar\Theta$ is the parameter's posterior-weighted mean, and the labels  $A$ and $B$ correspond to the baseline and alternative (contaminated) synthetic data vectors,  defined such that $\hat\Theta_A>\hat\Theta_B$. The parameter $\Theta_A^{\rm \tt low68}$ is the lower bound of the 68\% confidence interval for posterior $A$, while $\Theta_B^{\rm \tt up68}$ is the upper bound of the 68\% confidence interval for posterior $B$. Thus the denominator of \eq{eq:peakshift} is an effective $1\sigma$ error for parameter $\Theta$, accounting for possible asymmetry in the marginalized posteriors. 
We consider a parameter shift to be negligible if  $\Delta_{\Theta}<0.3$.

\begin{figure*}
  \centering
\includegraphics[width=\linewidth]{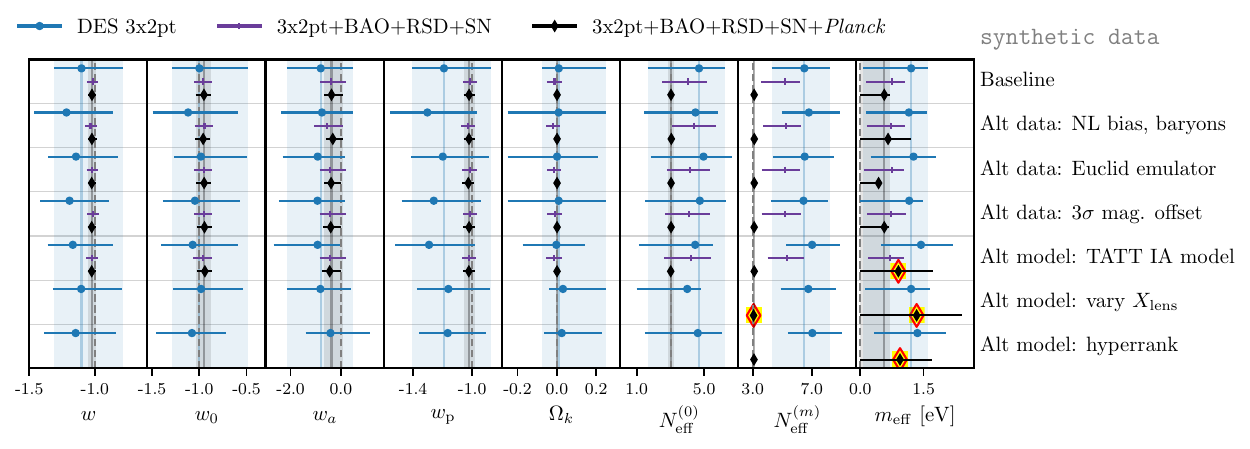}\\\vspace{-6pt}
\includegraphics[width=\linewidth]{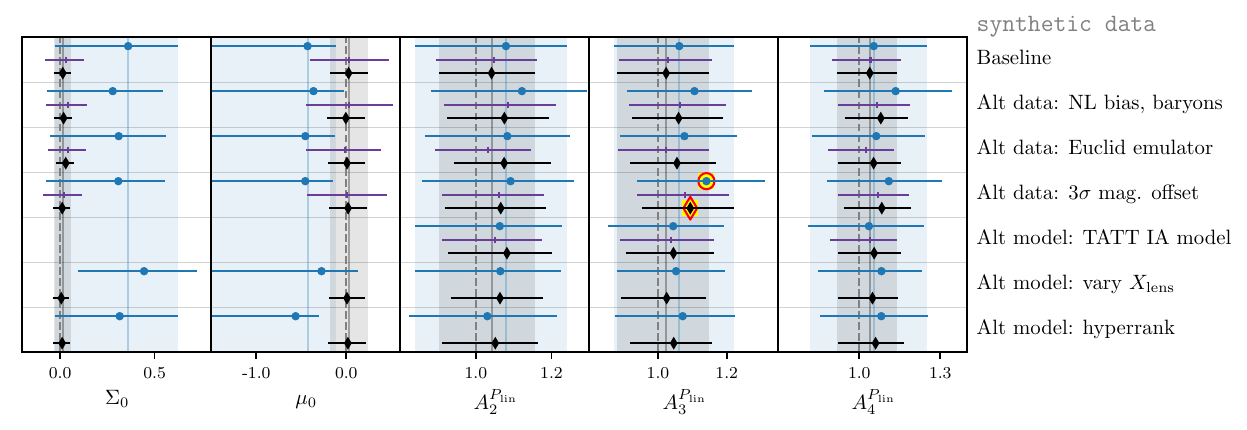}
  \caption{Simulated analysis constraints on beyond-\lcdm model parameters, showing robustness against systematics and model variations. Points and error bars show the mean and 68\% confidence interval for marginalized parameter constraints, and points that are offset from the baseline by more than $0.3\sigma$, according to \eq{eq:peakshift}, are highlighted in yellow and red.  To facilitate comparison between rows, solid vertical lines and shaded regions show the location and 68\% confidence interval of the baseline point for DES \mpp and DES 3$\times$2pt+BAO+RSD+SN+{\it Planck} results in blue and black respectively. Dashed vertical lines show the \lcdm values used to generate the synthetic data vectors. We use $\neff^{(0)}$ to identify the effective number of relativistic degrees of freedom when no mass is included in the model,  $\neff^{(m)}$ for that parameter in the \neffmeff model. For parameters where the combination of all data (DES \mpp and all external data) is much more constraining than DES \mpp, a version of this plot  with narrower axis ranges can be found in  \fig{fig:extparams_table_alldatonly} of \app{app:more_validation}.}
  \label{fig:extparams_table}
\end{figure*}

We perform these checks for DES \mpp and DES 3$\times$2pt+BAO+RSD+SN+{\it Planck} posteriors for each beyond-\lcdm model, as well as 3$\times$2pt+BAO+RSD+SN (leaving out {\it Planck}), using simulated external data likelihoods produced at the same fiducial  cosmology as the synthetic DES data. Results are shown in the ``Alt data" rows 
of \fig{fig:extparams_table}, with points and error bars indicating the mean and marginalized 68\% confidence interval for each parameter. Where error bars are not visible they are smaller than the size of the data point.  
In that plot the $\neff^{(0)}$ constraints are for the model which varies \neff only, while $\neff^{(m)}$ shows the $\Delta\neff$ constraint from the $\neffmeff$ model. Points with $\Delta_{\Theta}>0.3$ are highlighted. 

Nearly all shifts evaluated were below the $0.3\sigma$ threshold, meaning these beyond-\lcdm parameter estimates  are robust against each of the considered systematics. The only exception to this occurs for the \npg model's response to  changes in the assumed magnification parameter. For the \npg DES \mpp results we see $\Delta_{\Anpg_3}=0.31$, and for  3$\times$2pt+BAO+RSD+SN+\textit{Planck} \npg results we find $\Delta_{\Anpg_3}=0.37$.  We note that these numbers are close to the desired threshold, especially relative to our sampling uncertainty of$\Delta_{\Theta}\pm \sim 0.04$, and so are not very concerning.

\begin{figure*}
  \centering
\includegraphics[width=\linewidth]{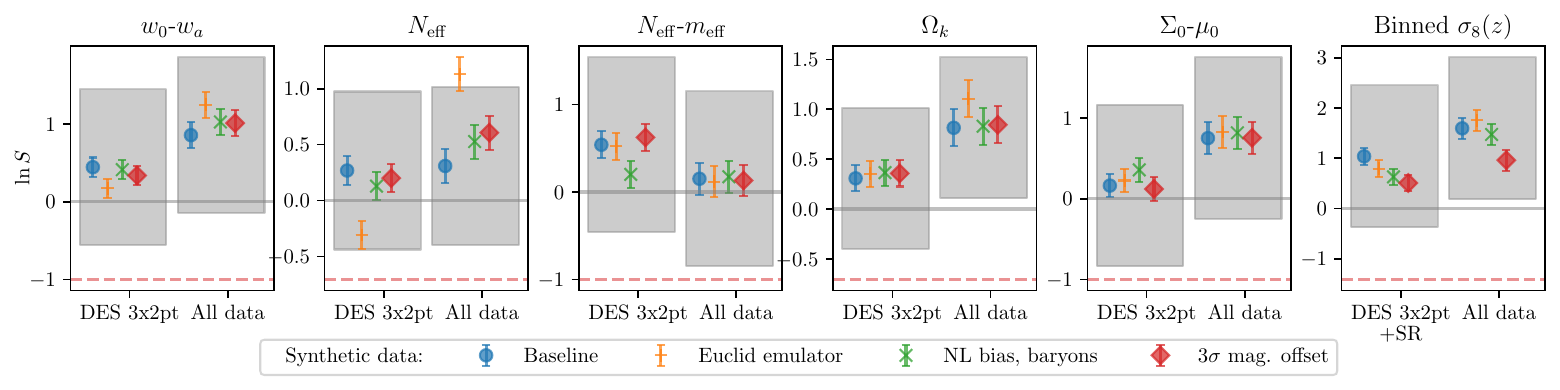}
  \caption{Impact of contamination by unmodeled systematics on the Suspiciousness model comparison metric evaluated for synthetic \lcdm{} data. Panels show Suspiciousness evaluated between different 
  beyond-\lcdm models and \lcdm{}, with the marker styles indicating different contaminations.  Blue circular markers indicate the baseline case where the synthetic data is produced using the same model as used for parameter estimation. Studying whether contaminations shift $\ln{S}$ towards more negative values  than the baseline measurement tells us whether they are likely to produce a false detection of beyond-\lcdm physics.
Shaded regions indicate the expected $\pm 1\sigma$ scatter around the baseline for Gaussian posteriors, 
and  the red dashed line denotes the value of $\ln{S}$ that would produce a $1\sigma$ preference for the beyond-\lcdm{} model (though as we discuss in \app{app:modcomp} $\ln{S}$ is expected to follow a $\chi^2$ distribution, such that the true uncertainty is more skewed toward negative values of $\ln S$ than the shaded regions, making the test conservative).  Error bars on points indicate sampling uncertainty.   Results are shown  for the analysis of DES \mpp alone and  3$\times$2pt+BAO+RSD+SN+{\it Planck} (All data). 
  }
  \label{fig:lnSsysshift}
\end{figure*}

We also check whether these contaminations affect our assessment of whether an extended model is favored relative to \lcdm. We do this by comparing the values of the Suspiciousness model comparison statistic $S$, defined in \app{app:modcomp}, evaluated in our contaminated and baseline simulated analyses.  
For a base model $\mathcal{M}_0$ (e.g. \lcdm) whose parameter space is a subspace of extended model $\mathcal{M}_X$, we define Suspiciousness so that  $\ln{S}<0$  indicates a preference for $\mathcal{M}_X$.  
Of the several model comparison statistics that we will ultimately report as part of our results, we use Suspiciousness here because it is readily calculable from importance-sampled posteriors.  
\fig{fig:lnSsysshift} shows the changes produced by systematic contamination in $\ln{S}$ relative to the expected amount of scatter, for DES \mpp and DES 3$\times$2pt+BAO+RSD+SN+{\it Planck} constraints.  
The largest shift occurs for \Neff{}-vs-\lcdm{}, where analyzing the \EuclidEmulator synthetic data shifts  $\ln{S}$ by $\sim1\sigma$ in the limit that posteriors are Gaussian. We should therefore interpret model comparison results for \Neff{} with some caution. 
Otherwise,  all systematics considered cause negligible changes in Suspiciousness and thus are unlikely to result in a false detection of tension with \lcdm.

\subsection{Robustness to model variations}\label{sec:changepipe}

We additionally study how parameter constraints respond to  changes in our model. By comparing results obtained using alternative modeling choices to those obtained from our fiducial model, we assess the robustness of our findings relative to those model variations.  As before, we quantify this comparison  in terms of the parameter shift $\Delta_{\Theta}$ defined in \eq{eq:peakshift}, and we consider any shifts with $\Delta_{\Theta}<0.3$  to be negligible. The model changes considered are:
\begin{itemize}
    \item \textbf{TATT}  - We use a five-parameter TATT intrinsic alignment model \cite{PhysRevD.100.103506} instead of the two-parameter NLA model used in the present baseline analysis. This model, which was the fiducial IA model in DES-Y3KP, has significantly more flexibility to describe IA scale and redshift dependence, allowing it to capture IA from tidal alignment and tidal torquing, as well as the response to the density-weighted tidal field. We use the same parameters and prior ranges as in DES-Y3KP. Comparison with an analysis using TATT allows us to test the robustness of our results against our choice of intrinsic alignment model. As explained in \sect{sec:2pcfmodel}, for each of the beyond-\lcdm models we perform a pre-unblinding model comparison between TATT and NLA to test whether there is tension with the choice of NLA as the fiducial IA model for that cosmology. Note that this set of tests are not performed  for \sigmu because the TATT modeling tools are not available  for our modified gravity calculations
    (see \sect{sec:sigmu}).
    
    \item \textbf{Varying \xlens}: We marginalize over a parameter \xlens which multiplies the galaxy bias terms appearing in $\gamma_t(\theta)$ calculations, thus allowing the galaxy--galaxy lensing observables to have a different bias parameter than galaxy clustering. 
    Such an effect was discovered after unblinding the \lcdm \mpp analysis using the \redmagic lens sample, and while it is still under investigation, it is thought to be due to an unaccounted for systematic related to lens sample selection~\cite{y3-2x2ptbiasmodelling}. This effect  motivated the choice of \maglim over \redmagic as the fiducial lens sample in DES-Y3KP. 
    While no evidence was found in \lcdm for $\xlens\neq1$ for our data vector, which uses the four-bin \maglim lens sample, we include results marginalizing over parameter to test the robustness of our beyond-\lcdm constraints to the presence of this kind of systematic in the real data analysis. 
    Note that a similar effect with independent \xlens values for each redshift bin is able to capture the issues with the fifth and sixth \maglim bins which led to their removal from the analysis. Studies in \lcdm have shown that the first four \maglim bins we analyze are consistent with  $\xlens=1$ in this redshift-dependent formulation as well~\cite{y3-2x2ptaltlensresults,y3-2x2ptbiasmodelling}. Thus to limit the parameter space of these tests, here we consider only a single redshift-independent \xlens parameter. As an additional exploration of this effect, in \app{app:simXlens}  we study the response of the extended models to synthetic data produced with \xlens $\neq 1$. 
    \item \textbf{Hyperrank}: Instead of using the four $\Delta z_s$ photo-$z$ bias parameters of our fiducial model, we use the hyperrank~\cite{y3-hyperrank} method  to marginalize over uncertainties in the source sample redshift distributions. This method uses three `rank' parameters to sample an ensemble of 1000 realizations of source $n(z)$ histograms that were generated using a three-step Dirichlet (3sDir) sampling method~\cite{y3-sompz, Sanchez:2020reg, Sanchez:2018wuh}. (Throughout this paper we will use ``hyperrank'' as shorthand for what might more properly be referred to as the 3sDir+hyperrank method.) Compared to the fiducial approach, hyperrank captures more information about uncertainties in the shapes of those redshift distributions as well as correlations between different source bins. Comparing constraints obtained using hyperrank to our fiducial model will indicate how sensitive our results are to our characterization of source photo-$z$ uncertainties. Note that in DES-Y3KP, it was found that using hyperrank for \lcdm produced a $0.53\sigma$ shift in \seight (see Appendix E1 of that Reference and our \app{app:hyperrank}).
    \end{itemize}
Because we are using the TATT comparisons to evaluate our choice of fiducial IA model (except for the \sigmu model),  we run TATT analyses for all three principal data combinations (DES \mpp alone, only low redshift probes: DES 3$\times$2pt+BAO+RSD+SN, and all data combined: DES 3$\times$2pt+BAO+RSD+SN+{\it Planck}). We run \xlens and hyperrank analyses for DES \mpp only, except for the \sigmu, \npg and $\neffmeff$ models. This choice was motivated by non-negligible shifts seen for   \npg and \neffmeff in the blinded real-data tests for DES \mpp-alone. Additional chains were run for \sigmu as well because that model also primarily affects structure growth.

\begin{figure*}
  \centering
\includegraphics[width=\linewidth]{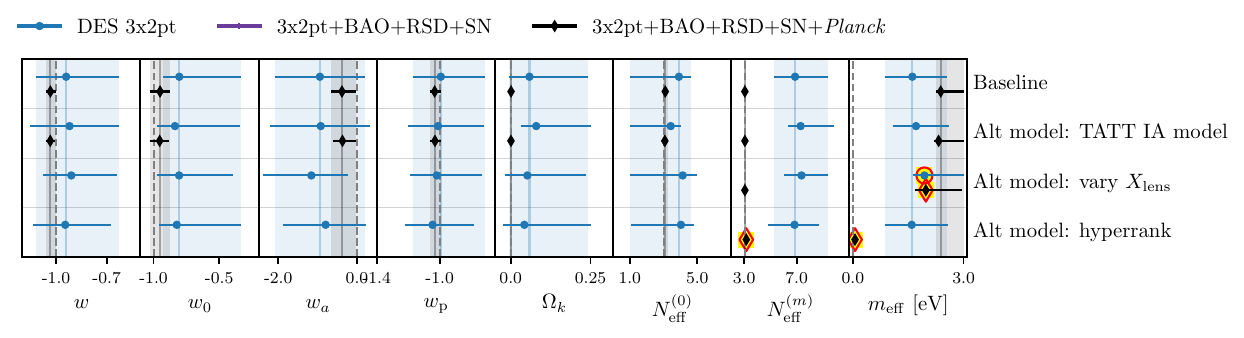}\\\vspace{-6pt}
\includegraphics[width=\linewidth]{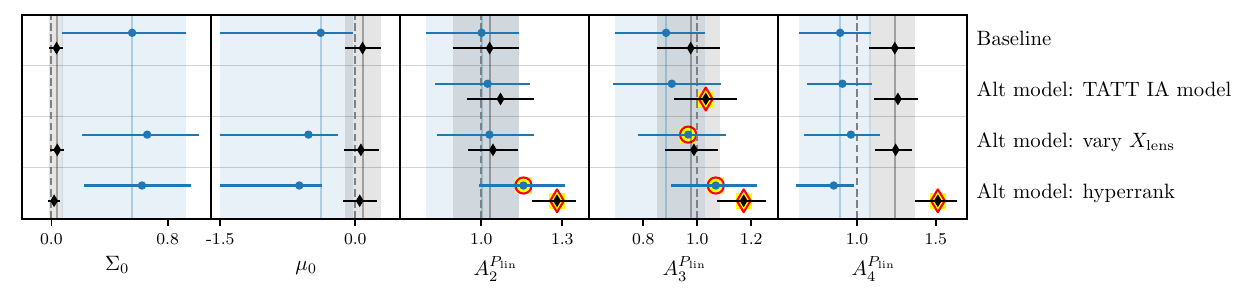}
  \caption{
  Beyond-\lcdm parameters constraints for real data and the impact of model variations.  
  Points and error bars show the mean and 68\% confidence interval for marginalized constraints on extended model parameters, and shifts larger than $0.3\sigma$ according to \eq{eq:peakshift} are highlighted in yellow and red. Dashed gray lines show parameter values that correspond to \lcdm.
  Solid vertical lines and shaded regions mark the DES \mpp and combination of all data, DES 3$\times$2pt+BAO+RSD+SN+{\it Planck} baseline results, in blue and black respectively, to facilitate comparison between rows.   
  No points are shown in second row of  \sigo and \muo constraints because of the lack of  TATT IA model implementation within the modified gravity pipeline. A version of this plot with narrower axis ranges can be found in \fig{fig:realdat_extparams_alldatonly} of \app{app:more_tableplots}.}
  \label{fig:realdat_extparams}
\end{figure*}

Results showing the impact of these model variations on simulated analyses are shown in the `Alt model' lines of \fig{fig:extparams_table}, and results for real data are shown in  \fig{fig:realdat_extparams}.
\footnote{Before unblinding cosmological parameter estimates or model comparisons with \lcdm, at this stage we studied only differences between the alternative and baseline chain constraints, working with a version of \fig{fig:realdat_extparams} where baseline means were subtracted from all numbers, causing all the points in the top row to be on zero. The Figure was updated after unblinding to show the actual parameter values of the constraints.}  
For synthetic and real data, other than for \neffmeff and \npg,  all shifts due to these model variations were below $0.3\sigma$. 
The non-negligible parameter shifts occurring for \neffmeff and \npg have motivated adjustments to analyses choices for those models. Further detailed discussion of those shifts  can be found for \neffmeff in \app{app:changepipe_neffmeff} and for \npg in \app{app:changepipe_npg}, respectively, which we briefly summarize here.
First, the sensitivity of the \neffmeff constraints to model variations seem to be caused by prior volume effects associated with an unconstrained part of parameter space at  small $\Delta\neff$. Since adjusting our prior to remove that part of parameter space restores robustness, this motivates our choice to report \neffmeff results using priors that require either $\Delta\neff>0.047$ or $m_{\rm th}<10$ eV, as noted at the end of \sect{sec:neffmeff}.

Second, for \npg the most concerning parameter shifts occur for hyperrank, which causes the $\Anpg_i$ amplitudes to change by $\sim 1-2\sigma$ relative to their baseline model estimates. Given this, we report \npg results for both the baseline and hyperrank versions of the analysis, using their comparison  as a rough estimate for the impact of systematic uncertainties related photometric redshift estimates. We note that model variations most severely impact inferences about \sigeight in the lowest redshift bin, which then propagates to affect $\Anpg_i$ parameters because they are defined relative to bin 1. Importantly, the derived parameters $\npgsig{i}\equiv \sigeight[\Anpg_i]^{1/2}$, which are more closely related to the observed amplitude of LSS, are more robust to modeling variations especially when we combine DES \mpp with external data.

\section{Results}\label{sec:res}

\begin{table}
\include{figures/realy3dat.extparams_textable}
\caption{Marginalized constraints on beyond-\lcdm parameters for DES Y3 \mpp, all external data ({\it Planck}+BAO+RSD+SN), and all data (\mpp{}+{\it Planck}+BAO+RSD+SN). Two-sided constraints report the mean and 68\% confidence interval for each parameter's marginalized posterior. One-sided constraints report 95\% bounds. 
} \label{tab:results_extparams}
\end{table}

We now show the principal results of our analysis. This section is organized as follows: In \tab{tab:results_extparams}, we show a summary of marginalized constraints on individual parameters. In \sects{sec:res_w0wa}-\ref{sec:res_npg}, we report and discuss constraints for each of the cosmological models studied in this work. In \sects{sec:res-S8} and~\ref{sec:res-mnu} we examine how these cosmological models and other model variations impact estimates of \seight and \mnu respectively.
While discussions of results for individual models touch on tensions and model comparisons, \sects{sec:res-tensions} and~\ref{sec:res-modcomp}, present more details about the definitions and determination of tension and model comparison statistics, respectively. 

In this section, parameter estimates shown with two-sided error bars report the mean and 68\% confidence interval of marginalized one-dimensional posteriors. One-sided errors report 95\% confidence bounds. Before unblinding, we decided that we would report results from combined datasets only if the $p$-value associated with the Suspiciousness tension metric, described in \app{app:tensions} and Ref.~\cite{y3-tensions}, is greater than 0.01. We evaluate that metric for two data combinations: DES \mpp versus the external low-redshift data combination BAO+RSD+SN, and {\it Planck} versus the combination of all non-CMB data, \mpp{}+BAO+RSD+SN.

\subsection{Results: \texorpdfstring{\wowa}{Lg}}\label{sec:res_w0wa}

\begin{figure*}
  \centering
\includegraphics[width=0.45\linewidth]{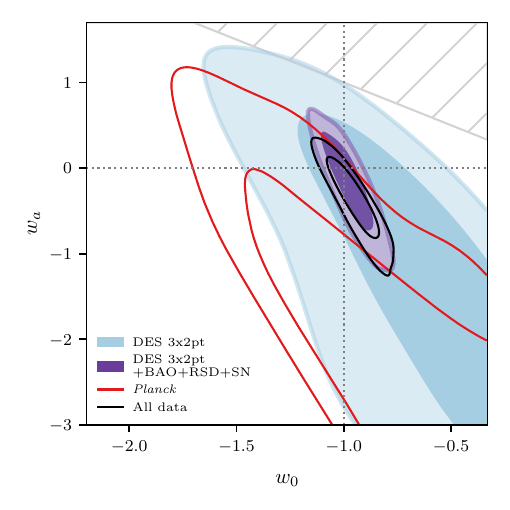}
\includegraphics[width=0.45\linewidth]{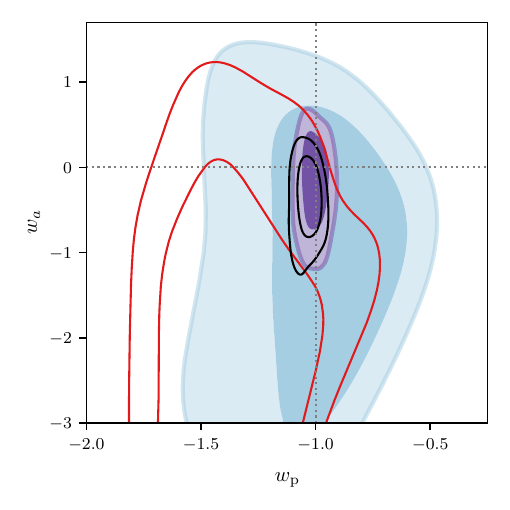} 
  \caption{Constraints on the dynamical dark-energy \wowa model. The left panel shows the constraints on $w_0$ and $w_a$, while the right panel shows constraints on the derived pivot value $w_{\rm p}$ and again $w_a$. Contours show 68\% and 95\% confidence regions. The pale blue contours are DES \mpp, the purple contours is the low-$z$ combination of \mpp{}+BAO+RSD+SN, the open red contours are for {\it Planck}, and the open black contours represent everything combined. The pivot redshifts derived for the $w_{\rm p}$--\wa are 0.24 for DES \mpp only, 0.21 for \mpp{}+BAO+RSD+SN, and 0.27 for all data constraints.  The gray hatched region shows the part of parameter space removed by the prior that requires $w_0+\wa<0$. }
  \label{fig:res_w0wa}
\end{figure*}

We start with dynamical dark energy described by parameters $w_0$ and $w_a$. The marginalized constraints are shown in the left panel of \fig{fig:res_w0wa}. We find
\begin{flalign} 
 w_0 &\geq -1.4,  &w_a &= -0.9\pm 1.2 && \text{DES\ Y3}\\[0.2cm]
 w_0 &= -0.95\pm 0.08, &w_a &= -0.4^{+0.4}_{-0.3}  &&\text{DES\ Y3 + External}.\nonumber
\end{flalign}
DES Y3 \mpp data alone weakly constrain\footnote{We report two-sided \mpp-only constraints on \wa because its one-dimensional marginalized posterior is not bounded by the \wa priors, but note that the constraint is strongly influenced by the intersection of the posterior with the upper \wo prior boundary.} $w_0$ and $w_a$, but are statistically consistent with the cosmological-constant values
of $(w_0, w_a)=(-1, 0)$. When combined with external data, the constraints tighten considerably. The combined constraints find $w_a<0$ at about $1\sigma$, and are thus consistent with the standard model. In the $\wowa$ plane,  the DES \mpp data alone have similar constraining power to {\it Planck} alone. Additionally,  \mpp{}+BAO+RSD+SN (without {\it Planck}) produce constraints comparable those from the combination of all data considered. 

The pivot equation-of-state  (see \sect{sec:w0wa}) is 
\begin{alignat}{2}
    \begin{aligned} 
 w_{\rm p}     &= -0.99^{+0.28}_{-0.17}\qquad &&\text{DES\ Y3}\\[0.2cm]
         &=  -1.03^{+0.04}_{-0.03}\qquad &&\text{DES\ Y3 + External},
    \end{aligned} 
\end{alignat}
where for the DES Y3 \mpp and all-data constraints, the pivot redshifts are $z_{\rm p} = 0.24$ and $z_{\rm p} = 0.27$, respectively. Note that DES alone does give a two-sided constraint on $w_{\rm p}$, unlike on $w_0$. 
The right panel of \fig{fig:res_w0wa} shows the constraints in the $(w_{\rm p}, w_a)$ plane. The DES Y3 \mpp data again qualitatively agree with CMB alone, but are  somewhat more centered on the \lcdm model values of $w_{\rm p}=-1$, $w_a=0$.  We note that the \mpp \wa constraints are slightly tighter than the latest Pantheon+ supernovae measurements~\cite{brout_2022}, which report a 68\% confidence interval width of 2.8 on $w_a$, compared to 2.3 from DES Y3 \mpp. 

\begin{figure}
  \centering
\includegraphics[width=\linewidth]{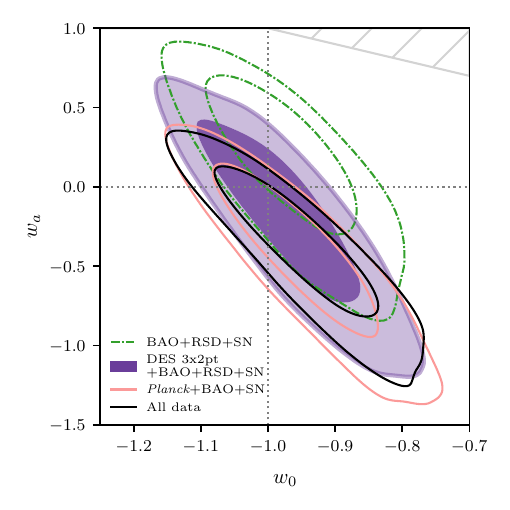}
  \caption{A more detailed look at how different data contribute constraining power to $\wo$ and $\wa$. The purple contours showing \mpp{}+BAO+RSD+SN and black contours showing all data constraints are the same as those in \fig{fig:res_w0wa}. The green contours show constraints from the external low-$z$ data alone (BAO+RSD+SN), while the pink contours show constraints from {\it Planck} combined with low redshift geometric probes only (BAO+SN). The gray hatched region shows the excluded region where $w_0+\wa>0$.}
  \label{fig:res_w0wa_zoomed}
\end{figure}

Fig \ref{fig:res_w0wa_zoomed} presents a more detailed view of the most powerful constraints in the \wowa plane, additionally showing constraints from the external low redshift data (BAO+RSD+SN) alone and {\it Planck} combined with low-redshift geometric data (BAO+SN). Looking at individual parameter constraints, we note that the marginalized posterior mean for \wo is essentially the same for all data combinations considered here, and that compared to constraints from BAO+RSD+SN alone, adding \mpp data causes the \wa estimate to shift downwards by about $1\sigma$. Adding {\it Planck} constraints to that moves the all-data \wa estimate slightly lower, but not by a significant amount, and produces constraints that are very similar to the {\it Planck}+BAO+SN data combination.  All constraints are statistically consistent with \lcdm parameter values. 

Overall, the DES \mpp data strengthen the case that the $(w_0, w_a)$ model parameters are in excellent agreement with the \lcdm values  $w_0=-1$, $w_a=0$.
We discuss the comparison between the present results and DES Y1 \mpp constraints on \wowa \cite{Abbott:2018xao} in \app{app:y1vy3}.

\subsection{Results: \texorpdfstring{\ok}{Lg}}\label{sec:res_ok}

\begin{figure}
  \centering
\includegraphics[width=\linewidth]{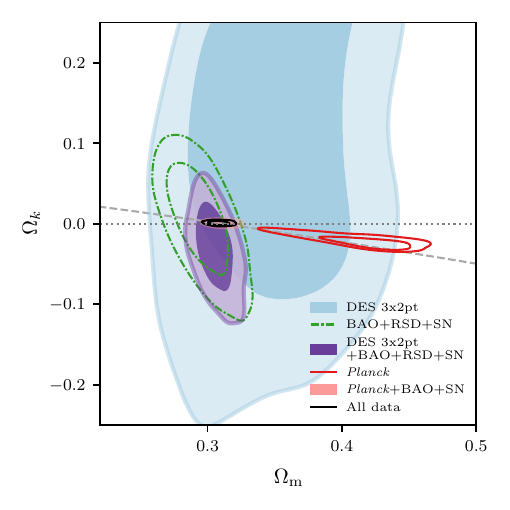}
  \caption{Constraints on the curvature density \ok and matter density \om. The diagonal dashed gray line shows the direction of the primary CMB anisotropies' geometric degeneracy~\cite{Efstathiou:1998xx}. 
  }
  \label{fig:res_ok}
\end{figure}

\fig{fig:res_ok} shows constraints on curvature  in the \ok--\om plane.  We see that curvature is not strongly constrained by the DES Y3 \mpp data alone, and that the constraints on curvature from all data (DES \mpp{}+BAO+RSD+SN+{\it Planck}) are identical to those from  {\it Planck} combined with low-redshift geometric probes (BAO+SN). The DES \mpp data do contribute constraining power when combined with other low redshift data by breaking degeneracies in the full parameter space. Specifically, when combined with BAO+RSD+SN, DES \mpp  data lower the upper bound on \ok by constraining \om and thus helping to break a degeneracy between \om and \ok. This decreases the width of the marginalized 68\% confidence interval on \ok by 20\%.

We recover the well-documented (see e.g. \cite{Planck:2018vyg,Park:2017xbl,Handley:2019tkm,DiValentino:2019qzk,DiValentino:2020hov,Efstathiou:2020wem,Glanville:2022xes,Hergt:2022fxk}) finding that constraints from {\it Planck} alone favor negative \ok at roughly three sigma 
and  are significantly offset from the low-redshift constraints. That offset is along the direction of the primary CMB anisotropies' geometric degeneracy~\cite{Efstathiou:1998xx}, indicated in \fig{fig:res_ok} with a gray dashed line\footnote{The line is drawn for constant shift parameter $\mathcal{R}\propto \sqrt{\Omega_m h^2 }(1+z_*)d_A(z_*)$ corresponding to the Planck+BAO constraints reported in Ref.~\cite{Planck:2018vyg}, where $z_*$ is the redshift of recombination.}, and when that degeneracy is broken with low-redshift observables the constraints shift to become consistent with \ok=0.  Quantifying the tension between the {\it Planck}-only and \mpp{}+BAO+RSD+SN posteriors, we find the Suspiciousness $p$-value to be exactly at our threshold of 0.01 for combining data.  
Given this, we report constraints for the combination of all the data, but additionally report the constraints from \mpp{}+BAO+RSD+SN (without {\it Planck}).  The marginalized constraints on \ok are
\begin{equation}
\begin{aligned}  
  \Omega_k &\geq -0.16\quad\quad\quad\quad \,\text{DES\ Y3}\\[0.2cm] 
 &= 0.001\pm 0.002\quad \text{DES\ Y3 + External}\\[0.2cm] 
 &= -0.03^{+0.04}_{-0.03} \qquad\text{DES\ Y3 + External, no CMB}
\end{aligned}
\end{equation}
Compared to the combination of all external data, the inclusion of \mpp in the DES Y3 + External constraints listed above narrows the 68\% confidence interval range by only 6\%. Compared to the eBOSS analysis of Ref.~\cite{Alam_2021},  in which \ok is measured from the combination of CMB, BAO, and SN, our constraints are about 10\% weaker, likely due to differences in analysis procedures (for example, we vary neutrino mass, while that analysis fixes it). Our ``no CMB'' constraints are slightly tighter than those reported in Ref.~\cite{Bel:2022iuf} from a different combination of low-redshift probes.

\subsection{Results: \texorpdfstring{\neff}{Lg}}\label{sec:res_neff}

Much like curvature, the number of relativistic species \neff is weakly constrained by DES \mpp alone, but combining the DES with external data leads to modest improvements. 
The constraints are 
\begin{alignat}{2}
  \begin{aligned}  
 \neff &\leq 7.8\quad &&\text{DES\ Y3}\\[0.2cm] 
&= 3.10^{+0.15}_{-0.16}\quad &&\text{DES\ Y3 + External}.
 \end{aligned}
\end{alignat}

\begin{figure}
  \centering
  \includegraphics[width=\linewidth]{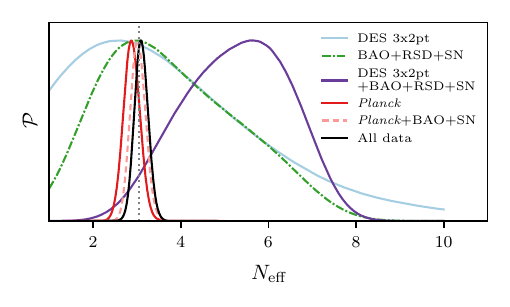} 
  \caption{Marginalized posterior on the number of relativistic species \neff. 
  The vertical dotted gray lines shows the standard model value of $\neff=3.044$.}
  \label{fig:res_neff}
\end{figure}

Constraints on the number of 
relativistic species are given in \fig{fig:res_neff}.
The DES \mpp and BAO+RSD+SN constraints both individually peak around $\neff\simeq 3$,  though we caution that these constraints are mainly shaped by prior projection effects. Both the \mpp and BAO+RSD+SN posteriors are unconstrained along $\neff-\Omega_b$ and $\neff-h$ degeneracy directions. Given our choice of priors, the upper bounds on \neff are shaped by the upper prior bound on $h$, while the lower bound on $\neff$ are determined by the lower prior bound for $\Omega_b$.  Both DES \mpp and RSD are sensitive to the amplitude of the power spectrum, which is affected by \neff through changes in the redshift of matter-radiation equality, while BAO is additionally sensitive to a small phase shift caused by \neff's impact on the Silk damping scale. 
These probes' posteriors have different degeneracies between \As and \neff, and the overlap between them rules out  small \neff values, while the upper bound is still primarily determined by where the posterior intersects the $\Omega_b$ prior. The constraints are still consistent with the standard-model value of \neff.  Once the \mpp{}+BAO+RSD+SN data are combined with {\it Planck}, the overall combined posterior shifts slightly compared to the {\it Planck}-only constraints, but  remain fully consistent with $\neff\simeq 3$.

\begin{figure*}
  \centering
\includegraphics[width=0.45\linewidth]{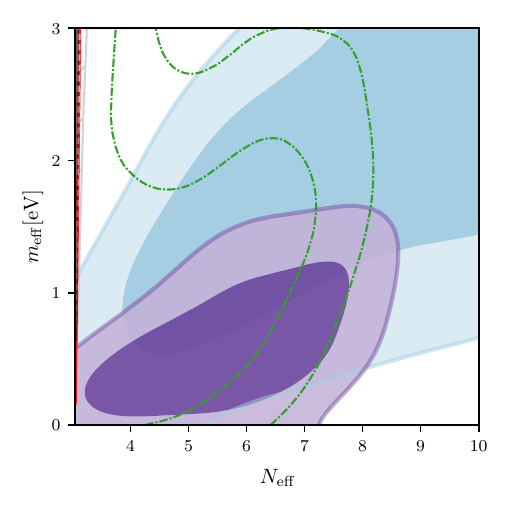} 
\includegraphics[width=0.45\linewidth]{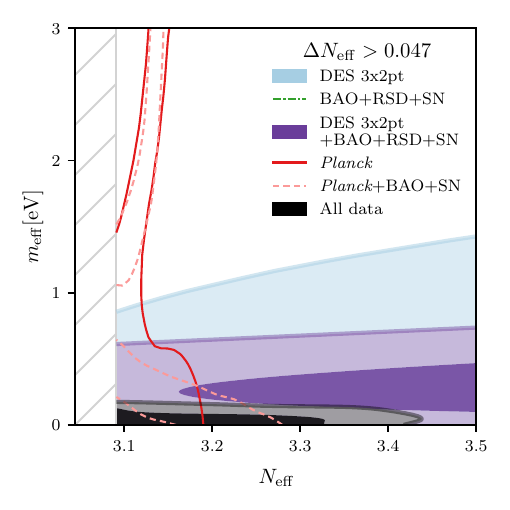}
\caption{Constraints on the  beyond-\lcdm parameters of the \neffmeff model. The left plot shows results obtained using the fiducial prior $3.044<\neff<10$, with axes spanning the full prior range. In it, the {\it Planck} and all-data   posteriors reside entirely inside the $m_{\rm th}>10$eV region where sterile neutrinos behave like CDM, shown in a very narrow gray hatched wedge along the left axis of the plot. The other contours show consraints from DES \mpp (blue), external low redshift data from BAO+RSD+SN (dashed green), and \mpp+BAO+RSD+SN (purple). The right panel shows chains run using an alternative prior where $\Delta\neff>0.047$, and with the plot range reduced to more clearly show the all-data (\mpp{}+BAO+RSD+SN+{\it Planck}) results. Because the DES \mpp and DES 3$\times$2pt+BAO+RSD+SN chains have very few samples in the excluded region, the alternative prior has no effect on the purple and blue contours. }
  \label{fig:res_neffmeff}
\end{figure*}
\subsection{Results: \texorpdfstring{\neffmeff}{Lg}}\label{sec:res_neffmeff}

We next consider the $\neffmeff$ model. Recall, following discussion in Sec.~\ref{sec:changepipe}, that this particular set of constraints is prone to parameter-space projection effects in the small-$\Delta\neff$ region, so our fiducial analysis imposes the constraint $\Delta \neff>0.047$, where $\Delta\neff\equiv\neff - 3.044$. 
Making that assumption, and reporting the constraints only for all data combined, we find 
\begin{equation}
    \begin{aligned} 
 \Delta\neff &<0.34, \quad\meff < 0.14\,\rm{eV}\quad \text{DES\ Y3 + External}.
    \end{aligned}
\end{equation}
The constraints are shown in \fig{fig:res_neffmeff}. 

We also consider constraints for a prior requiring $\Delta\neff>0$, but also $m_{\rm th}<10$ eV. For this prior, the constraints from all data combined are
\begin{equation}
    \begin{aligned} 
 \Delta\neff &<0.28, \quad\meff < 0.20\,\rm{eV}\quad \text{DES\ Y3 + External}.
    \end{aligned}
\end{equation}
This prior on the sterile neutrino's thermal mass matches the {\it Planck} 2018 analysis~\cite{Planck:2018vyg}, which used the same CMB likelihood as us, plus {\it Planck} lensing constraints (which we do not include) and BOSS DR12 BAO to constrain $\Delta\neff<0.23$ and $\meff<0.65$ eV. Thus, while we find slightly weaker constraints on \neff, the inclusion of DES \mpp and RSD effective measurements of the amplitude of structure at low redshifts allows us to tighten the upper bound on \meff by about a factor of three. 

Another interesting, if less direct, comparison can be made to the analysis of Ref.~\cite{Xu:2021rwg}. That work analyzed the {\it Planck} data including lensing, the full-shape BOSS DR12 measurement, and CFHTLens weak lensing measurements at fixed $\Delta\neff=0.047$, to find the constraint $\meff<1.6$ eV. They found that the weak lensing measurements were crucial to obtaining a tight constraint on $\meff$. While our results are not directly comparable (our \meff constraints would likely weaken if we performed our analysis at fixed, small $\Delta\Neff$), our findings lend further support to the idea that precise cosmic shear measurements of LSS can powerfully constrain the presence of light but massive relic particles produced in the early Universe.

\subsection{Results: \texorpdfstring{$\sigmu$}{Lg}}\label{sec:res_sigmu}

\begin{figure}
  \centering
\includegraphics[width=\linewidth]{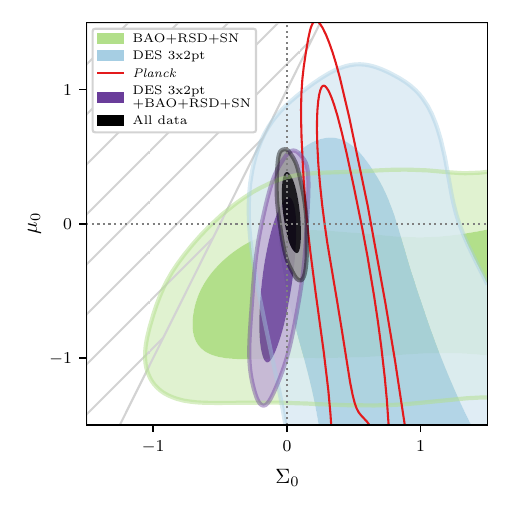}\\
  \caption{ Constraints on the $\sigmu$ modified gravity parameters, with axis ranges reflecting the parameters' prior ranges. The gray hatched region shows where samples are excluded by the $\muo<2\sigo +1$ requirement of {\sc MGCAMB}: any overlap of contours with that region is a reflection of KDE smoothing done by {\sc getdist}. }
  \label{fig:res_sigmu}
\end{figure}

Next, we show results of tests of gravity on cosmological scales parametrized by \sigmu in \fig{fig:res_sigmu}. As discussed in \sect{subsec:simfid}, \muo is not constrained by DES Y3 \mpp alone so will not be reported.
The \eq{eq:sigmu_prior} prior set on \sigmu is represented by a hatched area in the figure.
While the {\it Planck}-only contours are visibly offset from  the \mpp+BAO+RSD+SN combination, the  Suspiciousness tension metric comparing the two posteriors has $p=0.02$ (above the $0.01$ threshold), so we proceed with reporting results from the combination of all data. 
These constraints are 
\begin{equation}
  \begin{aligned} 
    \Sigma_0 &= 0.6\pm 0.4 \quad \text{DES\ Y3}\\[0.2cm]
    \Sigma_0 &= 0.04\pm 0.05,\quad \mu_0 = 0.08^{+0.21}_{-0.19} \quad \text{DES\ Y3 + External},
  \end{aligned}
\end{equation}
the latter of which can be compared to the external-only constraint, which is
$\Sigma_0=0.37^{+0.12}_{-0.09}, \mu_0=0.20\pm 0.22$.
Thus the addition of DES \mpp data both tightens the
constraints on $(\Sigma_0, \mu_0)$ and shifts them to be more consistent with  general relativity. 

The constraint on \sigo from DES Y3 \mpp data alone is limited by the linear scale cuts used for this model, which actually results in the \mpp-only bounds on \sigo presented here to be weaker than the comparable DES-Y1Ext result by 40$\%$. To understand this, recall from \sect{sec:fidscuts} that we define these linear scale cuts by iteratively removing small-angle measurements until the difference between  \mpp model predictions with and without nonlinear modeling are deemed insignificant. We assess the significance of that difference relative to the data covariance, which means that as measurements get more precise this method produces more stringent cuts. Quantitatively, \sigmu constraints in the present analysis are only based on 55\% of the fiducial data vector, while in  DES-Y1Ext the linear cuts retained 73\% of the data points.
The weakened \sigo constraints  imply that this method for protecting against the Y3 measurements' greater sensitivity to non-linear effects reduces the S/N available for cosmology inference and indicates the need for a more sophisticated method of accommodating for nonlinear modeling uncertainties as data get more precise.

Given the  offset of the {\it Planck} contour as well as complementary of growth measurements from RSD, which primarily constrain \muo,  and \mpp, which constrain \sigo, it is interesting to report modified gravity constraints from the combination of only low-redshift probes. This result is shown in purple in \figs{fig:res_sigmu}, leading to measurements of \sigmu independent of CMB observables:
\begin{equation}
\centering
  \begin{aligned} 
    \Sigma_0 = -0.06^{+0.09}_{-0.10},\quad \mu_0 = -0.4\pm0.4 \\
\text{DES\ Y3 + External, no CMB}.
  \end{aligned}
\end{equation}
Results from the combined constraints is partly driven by the BAO+RSD+SN measurement of $\mu_0= -0.5^{+0.4}_{-0.5}$, indicating $\muo<0$ at 1$\sigma$  significance. This is in contrast to the results reported in Ref.~\cite{Alam_2021}, in which analysis of the  same BAO+RSD data assuming a fixed background cosmology produces \muo constraints  centered on zero.

\begin{figure*}
  \centering
\includegraphics[width=6in]{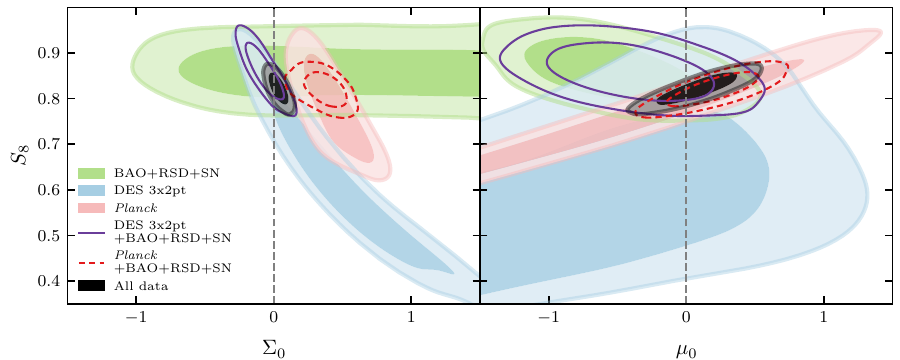}\\
  \caption{ Examination of how RSD growth information, as part of the BAO+RSD+SN external low redshift data combination, breaks degeneracies between \seight and modified gravity parameters \sigo and \muo when combined with either DES \mpp or {\it Planck}. }
  \label{fig:res_sigmu_vs_s8}
\end{figure*}

To aid in the interpretation of these results,  \fig{fig:res_sigmu_vs_s8} shows how different data combinations break degeneracies between the modified gravity parameters and \seight. The fact that lensing observables are sensitive to the product product $\seight\sigo$ limits the ability of DES \mpp alone to constrain \sigo. By constraining \seight and \muo, RSD measurements of the growth rate of structure are able to break that degeneracy and thus improve constraints on \sigo. 

The behavior of the {\it Planck}-only contours can be understood in terms of two relevant degeneracies. First, {\it Planck}'s constraints on \sigo mainly come from the impact of lensing, which smooths the high-$\ell$ peaks of the CMB power spectra.
(The integrated Sachs-Wolfe effect~\cite{Sachs:1967er,Hu:1993xh} also introduces some sensitivity to \sigo, but cosmic variance affecting the low multipoles where that occurs limits its constraining  power.) That lensing signal, like DES cosmic shear, is sensitive to the product $\seight\sigo$, and thus leads \seight-\sigo degeneracy parallel to that seen for \mpp.   Second, \muo  adds a degree of freedom to the relationship between CMB constraints on the primordial power spectrum amplitude $\As$ and the amplitude of density fluctuations in the late Universe. This leads to a positive degeneracy between \seight and \muo:  for a given value of  \As, which is tightly constrained by {\it Planck}, larger \muo will enhance structure growth and thus increase \seight. Combining {\it Planck} with the BAO+RSD+SN data allows the RSD observables' direct measurement of \seight break both of these degeneracies. 

The fact that the {\it Planck} contours, either from {\it Planck} alone or in combination with the BAO+RSD+SN low redshift data, are offset towards higher \sigo than \mpp follows a trend seen for DES Y1 data in both DES-Y1Ext and Ref.~\cite{Planck:2018vyg}. As noted in Ref.~\cite{Planck:2018vyg}, this preference is driven by the excess smoothing of  high-$\ell$ {\it Planck} measurements that that are captured by the phenomenological $A_{\rm L}$ parameter. These are the same features that pull the {\it Planck}-only \ok constraints towards negative values. When all data are analyzed together, the \sigo constraints are in agreement with those from \mpp, with the CMB measurements contributing to tightening constraints primarily by breaking the RSD posterior's weak degeneracy between \sigeight and \muo.

\subsection{Results:  \texorpdfstring{\npg}{Lg}}\label{sec:res_npg}

\begin{figure}
  \centering
\includegraphics[width=\linewidth]{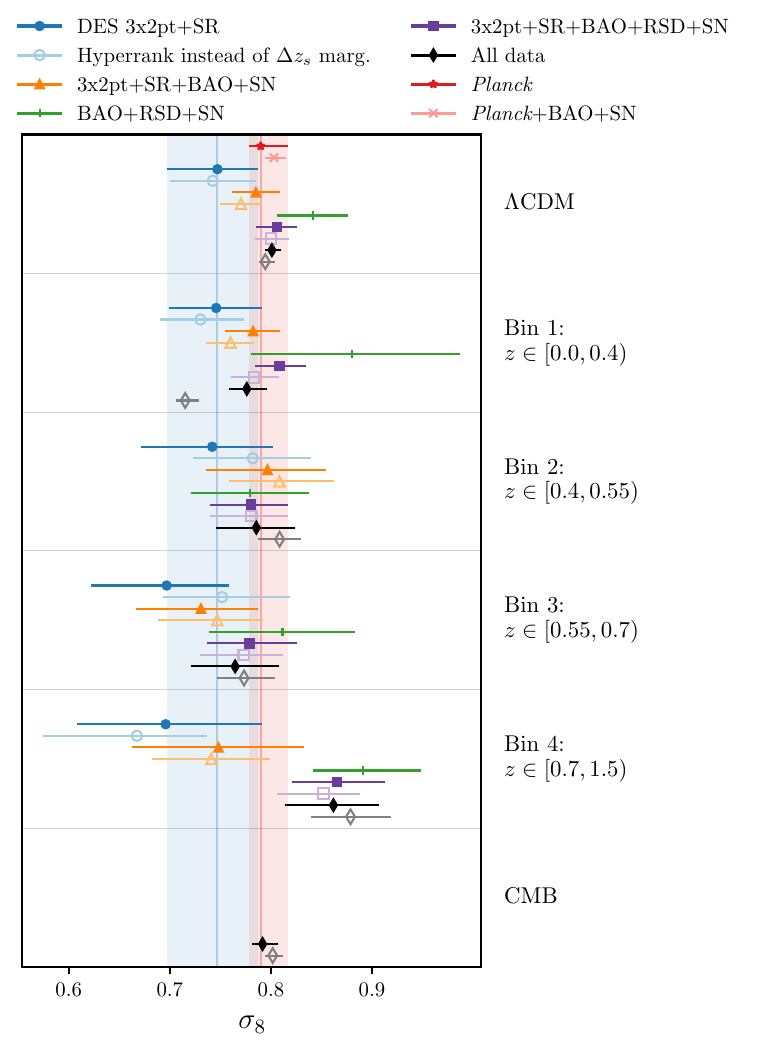}
  \caption{Constraints on  \npg  compared to \lcdm constraints in the top row. Points and error bars show the marginalized posterior mean and 68\% confidence intervals. Unfilled, lighter markers correspond to the same data as the darker points with matching colors and shapes, but were obtained using the hyperrank marginalization  over uncertainties in source galaxy redshift distributions.  Shaded bands highlight the location of the top row's \mpp and {\it Planck} \lcdm points for comparison.}
  \label{fig:npg_sig8_table}
\end{figure}

Finally, we report constraints on the \npg model. We begin by examining the  set of derived parameters $\npgsig{i}$ (see Eqs.~(\ref{eq:npg_params}-\ref{eq:news8})), which correspond to the values of \sigeight inferred from LSS observed in redshift bin $i$, and which  we showed in \sect{sec:changepipe} are more robust to model variations than the sampled $\Anpg_i$ parameters.  
These constraints are:
\begin{equation}
  \begin{aligned} 
  \npgsig{1} &= 0.75^{+0.05}_{-0.05}, \\[0.2cm]
    \npgsig{2}&= 0.74^{+0.06}_{-0.07}, \\[0.2cm]
    \npgsig{3} &= 0.70^{+0.06}_{-0.07}, \qquad
     \text{DES\ Y3\ \phantom{+ External}}\\[0.2cm] 
    \npgsig{4} &= 0.70^{+0.10}_{-0.09}, 
  \end{aligned}
\end{equation}  
and
\begin{equation}
  \begin{aligned} 
  \npgsig{1} &= 0.78^{+0.02}_{-0.02}, \\[0.2cm]
    \npgsig{2} &= 0.79^{+0.04}_{-0.04}, \\[0.2cm]
    \npgsig{3} &= 0.76^{+0.04}_{-0.04}, \qquad
     \text{DES\ Y3\ + External\ }\\[0.2cm]
    \npgsig{4} &= 0.86^{+0.04}_{-0.05}, \\[0.2cm]
     \npgsigcmb &= 0.792^{+0.015}_{-0.010},   \end{aligned}
\end{equation}  
\fig{fig:npg_sig8_table} presents these constraints in comparison to \lcdm constraints on \sigeight. In that figure, the set of lighter, unfilled data points show how the $\npgsig{i}$ constraints change when use the alternative hyperrank method of marginalizing over source galaxy photo-$z$ uncertainties. We find that hyperrank induces non-negligible but still small ($\sim 0.5\sigma)$ shifts in $\npgsig{i} $  for the \mpp-only $i\in\{2,3\}$ measurements, and all-data  $i\in\{2,\text{CMB}\}$ measurements, while a much larger, almost $3\sigma$ shift occurs for the all-data constraint on $\npgsig{1}$. As is discussed in more detail in \apps{app:changepipe_npg} and~\ref{app:hyperrank}, the lack of robustness of the lowest redshift is likely due to an interaction between the source $n(z)$ and IA modeling which is most significant at low redshifts. In the same Figure, we report additional results to facilitate interpretation of how different structure growth observables contribute to constraints. Namely, we show the combination of DES data with only geometric external data (\mpp+BAO+SN) shifts constraints to slightly higher \seight in both \lcdm and binned \sigeight, but not as much as the \mpp+BAO+RSD+SN data combination. Thus the combined analyis' shift towards higher \sigeight, especially in the highest redshift bin 4, seems to be primarily driven by the RSD likelihood.

\begin{figure*}
  \centering
\includegraphics[width=\linewidth]{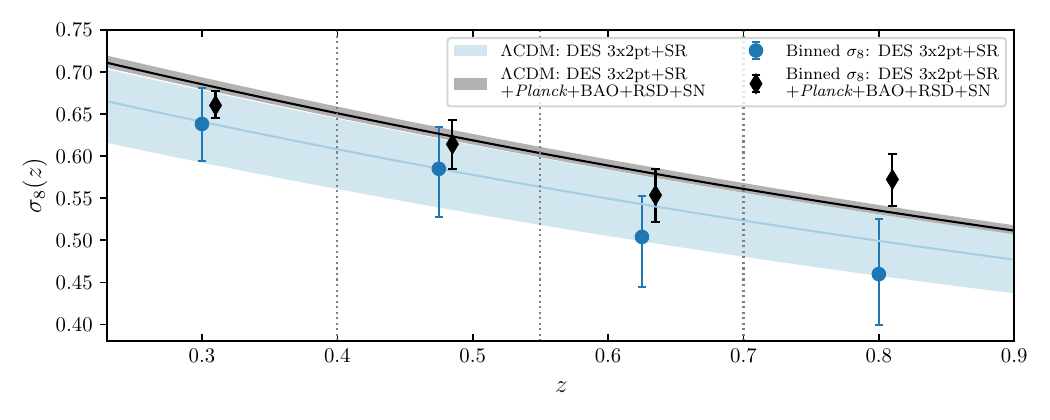}
  \caption{Inferred $\sigeight(z)$ from \lcdm (lines) and the \npg model (points). Lines and shaded bands show the mean and 68\% confidence interval inferred from \lcdm posteriors, with DES \mpp + shear ratio shown in blue and the combination of all data (\mpp{}+SR+Planck+BAO+RSD+SN) shown in black. 
  Points with error bars show \npg results for the same data combinations, plotted at an example redshift for each lens bin.  For readability, the all-data (black) points have been shifted by a small horizontal offset.  Vertical dashed lines show the bin divisions used to define  the \npg model's $\Anpg_i$ amplitudes. }
  \label{fig:npg_sig8z}
\end{figure*}

\fig{fig:npg_sig8z} translates these results to the inferred growth function $\sigeight(z)$. That figure compares marginalized constraints when we vary the \npg amplitude parameters shown with data points  at a few example redshifts, to the 68\% confidence bands obtained from \lcdm fits to DES \mpp and all data (3$\times$2pt+BAO+RSD+SN+{\it Planck}).

 All measurements are within approximately $1\sigma$ of the \lcdm \sigeight estimate. The fact that the DES-only constraints on $\sigeight(z)$ are consistently lower than {\it Planck} and that  our combined constraints find $\npgsig{4}$ to be  higher than \npgsig{i} in the other bins agrees with similar features seen in  Refs.~\cite{White:2021yvw,Brieden:2022lsd,Garcia-Garcia:2021unp}. In those works, analyses of DESI galaxies cross correlated with \textit{Planck} lensing, eBOSS QSO clustering, and both of those observables combined with DES Y1 \mpp measurements, respectively, suggest that the amplitude of structure at $z\sim 0.8$ may be slightly higher compared to lower redshift measurements, and thus hinting at a slower growth rate than expected in \lcdm. However the trends seen in these references, as well as that in our work, are only significant at the  $\sim$1$\sigma$ level and thus not strong enough to motivate any kind of firm conclusion. We also note that a similar trend is not found in Ref.~\cite{Garcia-Quintero:2020bac}'s binned modified gravity study using DES Y1 and BOSS DR12 LSS data. 
It will be interesting to monitor how new and more precise data constrain the time-evolution of the amplitude of density fluctuations.  

For completeness, we additionally report constraints on the sampled amplitude parameters used to implement this model,  $\Anpg_2$, $\Anpg_3$,  $\Anpg_4$, and $\Anpg_{\rm CMB}$ (see \eq{eq:npg_params}).  We emphasize that these constraints should be interpreted with caution, because in \sect{sec:changepipe} we found that they are not robust to a change in how we marginalize over our source photo-$z$ uncertainties. Particularly for the all-data constraints, this occurs because these sampled amplitude parameters are defined relative to bin 1, so the sensitivity to hyperrank seen for $\npgsig{1}$ propagates to $\Anpg_i$  inferences for higher redshift bins. Specifically, switching from our fiducial analysis to one with the hyperrank $n_s(z)$ marginalization scheme produces shifts between $0.5-2\sigma$ shifts in all of the amplitude parameters.  The hyperrank-model constraints on these parameters can be found in \tab{tab:results_extparams}.  
The constraints on these parameters using our baseline model are 
\begin{equation}
  \begin{aligned} 
    \Anpg_2 &= 1.00^{+0.14}_{-0.21}, \\[0.2cm]
    \Anpg_3 &= 0.88^{+0.14}_{-0.19}, \quad
     \text{DES\ Y3\ \phantom{+ External}}\\[0.2cm] 
    \Anpg_4 &= 0.90^{+0.20}_{-0.26}, 
  \end{aligned}
\end{equation}  
and
\begin{equation}
  \begin{aligned} 
    \Anpg_2 &= 1.03^{+0.11}_{-0.14}, \\[0.2cm]
    \Anpg_3 &= 0.98^{+0.11}_{-0.13}, \quad
     \text{DES\ Y3 + External\ }\\[0.2cm]
    \Anpg_4 &= 1.24^{+0.13}_{-0.16},\\[0.2cm]
        \Anpg_{\rm CMB} &= 1.04^{+0.04}_{-0.06}.
  \end{aligned}
\end{equation}

\subsection{Impact of model and analysis choices on \texorpdfstring{\seight}{Lg}}\label{sec:res-S8}

\begin{figure}
  \centering
\includegraphics[width=\linewidth]{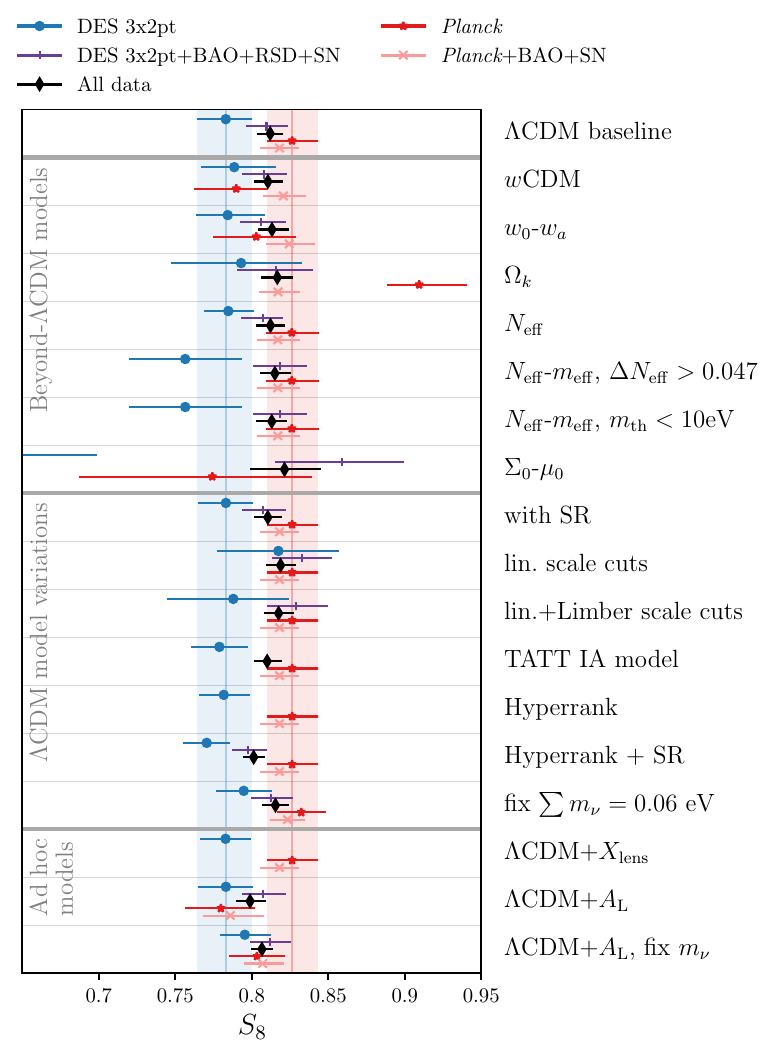}
  \caption{Constraints on \seight in different models and under different analysis assumptions. Points and error bars show the means and marginalized 68\% confidence, and the shaded bands mark the location of the top row's \lcdm baseline points for \mpp and {\it Planck}. Missing points simply indicate chains that were not run as part of other robustness tests. The rows with `SR' in the label and those varying $A_{\rm L}$ include shear ratio as part of the DES \mpp likelihood, all others do not. The  \sigmu \mpp constraint of $\seight=0.61^{+0.09}_{-0.16}$ is cut off to improve the dynamic range for other points in the plot. }
  \label{fig:S8table}
\end{figure}

In line with studies exploring whether beyond-\lcdm models  alleviate the tension between weak lensing and {\it Planck} \seight measurements (see \sect{sec:intro} and e.g.~Refs.~\cite{Abdalla:2022yfr,FrancoAbellan:2020xnr,Davari:2019tni,SolaPeracaula:2020vpg,SolaPeracaula:2019zsl,Camera:2017tws,DiValentino:2018gcu,Murgia:2020ryi,Archidiacono:2019wdp,DiValentino:2019dzu,Chudaykin:2022rnl,Akarsu:2021fol,Lucca:2021dxo,SolaPeracaula:2021gxi,Benevento:2022,Heimersheim:2020aoc}), in \fig{fig:S8table} we compare constraints on \seight obtained within a selection of cosmological models and, for comparison, analysis choices within the \lcdm model. In that Figure, points show the mean and 68\% confidence intervals of the marginalized \seight posterior, with different colors and marker styles corresponding to results  from different sets of observables. We report constraints for DES \mpp (blue), the combination of DES and other low redshift probes (\mpp{}+BAO+RSD+SN, purple), {\it Planck} alone (red), {\it Planck} combined with low-redshift geometric probes BAO+SN to break geometric degeneracies (pink), and the combination of all data (\mpp{}+BAO+RSD+SN+{\it Planck}, black). Blue and red vertical lines and bands mark the location of the baseline \lcdm DES \mpp and {\it Planck} constraints shown in the top row, to indicate the level of offset between those measurements and to facilitate comparisons with other rows. 

The first group of \seight constraints shown are for beyond-\lcdm models. These include \wcdm along with the extended models studied in this paper, except for the \npg model for which  $\sigma_8$ constraints were discussed above in \sect{sec:res_npg}. 
In these extended models, we see that the most overlap between {\it Planck} and \mpp \seight constraints occurs for dynamical dark energy described by \wcdm and \wowa. For both \lcdm and beyond-\lcdm models, we find that the combination of low redshift probes, i.e.\ combining DES Y3 \mpp with BAO, RSD and SN as shown in purple, measure \seight to be more consistent with {\it Planck} constraints than \mpp alone. This repeats the same finding of DES-Y3KP (see Figs.~14 and~15 of Ref.~\cite{y3-3x2ptkp}).  The behavior occurs  because the external geometric (BAO+SN) probes constrain \om to be at the higher end of the range allowed by the \mpp-only constraints. Given the \om--\sigeight degeneracy, this leads to a higher \seight value. 

For comparison, additional blocks of points in \fig{fig:S8table} show how constraints on \seight are affected by changes to the analysis choices while retaining the \lcdm model, and the impact of two  extensions to \lcdm which we label ad hoc  models. The \lcdm analysis choice variations include using the shear ratio likelihood, different scale cuts, the more general TATT IA model, the hyperrank method for marginalizing over source photo-$z$ uncertainties, and fixing neutrino mass. The ad hoc models include varying  \xlens, which introduces a mismatch between the galaxy bias affecting galaxy clustering and that affecting galaxy--galaxy lensing (see its description as part of the robustness tests of \sect{sec:changepipe}), and varying $A_{\mathrm{L}}$~\cite{Calabrese:2008rt}, which scales the amount of lensing-related smoothing affecting the CMB temperature and polarization power spectra (See \sect{sec:npg} and e.g. Ref.~\cite{Planck:2018vyg}). 
Both these ad hoc models correspond to the introduction of a parameter to explain features in the DES Y3 \mpp and {\it Planck} data respectively, and not new physics as opposed to the beyond-\lcdm models considered in this analysis.
While \xlens has little effect on the \mpp constraints and thus on the \mpp-{\it Planck} comparison, varying $A_{\mathrm{L}}$  leads to more  consistent estimates of \seight across all probe combinations shown (see also~\cite{DiValentino:2018gcu,DiValentino:2019dzu}). 

\subsection{Impact of model choice on \texorpdfstring{\summnu}{Lg}}\label{sec:res-mnu}

\begin{table}
\include{figures/realy3dat.mnu_textable_manual}
\caption{Impact of the cosmological model on the 95\% confidence upper bound  on the sum of neutrino masses. The ``All external'' column reports constraints from the combination of {\it Planck}+BAO+RSD+SN, while ``All data'' additionally includes DES \mpp constraints (\mpp{}+SR in the case of \npg and $A_{\rm L}$).} \label{tab:mnu}
\end{table}

\begin{figure}
  \centering
\includegraphics[width=\linewidth]{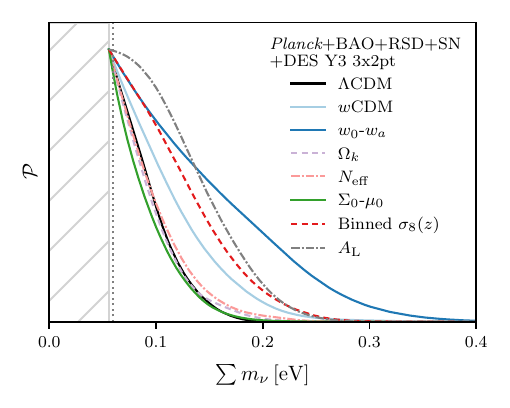}
  \caption{Marginalized posterior for the sum of neutrino masses obtained from the analysis of DES \mpp{}+BAO+RSD+SN+{\it Planck} in different cosmological models. The dotted vertical gray line shows the minimum mass allowed by neutrino oscillation measurements, $\summnu=0.06$ eV, and the hatched area shows the region excluded by the $\Omega_{\nu}h^2>0.006$ prior.}
  \label{fig:mnu}
\end{figure}

Next, we examine constraints on the sum of neutrino masses \summnu. While DES \mpp data are not particularly sensitive to this parameter, the fact that we vary $\Omega_{\nu}h^2$ as part of our baseline analysis allows us to study how its bounds are impacted by the assumed cosmological model. With this aim,   \tab{tab:mnu} reports the 95\% upper bound on \summnu from {\it Planck}+BAO+RSD+SN with and without DES \mpp for several of the models considered in this paper, while \fig{fig:mnu} shows the one-dimensional marginalized posterior from the combination of all data. For all models other than \sigmu and  \npg, 
including the DES \mpp likelihood either has no effect or slightly  weakens the bounds on \summnu. 

Neutrino mass constraints have a strong dependence on assumptions about the  time evolution of dark energy, with the all-data upper bound on \mnu increasing relative to the \lcdm value by a factor of 1.4 for \wcdm and 1.9 for \wowa. In contrast,  \ok and \neff have little impact. Modifying gravity with \sigmu weakens external data constraints on neutrino mass by external data, but when DES \mpp is included the resulting constraint  matches that of \lcdm.  Relaxing assumptions about the evolution of structure growth by binning $\sigeight(z)$ increases the bound by a factor of 1.4.

\subsection{Tension between DES and external data for extended models}\label{sec:res-tensions}
\begin{figure*}
 \centering
\includegraphics[width=0.8\linewidth]{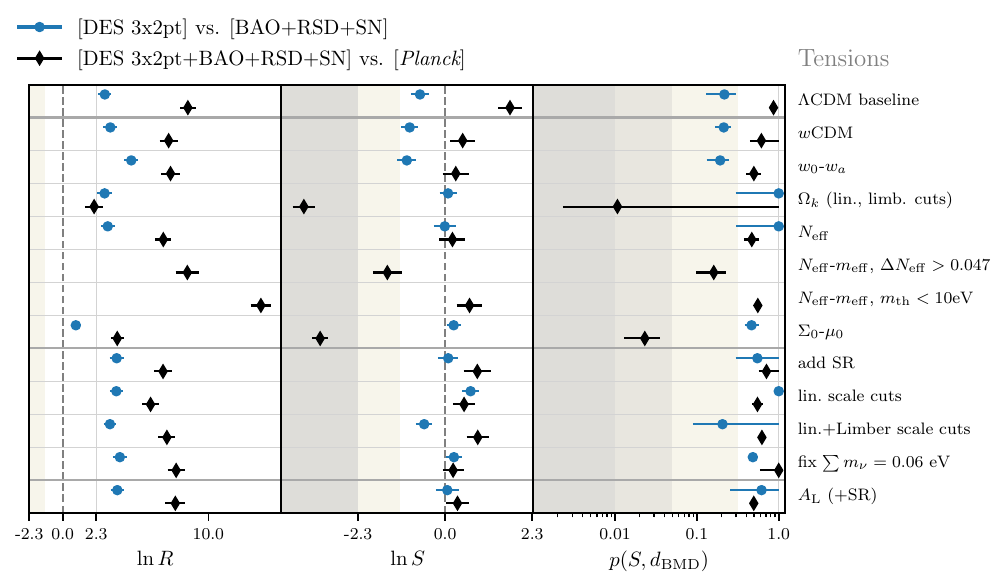}
  \caption{Measurements of tensions between datasets assuming various cosmological models and analysis variations. Blue circular points denote tensions between DES \mpp and low-$z$ BAO, RSD, and SN data, while black diamonds show tensions the combination of all of that  low-$z$ data (\mpp{}+BAO+RSD+SN) to Planck CMB constraints. For all panels, points further to the left indicate greater tension. The statistics $\ln{R}$ and $\ln{S}$ are the Bayesian evidence ratio and Suspiciousness, respectively. Their points and uncertainties reflect the mean and standard deviation of realizations estimating sample variance using the {\sc Anesthetic} software, and shaded regions correspond to substantial and strong evidence of tension according to the Jeffreys' scale. (The Jeffrey's scale is relevant for Suspiciousness because $\ln{S}$ can be interpreted as the value of $\ln{R}$ associated with the narrowest choice of uninformative prior.) The quantity $p$ measures the significance 
  accounting for the number of constrained parameters via the Bayesian model dimensionality $d_{\rm BMD}$. In the plot, $p$-value errors indicate the 0.16 and 0.84 quantiles of the sample variance realizations, points indicate the mean, and the  shaded regions highlight probabilities associated with $1\sigma$, $2\sigma$, and $3\sigma$ tension. If error bars are not seen they are smaller than the size of the marker. Further information and definitions of these tension metrics can be found in \app{app:tensions}.}
  \label{fig:tension_tableplot}
\end{figure*}

In \fig{fig:tension_tableplot} we report measures of tension between DES \mpp and external low-redshift probes (BAO+RSD+SN), as well as between the combination of all low redshift probes (\mpp{}+BAO+RSD+SN) and {\it Planck}.  We evaluate the significance of tension using three statistics, which are discussed in more detail in \app{app:tensions}: the Bayes ratio $R$, the Suspiciousness $S$, and a $p$ value that converts $S$ into a tension probability. We define these quantities such that $\ln{R}<0$ and $\ln{S}<0$ correspond to evidence of tension. The quantity $p(S,d_{\rm BMD})$, where BMD stands for ``Bayesian model dimensionality'' (see \eq{eq:dbmd_for_tension} for more details), approximates the probability, assuming a null hypothesis of agreement between datasets, that we will find a value of $\ln{S}$ as low or lower than the observed value. Thus, small $p$ corresponds to stronger tension. This probability is assessed using both $S$ and the quantity $d_{\rm BMD}$, which is a Bayesian estimate of the number of directions in parameter space in which a tension could be meaningfully detected --- that is, which are constrained by both datasets independently. 
We show multiple statistics here because while $R$ is likely to be the most familiar to readers, it has an undesirable sensitivity to the choice of flat prior ranges. In contrast, $S$ is insensitive to the prior range of well-constrained parameters, and its significance assessed via $p(S,d_{\rm BMD})$ is expected to agree with a number of other proposed tension metrics~\cite{y3-tensions}.  

We estimate $\ln{R}$, $\ln{S}$ and $d_{\rm BMD}$ using the  {\sc Anesthetic} software\footnote{\url{https://github.com/williamjameshandley/anesthetic}}~\cite{Handley:2019mfs}, which produces an ensemble of 200 realizations capturing the uncertainty introduced by sampling variance.  For $\ln{R}$ and $\ln{S}$, in \fig{fig:tension_tableplot} we report the mean of this ensemble, and use error bars (which are occasionally smaller than the datapoint) to show the standard deviation.  For the $p$-values, whose  ensemble distribution is significantly non-Gaussian, we report the median, and  approximate its one-sigma sampling variance errors using the 16 and 84\% quantiles. We use the threshold of $p(S,d_{\rm BMD})\geq0.01$ as a requirement for reporting combined constraints. 

For all models and statistics, there is no indication of any tension between DES \mpp and the external low-redshift probes (BAO+RSD+SN). This is also true for almost all evaluations of tension between the combination of all low-redshift probes  \mpp{}+BAO+RSD+SN and {\it Planck}. The only cases where we find significant tension are for the \ok  and \sigmu comparison of  \mpp{}+BAO+RSD+SN versus {\it Planck}, both of which have a significance between 2-$3\sigma$. 

As was noted above in \sect{sec:res_ok}, for \ok the $p$-value median is 0.010, exactly at our threshold for reporting combined constraints. This  merits further discussion, because in addition to being the most significant measure of tension reported, it is also the noisiest. The  16\% and 84\% quantiles are 0.002 and 1.0, respectively\footnote{As we will discuss in \app{app:tensions}, we assign $p=1$ to realizations with $d_{\rm BMD}<0$, reasoning that there can be no tension measured if  there are no shared parameters in the two datasets' independent constraints.}. This means that at an approximately $1\sigma$ level of certainty, our evaluation of tension between {\it Planck} and low-redshift \ok constraints  could plausibly be consistent with  both a  slightly-greater-than-$3\sigma$ tension and with there being no tension at all. This large scatter is driven by the small value of $d_{\rm BMD}=1.5\pm 1.6$ (reporting the mean and standard deviation from the sample variance estimate). This small $d_{\rm BMD}$ means that that there is limited overlap in the parameter directions constrained by {\it Planck} and \mpp{}+BAO+RSD+SN, making the assessment of tension  extremely sensitive to noise in the posterior estimates. To further contextualize this finding, we note that the   {\it Planck}-only preference for $\ok<0$ driving this tension signal has been the subject of extensive discussion in the literature (see e.g. \cite{Planck:2018vyg,Park:2017xbl,Handley:2019tkm,DiValentino:2019qzk,DiValentino:2020hov,Efstathiou:2020wem,Glanville:2022xes,Hergt:2022fxk}) which highlights the fact that the interpretation of this tension can depend on subtleties related the choice of priors, parameters sampled, {\it Planck} likelihood calculation method, as well as the relationship to features in the {\it Planck} power spectra also  captured by phenomenological parameter $A_{\rm L}$. 

For modified gravity, we see a tension between {\it Planck}  and the other data that is likely driven by the same $A_{\rm L}$-like features of the CMB power spectrum. For \sigmu, the tension measurement is less signficant and much less noisy than for \ok: the median Suspiciousness $p$-value is 0.024 with 16\% and 84\% quantiles  of 0.013 and  0.039. 

Note that \fig{fig:tension_tableplot} does not show tension results for the \npg model. This is because, as described in \sect{sec:npg}, in that model we sample different sets of parameters when fitting DES and {\it Planck} constraints separately and in combination. This makes tension metrics difficult to evaluate,  so for simplicity we will show combined `all-data' constraints on \npg without checking a tension metric. This should be a reasonably safe choice because that model's $\Anpg_i$ and $\Anpg_{\rm CMB}$ parameters introduce enough modeling freedom to capture any differences between observables.

\subsection{Assessing the preference for extended models relative to \texorpdfstring{\lcdm}{Lg}}\label{sec:res-modcomp}

\begin{figure*}
  \centering
\includegraphics[width=\linewidth]{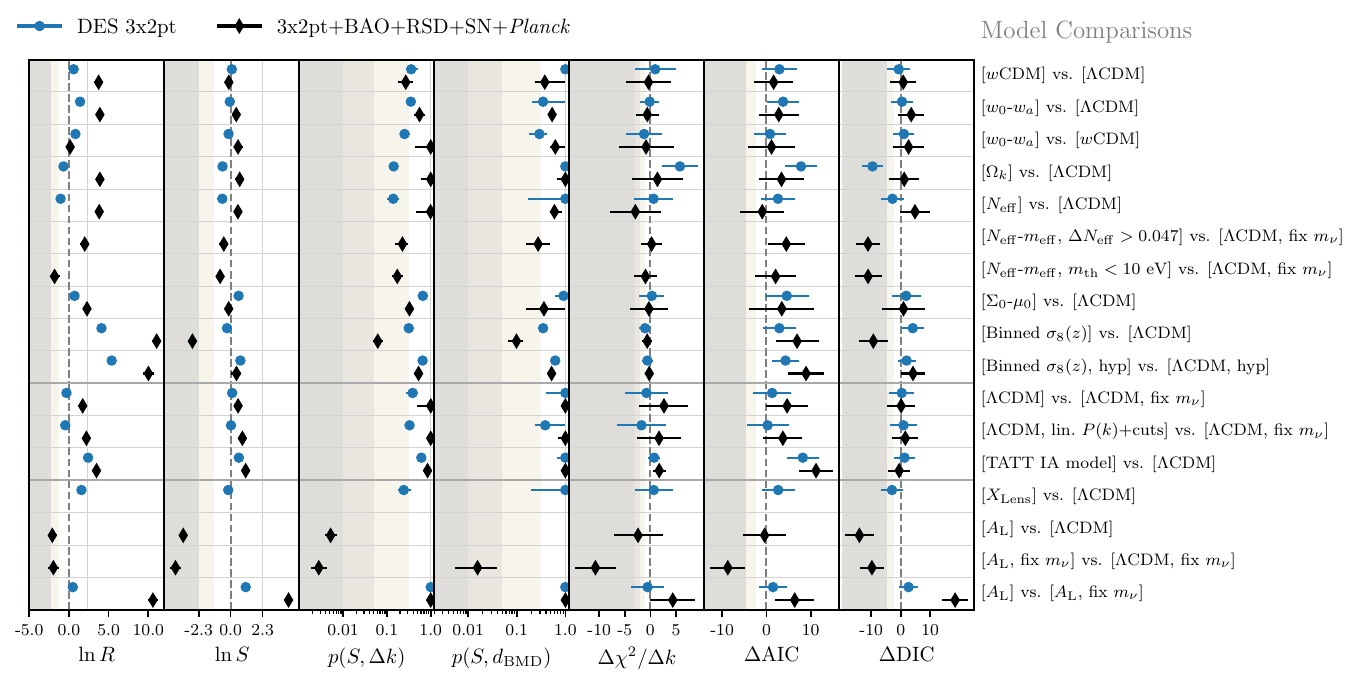}
  \caption{Model comparison metrics evaluated between pairs of nested models as listed on the right-hand side. All pairs are arranged so that the model listed first has more parameters. In all panels, points further to the left indicate more of a preference for the extended model. Blue points report metrics based on DES \ythree constraints alone, while black points are for DES \ythree combined with the {\it Planck}, BAO, RSD, and SN likelihoods. For the Bayes ratio $\ln{R}$ and Suspiciousness $\ln{S}$, error bars report the standard deviation associated with sampling variance and shaded regions show regions of  substantial and strong preference for the extended parameter space according to the Jeffreys' scale. The two sets of $p$ values evaluate the significance of the $\ln{S}$ results assuming a change in degree of freedom associated with the additional number of sampled parameters $\Delta k$ in the extended model and the Bayesian model dimensionality $d_{\rm BMD}$. The $p$-value points correspond to the mean estimate for sampling variance realizations, the uncertainties correspond to the 0.16 and 0.84 quantiles, and the shaded regions denote probabilities corresponding to $1\sigma$, $2\sigma$, and $\geq 3\sigma$. For the change in maximum posterior goodness-of-fit, $\Delta\chisq$, errors reflect the propagated standard deviations of 15 MAP estimates  performed for each chain, and shaded regions have boundaries at $\Delta\chisq/\Delta k=-1$, $-2$, and $-3$. For the information criterion statistics $\Delta$AIC and $\Delta$DIC, uncertainties reflect the same MAP estimate scatter, and the shaded regions are where model likelihoods quantified by Akaike weights (see e.g.\ Ref.~\cite{Liddle:2007fy}) match the probabilities associated with the $\ln{R}$ Jeffreys scale boundaries. Definitions and more information about these model comparison statistics can be found in Appendix~\ref{app:modcomp}.}
  \label{fig:modcomp_tableplot}
\end{figure*}

\fig{fig:modcomp_tableplot} shows several model comparison statistics.  We  show a variety of metrics here because it allows us to compare the results of different model comparison tests, and to account for the fact that readers may have different preferences regarding which of these tests are most familiar or interpretable. Points further to the left of each subpanel indicate \lcdm to be more  disfavored with respect to the extended model, while those on the right side of the panels favor \lcdm and disfavor the extension. 
Definitions of these metrics and details about how they are computed can be found in \app{app:modcomp}, though we will summarize them here.
The metrics include the  Bayesian evidence ratio $R$ and Suspiciousness $S$, both defined so that $\ln{R}<0$ and $\ln{S}<0$ indicates that the data disfavour \lcdm. 
We also report the ratio of the change in the maximum posterior goodness-of-fit to the number of added parameters, $\dchisq/\Delta k$, $\Delta$AIC which is an information-theory derived metric based on $\Delta\chisq$ with a penalty for adding parameters, and $\Delta$DIC which is related to $\Delta$AIC but adjusted for the number of parameters constrained by the data. 

For Suspiciousness, we report two $p$-values converting $\ln{S}$ to  probabilities: $p(S,d_{\rm BMD})$, in which $d_{\rm BMD}$ uses Bayesian model dimensionality to quantify the number of additional parameters constrained in the beyond-\lcdm analysis, and $p(S,\Delta k)$ which instead simply uses the number of additional sampled parameters $\Delta k$ (e.g. $\Delta k=1$ for \ok, $\Delta k = 2$ for $\wowa$, etc.). The model comparison definition of $S$ can be shown to be equivalent to the change in posterior-averaged log-likelihood for different models (see Ref.~\cite{fastismore}, and for the analogous relation in the case of data tensions, Ref.~\cite{Heymans:2020gsg}). This means $p(S,\Delta k)$ can be viewed as a Bayesian analog to evaluating the probability of the best-fit $\dchisq/\Delta k$. 

Like $\dchisq/\Delta k$, the information provided by $\ln{S}$ for model comparison is inherently asymmetric: the ``hardening'' of $\ln{S}$  against prior choice means that while it can be used to quantify the significance of preference for an extended model, it will never definitively favor the model with fewer parameters. This is in contrast to $\ln{R}$, for which tight constraints around \lcdm parameter values relative to the prior range will cause the extended model to be definitively disfavored.

Examining $R$,  
we find that none of the extended models are significantly preferred against \lcdm, with the combined data usually mildly or definitively favoring \lcdm. The only exception to this is for \neffmeff when we impose an upper bound on the sterile neutrino's thermal mass, where according to the Jeffreys scale we find a substantial (but not strong)   preference for the extended model. The preference for \lcdm reaches a particularly decisive level in the \npg case, unsurprisingly as it is a phenomenological model adding several parameters without much enhancing the overall fit to data.   
Considering model variations while assuming \lcdm, we find that there is neither a preference for varying neutrino mass nor for varying the additional TATT IA parameters.  Relative to the \neffmeff results, the preference against varying active neutrino mass is comparable to that against \neffmeff  when the $\Delta \neff >0.047$ prior is used, but that the sterile neutrino model is more favored when we use the $m_{\rm th}<10$ eV prior. For the ad hoc models,  fixing \xlens is favored over varying it, in line with previous studies of the \maglim lens sample, and that there is strong but not significant preference for varying  $A_{\rm L}$, in line with previous {\it Planck} analyses~\cite{Planck:2018vyg,BOSS:2016wmc,DiValentino:2019dzu}. 
 
Turning our attention to $\ln{S}$, we see that panel generally reports values lower than $\ln{R}$, reflecting the expectation that Suspiciousness is not able to significantly favor \lcdm over extended models.  
When assessing the significance of $S$ measurements using the associated $p$-values,  we find qualitative agreement with the evidence ratio findings. For both the main set of beyond-\lcdm models and model variations within \lcdm, most preferences for the extended models over \lcdm are less than $1\sigma$ significance, and all are less than $2\sigma$.  The strong $\ln{R}$ evidence for \neffmeff with  $m_{\rm th}<10$ eV translates via $p(S,\Delta k)$ to a preference of slightly over $1\sigma$ significance.\footnote{For that model, $p(S,d_{\rm BMD})$ is not reported because the prior on $m_{\rm th}$ restricts the parameter space in a way that causes the Bayesian model dimensionality of the  \neffmeff posterior to become smaller than that of \lcdm, causing the $p$-value to become undefined.}
The largest shift between  $R$ and $S$ values occurs the \npg model, particularly for the baseline model's all-data result. However, given the large number  of added  parameters, the $p$-values report that  preference for  \npg  remains insignificant, at less than $2\sigma$. 

The $\Delta$AIC roughly tracks the evidence ratio, and the $\Delta \chi^2$ and $\Delta$DIC the Suspiciousness, as  expected from the respective definitions. 

The overall model comparison conclusion is that none of the models considered  offers a compelling alternative to \lcdm in explaining the data.

\section{Conclusions}
\label{sec:ccl}

We have presented constraints on extensions to the \lcdm cosmological model from DES Y3 measurements of cosmic shear, galaxy--galaxy lensing, and clustering (\mpp summary statistics) in addition to state-of-the-art external data. We investigated how such extensions affect the modeling of \mpp observables, and validated  the analysis using simulated and blinded real data to ensure that known sources systematic error cannot lead to a false detection of beyond-\lcdm cosmology.
This work allows us to obtain robust and precise constraints on beyond-\lcdm cosmology thanks to the unprecedented statistical power of the DES Y3 galaxy \metacal shape and \maglim lens catalogs. Our analysis indicates no significant deviations from \lcdm and its precision is primarily limited by the need for further theoretical developments. 

We first expand the exploration of dark energy properties by constraining time dependence of its equation of state. While constraints from DES \mpp alone do not contribute significantly to \wowa information compared combination of all external data, their precision are comparable to those from other individual cosmological probes. The precision of constraints from  DES Y3 \mpp on \wowa is comparable to that of \textit{Planck} alone, and our \wa constraints are slightly tighter than the latest measurements from Pantheon+ alone~\cite{brout_2022}.
Combining datasets yields precise estimates for \wowa which are consistent with a cosmological constant, with the constraining power from measurements of only low redshift \mpp+BAO+RSD+SN probes comparable to data combinations including CMB observables.  

DES \mpp measurements contribute to  constraints on the curvature density of the Universe \ok mainly by constraining \om, which helps break a degeneracy between \om and \ok when \mpp data are combined with  BAO+RSD+SN, leading to a 20\% improvement on curvature constraints that can be obtained from  low redshift probes. While this low-redshift measurement of \ok is an order of magnitude weaker than constraints including CMB observables, it is an interesting independent check, given the much-discussed $\sim3\sigma$ tension between {\it Planck} and BAO curvature constraints, which we recover. We find combined constraints to be compatible with flatness, whether or not {\it Planck} likelihoods are included in the analysis. The constraining power contributed by DES measurements to this study  is limited by the lack of validated non-linear LSS modeling and non-Limber projection calculations for non-flat geometry.

Next, we constrain two models sensitive to additional relativistic particle species in the early Universe. We find that DES \mpp measurements have little impact on inferences about changes to the number of relativistic species parameterized by \neff, but that they are  a powerful tool for constraining  the impact of light relic particles with non-zero mass on the evolution of large scale structure.  We explore this by constraining a species of 
sterile neutrinos with effective mass \meff and a temperature set by \neff.  As in the case when the parameter is varied alone, \neff is primarily constrained by CMB observables, while growth information from DES Y3 \mpp and external RSD data allows us to tightly constrain \meff. In doing this, our combined analysis of all data improves upon the best available constraints on  \meff by a factor of three, finding $\meff < 0.20$eV. This constraining power is limited by a lack of validated small scale modeling, which requires us to apply conservative linear scale cuts to the \mpp measurements. Given this, we stress that modeling advances such as those being developed in e.g.\ Refs.~\cite{Xu:2021rwg,Munoz:2018ajr,Brandbyge:2017tdc, Banerjee:2022} will be key to enabling more precise constraints, and that, excitingly, more powerful constraints are attainable even with existing data. 
 
We also test gravity on cosmological scales measuring the \sigmu parameters. The most interesting constraints for this model come from the combination of multiple observables. 
In particular, the complementary approach of DES \mpp and external RSD measurements to measuring large scale structure allows us to break degeneracies between the modified gravity parameters and \seight. 
 Their combination thus gives tight constraints, particularly on \sigo,  resulting in consistency with general relativity. We find that {\it Planck} temperature and polarization constraints prefer slightly higher values of \sigo than either \mpp alone or the combination of all low-redshift probes, likely driven by high-$\ell$ feature of the CMB power spectra that drives the offset seen in the \ok analysis. When all data are analyzed together the resulting constraints remain consistent with GR. As with sterile neutrinos, the constraining power contributed by DES \mpp is also limited by our use of linear scale cuts. 
In fact, because the increased precision \mpp measurements causes our procedure for defining linear scale cuts to remove  larger fraction of data points in Y3 compared to Y1, constraints placed on \sigo by \mpp alone to actually weaken compared to the similar analysis in DES-Y1Ext.
Looking ahead to DES Y6 and next-generation surveys, this underlines the need for more sophisticated methods of accounting for non-linear modeling uncertainties when performing cosmological tests of gravity.

Finally, we perform a more generic test of \lcdm's predictions for structure growth via a \npg model, in which we introduce amplitude parameters that allow the normalization of the matter power spectrum to vary independently in four redshift bins defined by the lens galaxy sample. While constraints on the sampled amplitude parameters are not robust to  changes in how we account for source galaxy photometric redshift uncertainties, the \sigeight values inferred separately for each redshift bin are more robust, especially for the higher redshift bins when \mpp constraints are combined with external data. 
This analysis finds no significant deviation from the prediction of \lcdm, and highlights the importance of carefully accounting for the impact of photo-$z$ uncertainties when investigated beyond-\lcdm parameterizations which affect the growth of structure. 

In summary,  we have conducted  robust tests of extensions to \lcdm using the unprecedentedly precise DES Y3 \mpp measurements in combination with other state-of-the-art cosmological data,  while underlining challenges that will need to be addressed for future wide field galaxy surveys to further test the laws and contents of the Universe. 
We ultimately detect no significant preference for any of the extended models studied in our analysis. Thus, \lcdm remains the favored model to describe our data. 

\section*{Acknowledgements}
The analysis made use of the software tools {\sc SciPy}~\cite{Jones:2001}, {\sc Astropy}~\cite{astropy:2013,astropy:2018}, {\sc NumPy}~\cite{numpy:2020,Oliphant:2006},  {\sc Matplotlib}~\cite{matplotlib:2007}, {\sc CAMB}~\cite{Lewis:1999,Howlett:2012}, {\sc MGCAMB}~\cite{Zucca:2019,Hojjati:2011ix,Zhao:2008bn}, {\sc GetDist}~\cite{getdist:2019}, {\sc Multinest}~\cite{Feroz:2007,Feroz:2008,Feroz:2013},  {\sc Polychord}~\cite{Handley:2015a,Handley:2015b}, {\sc Anesthetic}~\cite{Handley:2019mfs}, \cosmosis~\cite{Zuntz:2014csq}, and {\sc Gnu Parallel}~\cite{gnu_parallel}. Elements of the DES modeling pipeline additionally use {\sc Cosmolike}~\cite{Krause:2016jvl} including {\sc CosmoCov}~\cite{Fang:2020vhc}, {\sc Halofit}~\cite{Takahashi:2012em,Bird:2011rb}, {\sc Fast-PT}~\cite{McEwen:2016fjn,Fang:fpt2}, and {\sc Nicaea}~\cite{Kilbinger:2008gk}.

This work was supported  through computational resources and
services provided by the National Energy Research Scientific Computing Center (NERSC), a U.S. Department of Energy Office of Science User Facility operated under Contract No. DE-AC02-05CH11231; by the Sherlock cluster, supported by Stanford University and the Stanford Research Computing Center; by the Center for Scientific Computing (NCC/GridUNESP) of the São Paulo State University (UNESP) and by the JPL ITSD High Performance Computing group.

Funding for the DES Projects has been provided by the U.S. Department of Energy, the U.S. National Science Foundation, the Ministry of Science and Education of Spain, 
the Science and Technology Facilities Council of the United Kingdom, the Higher Education Funding Council for England, the National Center for Supercomputing 
Applications at the University of Illinois at Urbana-Champaign, the Kavli Institute of Cosmological Physics at the University of Chicago, 
the Center for Cosmology and Astro-Particle Physics at the Ohio State University,
the Mitchell Institute for Fundamental Physics and Astronomy at Texas A\&M University, Financiadora de Estudos e Projetos, 
Funda{\c c}{\~a}o Carlos Chagas Filho de Amparo {\`a} Pesquisa do Estado do Rio de Janeiro, Conselho Nacional de Desenvolvimento Cient{\'i}fico e Tecnol{\'o}gico and 
the Minist{\'e}rio da Ci{\^e}ncia, Tecnologia e Inova{\c c}{\~a}o, the Deutsche Forschungsgemeinschaft and the Collaborating Institutions in the Dark Energy Survey. 

The Collaborating Institutions are Argonne National Laboratory, the University of California at Santa Cruz, the University of Cambridge, Centro de Investigaciones Energ{\'e}ticas, 
Medioambientales y Tecnol{\'o}gicas-Madrid, the University of Chicago, University College London, the DES-Brazil Consortium, the University of Edinburgh, 
the Eidgen{\"o}ssische Technische Hochschule (ETH) Z{\"u}rich, 
Fermi National Accelerator Laboratory, the University of Illinois at Urbana-Champaign, the Institut de Ci{\`e}ncies de l'Espai (IEEC/CSIC), 
the Institut de F{\'i}sica d'Altes Energies, Lawrence Berkeley National Laboratory, the Ludwig-Maximilians Universit{\"a}t M{\"u}nchen and the associated Excellence Cluster Universe, 
the University of Michigan, NSF's NOIRLab, the University of Nottingham, The Ohio State University, the University of Pennsylvania, the University of Portsmouth, 
SLAC National Accelerator Laboratory, Stanford University, the University of Sussex, Texas A\&M University, and the OzDES Membership Consortium.

Based in part on observations at Cerro Tololo Inter-American Observatory at NSF's NOIRLab (NOIRLab Prop. ID 2012B-0001; PI: J. Frieman), which is managed by the Association of Universities for Research in Astronomy (AURA) under a cooperative agreement with the National Science Foundation.

The DES data management system is supported by the National Science Foundation under Grant Numbers AST-1138766 and AST-1536171.
The DES participants from Spanish institutions are partially supported by MICINN under grants ESP2017-89838, PGC2018-094773, PGC2018-102021, SEV-2016-0588, SEV-2016-0597, and MDM-2015-0509, some of which include ERDF funds from the European Union. IFAE is partially funded by the CERCA program of the Generalitat de Catalunya.
Research leading to these results has received funding from the European Research
Council under the European Union's Seventh Framework Program (FP7/2007-2013) including ERC grant agreements 240672, 291329, and 306478.
We  acknowledge support from the Brazilian Instituto Nacional de Ci\^encia
e Tecnologia (INCT) do e-Universo (CNPq grant 465376/2014-2).

This manuscript has been authored by Fermi Research Alliance, LLC under Contract No. DE-AC02-07CH11359 with the U.S. Department of Energy, Office of Science, Office of High Energy Physics.


\bibliography{y3extkp}
 
\appendix
 

\section{\halofit validation for \texorpdfstring{$\wowa$}{Lg}}\label{sec:w0wa_casarini_check}

We model the matter power spectrum by using {\sc CAMB}~\cite{Lewis:1999,Howlett:2012} to compute the linear matter power spectrum and the \halofit semi-analytic fitting formula from Ref.~\citep{Takahashi:2012em} (with the  prescription from Ref.~\citep{Bird:2011rb} for massive neutrinos)  to compute non-linear corrections.  The \halofit fitting formula depends on the linear matter power spectrum and a subset of cosmology parameters, and was developed based on fits to \wcdm $N$-body simulations. While those simulations include cosmological models with $w\neq 1$, they do not include cases where the dark energy equation of state varies with time. 

\begin{figure}
  \centering
  \includegraphics[width=\linewidth]{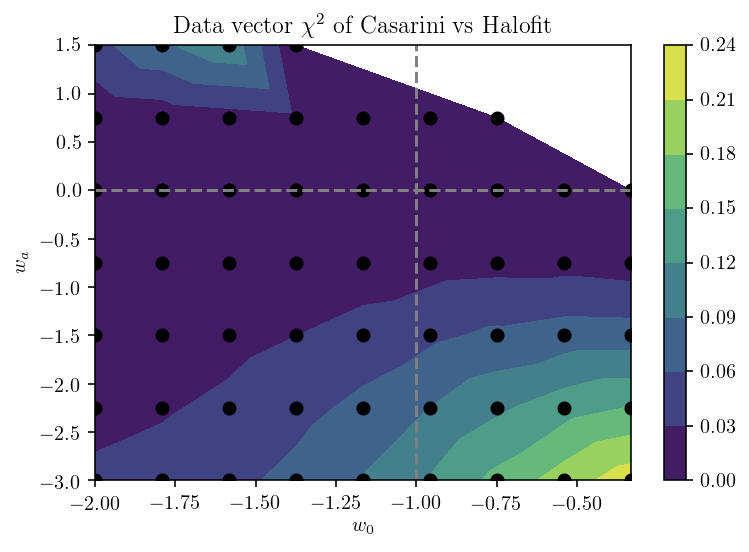}
  \caption{Validation of \halofit{} for nonlinear modeling in $\wowa$, showing the  $\Delta\chi^2$ difference between the \mpp data vector computed with the fiducial pipeline versus one where the Casarini~\cite{Casarini:2016ysv} method has been used to compute the nonlinear matter power spectrum. The colored heatmap shows the interpolation between points sampled in a grid (black circles) in the $\wowa$ plane. The white region in the upper right reflects the excluded region where $\wo +\wa > 0$. $\Delta \chi^2 \ll 1$ across the full space, indicating that the modeling of \halofit{} is sufficient.
  }\label{fig:w0wagrid}
\end{figure}

To validate the use of \halofit to model the non-linear matter power spectrum in our \wowa model, we note that \citet{Casarini:2016ysv} provides a recipe for computing the nonlinear power spectrum $P(k,z)$ for \wowa model given modeling ingredients for \wcdm. That scheme works by identifying an effective \wcdm model for each redshift $z$ for a given \wowa cosmology, chosen so $w_{\rm eff}(z)$ matches the distance from redshift $z$ to that of last scattering. 
Using $N$-body simulations, \citet{Casarini:2016ysv} shows that this mapping can be used to accurately compute nonlinear matter power spectra in \wowa cosmologies. While employing the Casarini mapping is not practical for our full analysis because it is too computationally intensive, we can use it for validation. To do this, we use the Casarini prescription to compute the non-linear matter power spectrum for a grid of cosmologies in spanning our two-dimensional $\wowa$ prior range. We then compare predictions for the \mpp data using our fiducial pipeline to those using the Casarini nonlinear power spectra. For each \wowa gridpoint, we evaluate $\Delta\chi^2$ between our fiducial and the Casarini model predictions and show the results in \fig{fig:w0wagrid}. Differences between these calculations have $\dchisq<0.24$ for all allowed values of $\wo$ and $\wa$, with the largest differences occurring in the high-$\wo$, low-$\wa$ part of parameter space. For all regions where $\wa>-2.5$, this difference is $\dchisq<0.15$. Based on these results, we conclude that our fiducial model using \halofit is accurate enough to perform our \wowa analysis with our fiducial scale cuts. 

We posit that a significant driver of the  Casarini $w_{\rm eff}(z)$  mapping's success in modeling non-linear power for \wowa comes from the fact that it correctly accounts for the impact of dynamic dark energy on the linear growth factor. Thus, the fact that we are correctly computing the linear matter power spectrum for the \wowa cosmology allows us to reach this accuracy even if we are not explicitly accounting for the $w_a$ parameter in \halofit.

\section{Validation of Weyl potential pipeline}
\label{app:Weyl-validation}

\begin{figure}
  \centering
  \includegraphics[width=\linewidth]{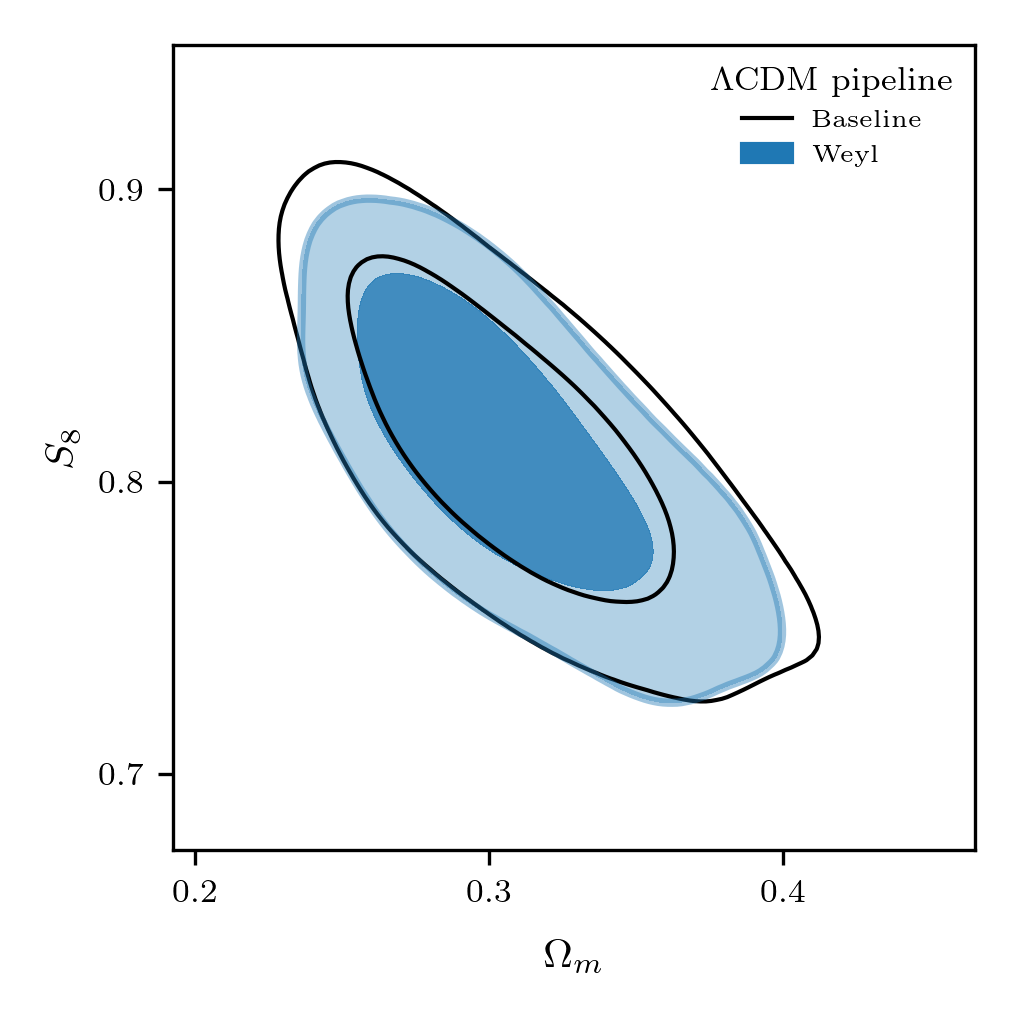}
  \caption{Constraints on $\Omega_{\rm m}$ and $S_8$ in \lcdm, showing 68 and 95\% confidence limits from the baseline DES Y3 \cosmosis pipeline in black and the modified version using the Weyl potential for the lensing predictions in blue. These results are obtained using \mpp measurements for the same linear scale cuts used to obtain \sigmu constraints.}
  \label{fig:modweyl}
\end{figure}

As an additional validation of the modeling pipeline used to constrain the modified gravity \sigmu model, we compare parameter estimates in \lcdm for the baseline \cosmosis pipeline used in DES Y3 \mpp analyses to the pipeline modified to use the Weyl potential to model lensing-related quantities (see Sec.\ \ref{sec:sigmu} for more details). 
Figure \ref{fig:modweyl} shows \lcdm constraints obtained by analyzing \mpp data using the same linear scale cuts applied for the \sigmu analyses.  That figure 
shows the results for the baseline pipeline in black and the Weyl pipeline in blue.
Differences are negligible for the estimated cosmological parameters \om and \seight.  

\section{The \fastismore framework for robustness tests}\label{app:fastismore}

The \fastismore framework used for the model robustness tests described in \sect{sec:simdat_changeDV} is at its core an application of importance sampling for performing validation tests of the robustness against systematics. Importance sampling (IS) is a method to quickly estimate a target distribution $p(\mathbf\Theta)$ by using samples from a proposal distribution $q(\mathbf\Theta)$ reweighted by the ratio $p(\mathbf \Theta)/q(\mathbf \Theta)\propto\mathcal{L}_p/\mathcal{L}_q$ (see e.g.~Ref.~\cite{Trotta:2017wnx}). 
The \fastismore framework consists of several parts: the infrastructure for performing fast IS posterior estimates, code for computing quality statistics for that estimate as well as for how the IS procedure contributes to sampling variance, and guidelines for using those metrics based on a number of simulated analyses. These tools will be made publicly available and documented in more detail in an upcoming publication, Ref.~\cite{fastismore}.

The accuracy of an IS estimate depends on whether the samples drawn from the proposal distribution cover the relevant parameter space of the target distribution with high enough density. In our application for robustness tests, the proposal distribution is the posterior of our baseline simulated analysis and the target is the posterior for a synthetic data vector contaminated with a systematic that is not included in the analysis model, as described in \sect{sec:simdat_changeDV}. 
Given this, chains will only be required if a contamination produces a significant change in the posterior.  
For cases where IS can be used, this method allows the impact of a systematic to be assessed in seconds, as opposed to the thousands of core-hours required to run a chain.

We quantify the performance of IS posterior estimates using  Kish's effective sample size: 
\begin{equation}
{\rm ESS} = \frac{\left(\sum_i w_i\right)^2}{\sum_i w_i^2},
\end{equation} 
\noindent where $w_i$ is the total weight of each sample. Assuming this approximates the true effective sample size (for a discussion on the validity of this approximation, see~\cite{Elvira:2022}), one can estimate the standard error on the mean $\bar\theta$ of the parameter of interest $\theta$ by using $\sigma_{\bar{\theta}}\approx\sigma_\theta / \sqrt{{\rm ESS}}$, where $\sigma_\theta$ is the standard deviation of the posterior. 
The uncertainty on the parameter mean shift is then estimated as:
\begin{equation}
\sigma_{\Delta\bar\theta} \lesssim \sigma_\theta \sqrt{\frac{1}{{\rm ESS}_{\rm base}} + \frac{1}{{\rm ESS}_{\rm cont}}},
\end{equation}
\noindent where ${\rm ESS}_{\rm base}$ is the effective sample size of the baseline chain  
and ${\rm ESS}_{\rm cont}$ is the effective sample size of the IS-estimated contaminated samples.  
For the analysis of contaminated synthetic data described in \ref{sec:simdat_changeDV} of this paper, we found ${\rm ESS_{\rm cont}} > 150$ for all validation tests such that the uncertainty on the parameter mean shifts are at most $0.08\sigma$.  
Thus all of our systematics tests pass the IS quality requirements.

We also employ the \fastismore framework to assess the impact of systematics on model comparison metrics  
between \lcdm and a beyond-\lcdm models. 
To assess how Suspiciousness (see \app{app:modcomp}) $\ln{S}$ is affected by contamination from systematics, we must compare quantities derived from four chains: a baseline ($B$) and alternative (contaminated) synthetic data ($A$) chain for both the extended model ($X$) and \lcdm ($0$). As will be shown in Ref.~\cite{fastismore}, the change in $\ln{S}$ due to systematic contamination can be written in terms of the within-model differences between the baseline ($B$) and alternative ($A$) synthetic data,
\begin{eqnarray}\label{eq:SfromIS}
   \Delta \ln{S}_{X0}&= \ln{S}_{X0}^{A} - \ln{S}_{X0}^{B}\\
   &=\ln{S}^X_{AB} -\ln{S}^0_{AB}.
\end{eqnarray}
Here, we use $S_{ab}^{c}$ to denote Suspiciousness between $a$ and $b$, keeping $c$ fixed, defined so $\ln{S}<0$ indicates a preference for $a$. This rearrangement makes the calculation more tractable, as the quantities $S^X_{AB}$ and $S^0_{AB}$ are easily computed using importance sampling.  

\section{Additional validation plots and discussion}
\label{app:more_validation}
Here we provide additional information to supplement the validation test results presented in \sect{sec:analysis}.

\subsection{All-data robustness plots}\label{app:more_tableplots}
\begin{figure}
  \centering
\includegraphics[width=\linewidth]{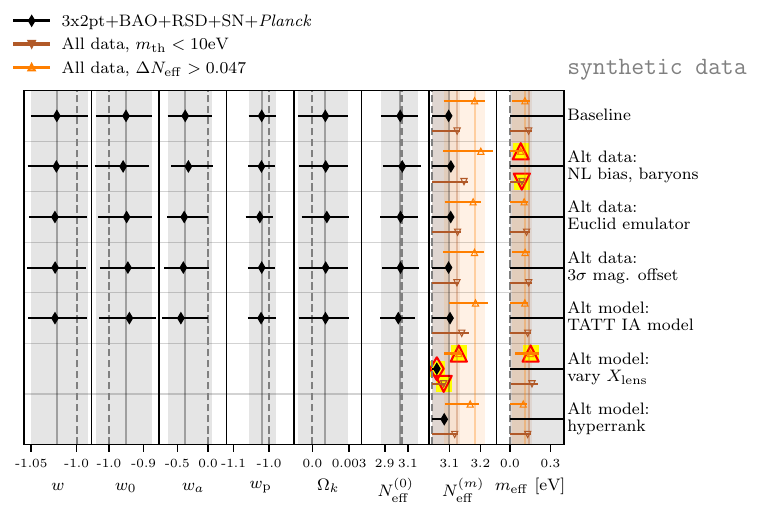}
  \caption{Constraints on beyond-\lcdm model parameters for the same simulated analyses studied in \fig{fig:extparams_table} of \sect{sec:simdat_changeDV}, but with narrower axis ranges for models where All-data is much more constraining than DES \mpp. Points and error bars show the mean and 68\% confidence interval for marginalized constraints on extended model parameters. Yellow and red highlights indicate shifts larger than $0.3\sigma$ according to \eq{eq:peakshift}. Vertical dashed lines show the input, \lcdm parameter values, while solid vertical lines and shaded regions show the location of the baseline results. For the effective number of radiative degrees of freedom, $\neff^{(0)}$ is the constraint for the model with no sterile neutrino mass, and   $\neff^{(m)}$ shows the constraints for the \neffmeff model. For \neffmeff we additionally show  results using the alternative prior $\Delta\neff>0.047$, which are less subject to projection effects and thus more robust than posteriors produced using the fiducial \neffmeff prior.}
  \label{fig:extparams_table_alldatonly}
\end{figure}

\begin{figure}
  \centering
\includegraphics[width=\linewidth]{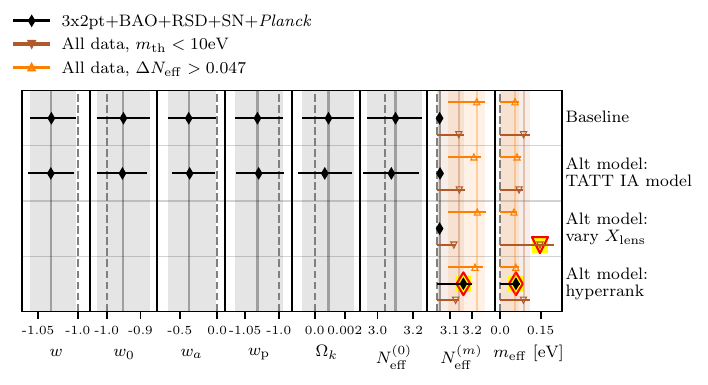}
  \caption{ Real All-data constraints for the same analyses presented in \fig{fig:realdat_extparams} of \sect{sec:changepipe}, but with narrower axis ranges for parameters where All-data is much more constraining than DES \mpp.  Points and error bars show the mean and 68\% confidence interval for marginalized constraints on extended model parameters, yellow and red highlights indicate shifts larger than $0.3\sigma$ and shaded regions indicate the location of the baseline results.}
  \label{fig:realdat_extparams_alldatonly}
\end{figure}

We begin by showing plots complementing \fig{fig:extparams_table} of \sect{sec:simdat_changeDV} and   \fig{fig:realdat_extparams} of \sect{sec:changepipe}, which show the impact of data vector contamination and model variations on the combined \mpp{}+BAO+RSD+SN+{\it Planck} constraints. Since those all-data   constraints are much more constraining than the DES \mpp alone for some models, the effects of data vector contamination and model variations on the combined constraints are not clearly visible in plots in the main body of the text, whose ranges are set to show the \mpp-only constraints. To facilitate closer examination, here we  include versions of those plots showing only the all-data constraints.  \fig{fig:extparams_table_alldatonly} shows an All-data only version simulated analysis plot, while \fig{fig:realdat_extparams_alldatonly} shows the real data response to model variations. 

In both  plots, we additionally show the results of the robustness tests for \neffmeff when the priors requiring $\Delta\neff>0.047$ or $m_{\rm th}<10$ eV are applied. We set the \meff axis range to show those results clearly, cutting off the fiducial prior points which were shown in \sect{sec:changepipe} to have non-negligible shifts. For these alternative \neffmeff priors, the nonlinear bias and baryon contamination produces a shift $\Delta_{\meff}=-0.33$, which is only slightly above and within sampling variance of the 0.3 threshold.

\subsection{Robustness investigation: \texorpdfstring{\neffmeff}{Lg}}\label{app:changepipe_neffmeff}

In the robustness tests of  \sect{sec:changepipe}, we noted that \neffmeff parameter constraints shifted significantly in response to several model variations. Here we provide further description of these non-negligible parameter shifts and argue that they are likely the result of prior volume effects associated with an unconstrained region of parameter space at small $\Delta\neff$. For $\neffmeff$, recall that we will be primarily focused on results from all data (DES 3$\times$2pt+BAO+RSD+SN+{\it Planck}) rather than DES \mpp only results. %

Of the parameter shifts that are above our desired threshold for \neffmeff, those produced by the TATT model variation are the least significant, and are not very concerning. For the simulated analysis, switching to TATT causes  $\Delta_{\meff}=0.37$. This change in \meff is not concerning, both because it is only marginally over our threshold, and because our simulated  constraints on \meff are one-sided upper bounds so that this shift can be interpreted as being simply due to weakened constraining power in a more complicated IA model. When these tests were repeated for real data, shifts in the constraints on both \neff and \meff were negligible. 

The \xlens and hyperrank tests have a more dramatic impact on the $\neffmeff$ constraints from all data.
For the analysis of real data, varying \xlens causes $\Delta_{\meff}=-0.41$, while using hyperrank causes $\Delta_{\neff}=0.91$ and, strikingly, $\Delta_{\meff}=-18$. To understand this behavior, we note that the upper bound on \meff for all data is largely determined by how the posterior is shaped in the low-$\Delta\neff$ region of parameter space where \meff has no impact on observables because the sterile neutrinos are indistinguishable from cold dark matter. Because CMB measurements provide tight upper bounds on $\Delta\neff$, a significant fraction of the all data posterior occupies this region of parameter space. The flatness of the likelihood in the \meff direction when $\Delta\Neff$ is low (independent of what data are considered) implies that the marginalized  constraints on both \neff and \meff  are  highly susceptible to the details of how the higher-dimensional posterior projects those parameter directions. Given this, our all-data $\neffmeff$ constraints may be highly sensitive to the choice of nuisance parameters,  and the data's noise realization may produce  large shifts in \meff that do not necessarily carry physical information.  As we describe  in \app{app:hyperrank}, hyperrank causes hard-to-characterize changes to the posterior shape even in \lcdm, so it is plausible that the dramatic $\Delta_{\meff}$ produced by hyperrank is related to this kind of parameter-space projection.

This prior-volume-effect hypothesis is supported by the fact that our results become more robust when we apply priors  to remove the unconstrained  low-$\Delta\neff$ region. We consider two such priors. We run additional (real data) chains raising the lower bound on \neff to require $\Delta\neff>0.047$, and additionally we consider a cut on the physical sterile neutrino mass, assuming a thermal relic model, $m_{\rm th}<10\,$eV.  This $m_{\rm th}$ cut matches the prior used for  \neffmeff in the {\it Planck} 2018 analysis~\cite{Planck:2018vyg}. Plots showing these shifts for all-data constraints with the alternative $\Delta\neff$ priors can be found above in \figs{fig:extparams_table_alldatonly} and~\ref{fig:realdat_extparams_alldatonly}. 

Requiring $\Delta\neff>0.047$ leads to negligible shifts between the all-data constraints from the hyperrank and the baseline models, as well as between the baseline and varying \xlens. Requiring $m_{\rm th}<10$eV, which is the prior matching the {\it Planck} 2018 analysis, results in neglible shifts due to \xlens, and hyperrank-vs-baseline shifts of   $\Delta_{(\neff,\meff)}=(-0.26, -0.39)$, only slightly above our $0.3\sigma$ threshold.  
For comparison, we also consider simulated $\neffmeff$ analyses of all data combined, post-processing the chains to enforce the alternative priors rather than running new chains.  For simulated all-data analyses with fiducial priors, \xlens produces $\Delta_{(\neff,\meff)}=(-0.71,+0.57)$ and hyperrank produces $(-0.27,+0.39)$, while with the alternate priors hyperrank causes negligible shifts and \xlens produces a shift of $\Delta_{\neff}\sim 0.5$. While this latter shift is non-negligible, it is not concerning given the robustness of the real data results, especially noting that the post-processing used for these synthetic-data tests results in significant sampling uncertainty. 
 
\begin{figure}
  \centering
\includegraphics[width=\linewidth]{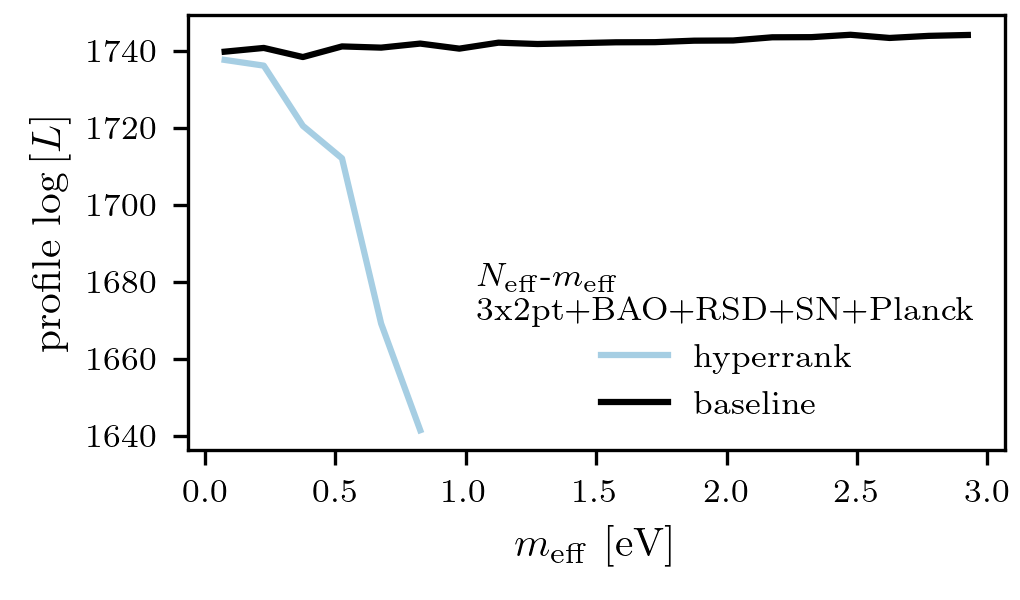}\\
\includegraphics[width=\linewidth]{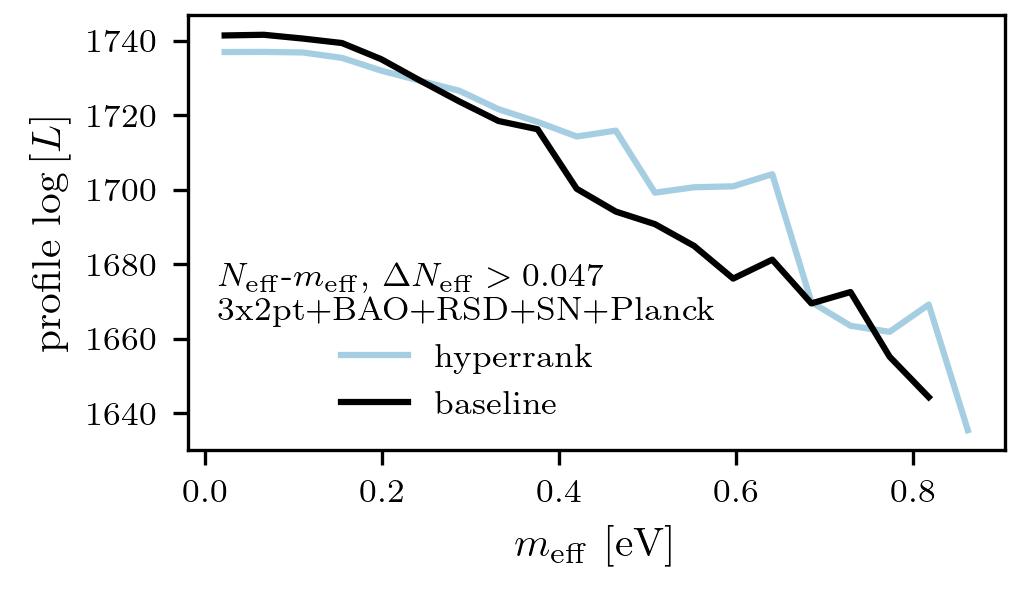}\\
  \caption{ Profile likelihood for All-data \neffmeff chains, comparing of baseline and hyperrank. Lines show the maximum likelihood found for a chain sample in each of 20 bins of \meff values. The upper panel shows results for the fiducial \neffmeff priors, and the lower panel shows resutls for chains run with the alternative $\Delta\Neff>0.047$ prior. The fact that hyperrank does not produce a better fit to the data than the baseline at small \meff supports the idea that shift in \meff constraints for hyperrank is due to parameter space projection effects  rather than an actual strong  preference for small \meff values.}
  \label{fig:neffmeff-hyp-profilelike}
\end{figure}

To lend further support to the idea that the dramatic shift in the all-data \neffmeff constraints (particularly \meff) when using hyperrank is a prior volume effect, we study the profile likelihood for \meff in \fig{fig:neffmeff-hyp-profilelike}. The profile likelihood shows the maximum likelihood in 20 bins of sampled \meff values. For the fiducial \neffmeff priors, we see in the top panel that while the hyperrank maximum sampled likelihood drops at large \meff,  the hyperrank model does not produce a better fit  to the data than the baseline chain in the small \meff regime. The lower panel shows the same profile for chains run with a prior requiring $\Delta\neff>0.047$. In that panel, the hyperrank and baseline profile likelihoods remain fairly similar as \meff increases. This suggests that that drop seen for the hyperrank chain in the top panel is due to a lack of chain samples exploring the high \meff, rather than a dramatically worse fit. This supports the idea that the differences between the marginalized  hyperrank and baseline posteriors are driven by parameter space projection effects.

Overall these studies support the conclusion that our \neffmeff constraints are significantly more stable against model changes when a prior is applied to remove the unconstrained small-$\Delta$\neff region. This motivates our choice to focus on results reported for \neffmeff constraints obtained with the $\Delta\neff>0.047$ and $m_{\rm th}<10$eV  priors rather than those obtained with the fiducial prior.

\subsection{Robustness investigation: \texorpdfstring{\npg}{Lg}}\label{app:changepipe_npg}

Here we describe and investigate non-negligible shifts in the \npg parameters due to the hyperrank and \xlens model variations studied in \sect{sec:changepipe}. The only  $>0.3\sigma$ shift produced by varying \xlens, $\Delta_{\Anpg_{3}}=0.36$ for DES \mpp-alone, is not far above our threshold so is not concerning. We thus focus primarily on the impact of the hyperrank approach to parameterizing source galaxy redshift uncertainties. 
For the real-data \mpp-only analysis, using hyperrank leads to parameter shifts of $\Delta_{(\Anpg_{2},\Anpg_{3})}=(+0.72,+0.84)$, while analyzing all data produces even more significant  parameter shifts:  $\Delta_{(\Anpg_{2},\Anpg_{3},\Anpg_{4})}=(+1.73,+1.36,+1.42)$. 

A detailed characterization of what is driving the low-redshift amplitude shifts is beyond the scope of this paper, though we  present some exploratory investigation in \app{app:hyperrank}. What these findings clearly highlight is that source photo-$z$ uncertainties and our  method of accounting for them can have a significant impact on inferences about the growth of LSS over time, especially when constraining models that, like our \npg parameterization, add  degrees of freedom beyond what is expected for \lcdm. 
While in principle the hyperrank method should provide a more complete description of photo-$z$ uncertainties than the baseline $\Delta z_s$ nuisance parameters, without additional validation  we are not confident in switching our main analysis to use it for \npg. 
Given this ambiguity, we choose to report for \npg constraints and model comparisons  for both our baseline model and for hyperrank. Showing results from both analyses will roughly quantify the size of the photo-$z$-related systematic uncertainties. Note that this means that if we find tension with \lcdm in one but not both of these \npg analyses, we will not be able to definitively claim a discovery of non-standard large scale structure growth.

\begin{figure}
  \centering
\includegraphics[width=\linewidth]{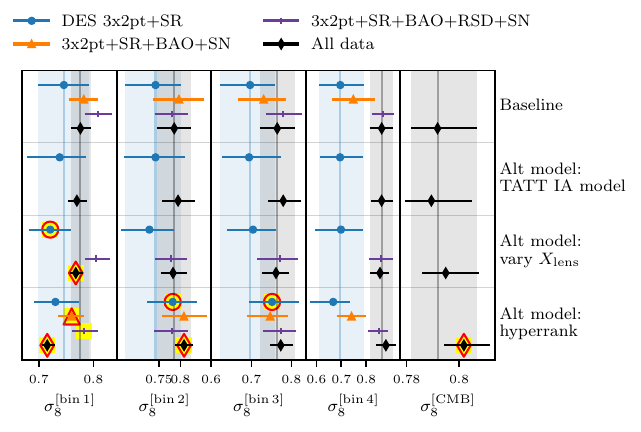}
  \caption{Impact of model variations on the derived parameters $\npgsig{i}\equiv \sigeight[\Anpg_i]^{1/2}$. Plot formatting matches that of \fig{fig:realdat_extparams} and All data (in black) refers to DES 3$\times$2pt+SR+BAO+RSD+SN+{\it Planck}. } 
  \label{fig:npgalt_paramshifts}
\end{figure}

To further characterize what we can or cannot say robustly about our \npg inferences, we repeat these validation tests\footnote{Note that these redefined amplitude parameters, and thus these additional robustness tests, were explored after unblinding.} for the derived parameters $\npgsig{i}$  defined in \eq{eq:news8}.  
Recall that $\npgsig{i}$ are more closely related to the amplitude of large scale structured observed separately in each redshift bin than the sampled $\Anpg_i$ amplitudes, which are defined relative to the  amplitude in the lowest redshift bin. 
When real data are analyzed, model variations still produce non-negligible $\npgsig{i}$ shifts, as can be seen in  \fig{fig:npgalt_paramshifts},  but these are smaller than those found for the sampled $\Anpg_i$ parameters. Theshift  for the lowest redshift bin shifts in response to varying \xlens are only slightly above the $0.3\sigma$ threshold, with  $\Delta_{\npgsig_1}=-0.42$ and $-0.39$ for DES \mpp and all data respectively. For hyperrank, the non-negligible shifts for \mpp-only are $\Delta_{(\npgsig{2},\npgsig{3})}=(+0.47,+0.64)$, while for all data they are $\Delta_{(\npgsig{1},\npgsig{2},\npgsigcmb)}=(-2.72,+0.52,+0.60)$. For \mpp{}+BAO+RSD+SN (leaving out {\it Planck}), the only non-negligible shift is $\Delta_{\npgsig{1}}=-0.73$, and similarly \mpp+BAO+SN (DES combined with only geometric external data) it is $\Delta_{\npgsig{1}}=-0.62$.

Thus, we see that when we combine \mpp with external constraints, the source photo-$z$ marginalization scheme  primarily contributes systematic uncertainty to the LSS amplitude measured for the lowest redshift bin. The fact that the sampled $\Anpg_i$ parameters are defined relative to bin 1 is thus why those amplitudes are strongly affected by the change to hyperrank. Our study of the $\npgsig{i}$ derived parameters show that when we combine DES \mpp measurements with other low redshift geometric probes, we are in fact able to make fairly robust inferences of \npg for the redshift ranges corresponding to bins 2-4.

\subsection{Hyperrank discussion}
\label{app:hyperrank}

In our model robustness validation tests, for both \neffmeff and for \npg we found that using the hyperrank~\cite{y3-hyperrank} method to marginalize over source photo-$z$ uncertainties, as opposed to the fiducial $\Delta z_s$ mean-shift nuisance parameters, caused non-negligible shifts in the beyond-\lcdm parameter posteriors. Here we present some additional investigation into that behavior.  To place these studies in context, we illustrate that switching to hyperrank can have non-negligible impacts on parameter estimation even in \lcdm.  The method's impact on shear-only analysis has been thoroughly studied and the DES Y3 cosmic shear results were found to be robust to this model  variation~\cite{y3-cosmicshear1}. However, in DES-Y3KP it was found that switching to hyperrank produces a $0.53\sigma$ shift in \seight. 

\begin{figure}
  \centering
\includegraphics[width=\linewidth]{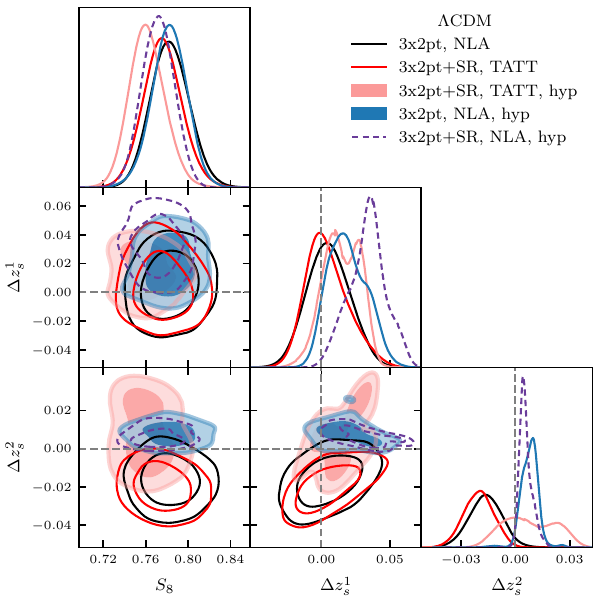}
  \caption{Comparison of baseline and hyperrank chain posteriors for $\seight$ and shifts in mean source bin redshift around the fiducial $n_s(z)$ means for bins 1 and 2. This is a \mpp version of the plot shown for shear-only in Fig.~17 of Ref.~\cite{y3-cosmicshear1}.}
  \label{fig:lcdm-hyp-check}
\end{figure}

\fig{fig:lcdm-hyp-check} further illustrates this behavior by showing the \lcdm  constraints on \seight and the mean redshift of a subset of source bins for various iterations of our analysis choices. In that figure, solid red and black lines show \lcdm posteriors on the photo-$z$ bias nuisance parameters for the baseline settings in this work (black, using NLA as the IA model and not including shear ratio) and in the \lcdm analysis of DES-Y3KP (red, TATT IA model, including shear ratio). The shaded pink contours show the results for the hyperrank chain that was run as part of robustness tests in DES-Y3KP (with TATT and shear ratio), while  dashed purple contours  and shaded blue contours show hyperrank chains run as part of this paper's beyond-\lcdm studies (with NLA), with and without including the shear ratio likelihood, respectively. We see that all of the hyperrank chains have multi-modal posteriors, and that there are significant qualitative difference between the various hyperrank chains compared. The choice of IA model has the largest impact on the shape of these posteriors, but even comparing the two NLA chains (blue-shaded and purple-dashed), including or not including shear ratio can also cause non-negligible posterior shifts. Notably, we find that these model variations have a more significant impact on \seight estimates when hyperrank is used than in the baseline $\Delta z$ approach to marginalizing photo-$z$ uncertainties.  
 
These effects can be understood in terms of interactions between intrinsic alignment (IA) parameters and details of the shape of the source redshift distribution. The contribution of IA to measured cosmic shear become fractionally more important at low source galaxy redshifts  where lensing contributions are smaller. Additionally, the IA signal depends on the projection of $n_s^2(z)$ onto the sky, while lensing depends on the square of the projected $n_s(z)$. Together, these two effects mean that  IA calculations are more sensitive to the detailed shape of the source galaxy redshift distribution, and thus that different hyperrank  $n_s(z)$  realizations can have very  different IA kernels, especially depending on their low-$z$ features. While  including or not including the shear ratio likelihood doesn't significantly impact the cosmology constraints in the baseline \mpp analysis, this choice does have an impact on the IA parameter constraints, and thus affects the posterior when hyperrank is used. 

If this behavior is occurring in \lcdm, it perhaps not surprising that hyperrank causes significant shifts in the \Anpg\ amplitudes when we allow \sigeight to vary independently in different redshift bins. We explored whether  specific features in the $n(z)$ distributions sampled by hyperrank correlate with the \npg amplitude parameters, and thus might be driving the shifts we see. Unfortunately, this investigation did not yield any further insights. 

\section{Metrics for assessing tension between datasets}\label{app:tensions}

In \sect{sec:res-tensions}, we employ three different tension metrics to assess agreement between datasets: the Bayes ratio $R$, Suspiciousness $S$, and a $p$-value computed from $S$ and the Bayesian model dimensionality $d_{BMD}$.  
The Bayes ratio $R$ is defined for independent datasets $A$ and $B$ and for their combination $AB$ as \citep{Marshall:2004zd}: 
\begin{equation}
    R\equiv \frac{\mathcal{Z}_{AB}}{\mathcal{Z}_{A}\mathcal{Z}_{B}},
\end{equation}
where 
\begin{equation}
    \mathcal{Z}_D \equiv P(\mathbf{D} | \mathcal{M}) = \int d \mathbf \Theta \ \mathcal{L}(\mathbf{D} |\mathbf \Theta, \mathcal{M}) \pi(\mathbf \Theta | \mathcal{M}).
\end{equation}
is the Bayesian Evidence.
In that expression, $\mathcal{L}$ is the likelihood of observing the data given model $\mathcal{M}$ and  parameter values $\mathbf \Theta$, and $\pi$ is the prior probability of those parameters given the model. 
In effect $R$ can be viewed as a hypothesis test assessing 
the odds of both datasets being described with a single set of parameters ($\mathcal{Z}_{AB}$) as opposed to two independent sets of parameters ($\mathcal{Z}_{A}\mathcal{Z}_{B}$). Smaller values of $R$  
indicate stronger evidence of tension between measurements from datasets $A$ and $B$.  
When using $\ln{R}$ the strength of the tension is usually interpreted using the Jeffreys' scale~\cite{jeffreys61}, where $\ln{R}<-2.3$ is considered `strong' tension with approximately $10:1$ odds, $-2.3<\ln{R}<-1.2$ is considered `substantial' tension with approximately $3:1$ odds, and $\ln{R}>-1.2$ indicates the datasets are in agreement.
This  interpretation of odds is only correct in the context where one of the models being considered is correct and where the priors accurately characterize prior beliefs on the parameters. As is discussed in e.g.~Refs.~\cite{y3-tensions,Handley:2019wlz,Lemos:2019txn}, the value of $\ln{R}$ depends strongly on the choice of parameter prior ranges, so when interpreting tension assessed with the Bayes ratio, one  should check the robustness of conclusions under reasonable changes to those priors. The fact that we use wide, uninformative priors therefore makes $\ln{R}$ somewhat ambiguous to interpret as a tension metric for this work.

Given this, we additionally report tension using the Bayesian Suspiciousness  $S$~\cite{Handley:2019wlz}. Like the Bayes ratio, Suspiciousness measures tension between posteriors in our full sampled parameter space, but removes the prior dependence by dividing $R$ by the information ratio $I$,
\begin{equation}\label{eq:Sfortension}
    \ln{S}  = \ln{R} - \ln{I}.
\end{equation}
The information ratio quantifies the probability of data $A$ and $B$ given the prior width, and is defined
\begin{equation}
    \ln{I} \equiv \mathcal{D}_A+\mathcal{D}_B - \mathcal{D}_{AB},
\end{equation}
where $\mathcal{D} \equiv \int {\cal P} \ln \left( {\mathcal{P}/ \pi} \right) \ d\theta$ 
is the Kullback--Leibler Divergence~\cite{Kullback:1951} from the prior $\pi$ to the posterior $\mathcal{P}$ for a given dataset. One can interpret the Suspiciousness as a posterior-averaged goodness-of-fit statistic between the combined vs. independent datasets A and B~\cite{Heymans:2020gsg}. 

As with the Bayes ratio, more negative values of $\ln{S}$ indicate stronger evidence of tension, while positive values indicate agreement between datasets. To further quantify the strength of tension or agreement, we can use the fact that if both datasets come from the same set of parameters and there is some choice of parameters in which the posterior is roughly Gaussian, the quantity $d_{\rm BMD}-2\ln{S}$ follows approximately a $\chi^2_{d_{\rm BMD}}$ probability distribution, where $d_{\rm BMD}$ counts the number of parameter dimensions constrained by both posteriors $A$ and $B$ \cite{Handley:2019wlz}. We use this information to compute a $p$-value estimating the probability of finding a value of $\ln{S}$ as small or smaller than the measured value if the two datasets are actually in agreement.

For each chain we assess $d_{\rm BMD}$ using the Bayesian Model Dimensionality \cite{Handley:2019pqx}, which estimates the number of parameters constrained by a given posterior. It is equal to  
\begin{equation}\label{eq:dbmd}
    d = 2\left(\langle (\ln{\mathcal{L}})^2\rangle - \langle \ln{\mathcal{L}}\rangle^2\right),
\end{equation}
where again angled brackets refer to the posterior-weighted average. We compute the number of parameters that are independently constrained by both datasets $A$ and $B$ as  
\begin{equation}\label{eq:dbmd_for_tension}
    d_{\rm BMD} = d_A + d_B - d_{AB},
\end{equation}
where the $d_{AB}$ term avoids double-counting parameters. We evaluate the survival function, 
\begin{equation}\label{eq:pbmd}
p(S,d_{\rm BMD}) = \int_{d_{\rm BMD} - 2\ln{S}}^{\infty} \chi^2_{d_{\rm BMD}}(x)\,{\rm d}x
\end{equation}

There are several other metrics one could use to assess tensions between datasets, including but not limited to parameter difference distributions~\cite{Raveri:2021wfz,Raveri:2018wln} and eigentensions~\cite{ParkRozo:2019}. It was shown in Ref.~\cite{y3-tensions} that all of these metrics give results in agreement with $\ln{S}$ when quantifying tensions. We therefore focus on $S$, which has the advantage of requiring no additional computation after running the chains. 

\section{Model comparison metrics}\label{app:modcomp}
 
In \sect{sec:res-modcomp} we perform a number of model comparison tests to assess whether data favor any extensions to \lcdm over that cosmological standard model. Here we present definitions of the metrics used to assess those comparisons, divided into two categories: Bayesian quantities depending on the posterior distribution in the full parameter space, and those depending on maximum {\it a posteriori} probability (MAP) estimates.  To define these metrics, let us consider two models: a baseline model $\mathcal{M}_0$ and a model which extends it, $\mathcal{M}_X$, such that the parameter space of $\mathcal{M}_0$ is a subspace of $\mathcal{M}_X$'s parameter space.

\subsection{Bayesian quantities}
The Bayesian quantities are analogous to those used to evaluate tensions between datasets.
We define the Bayes ratio by comparing the Bayesian evidence evaluated in these different models but with matching data via,
\begin{equation}\label{eq:modcompR}
R = \frac{\mathcal{Z}_0}{\mathcal{Z}_X}.
\end{equation}
Suspiciousness $S$ and an associated $p(S,d_{\rm BMD})$ are defined in the similar way to \eqs{eq:Sfortension}-\ref{eq:pbmd}, but with $\mathcal{M}_0$ replacing all quantities associated with the $AB$ joint constraints, and $\mathcal{M}_X$ replacing the sums where contributions from datasets $A$ and $B$ are included separately. That is to say, the definition of $S$ in \eq{eq:Sfortension} still holds, but with a single dataset used in all parts of the evaluation and
\begin{equation}
    \ln{I} \equiv \mathcal{D}_X - \mathcal{D}_0,
\end{equation}
and similarly,
\begin{equation}\label{eq:dbmd_for_model}
d_{\rm BMD} \equiv d_X - d_0.
\end{equation}

One challenge of using $\ln{S}$ for model comparison is that number of additional constrained parameters $d_{\rm BMD}$ is typically small, such that noise in the estimate of $d_{\rm BMD}$ can have a large impact on $p(S,d_{\rm BMD})$. Specifically, if we let $\Delta k$ be the number of parameters that are in $\mathcal{M}_X$  but not in $\mathcal{M}_0$, then $d_{\rm BMD}\leq \Delta k$. Most of the models we consider have $\Delta k=1$ or 2, meaning that if the additional extended model $X$ parameters are not well constrained, $d_{\rm BMD}$ can be very small. If we could make a noiseless estimate of $d_{\rm BMD}$, this quantity would approach zero in the limit of completely unconstrained beyond-\lcdm parameters. In practice, sampling variance can cause the estimates made in \eqs{eq:dbmd} and~\ref{eq:dbmd_for_tension} to return values of $d_{\rm BMD}<0$.  This complicates interpretation of  $p(S,d_{\rm BMD})$ (\eq{eq:pbmd}), because the $\chi^2$ probability distribution is undefined for negative numbers of degrees of freedom. For estimates where this happens we report $p(S,d_{\rm BMD})=1$, reasoning that if the added parameters of $\mathcal{M}_X$ are unconstrained, no tension with $\mathcal{M}_0$ can be found. 

Given the ambiguity of this determination, we additionally report a  tension probability using the number of added parameters $\Delta k$ instead of $d_{\rm BMD}$: $p(S,\Delta k)$.
This ensures we avoid scenarios where the probability distribution for $S$ is undefined.
In addition to being less subject to sampling error, $p(S,\Delta k)$ has a benefit for interpretation, since it can be viewed as a Bayesian  likelihood ratio test. 
As will be shown in Ref.~\cite{fastismore}, we can interpret the model-comparison formulation of Suspiciousness as the change in \textit{posterior-averaged} goodness-of-fit, $2\ln{S} = \langle\chi^2_X\rangle - \langle\chi^2_0\rangle$. This means $p(S,\Delta k)$, which evaluates this change relative to the expected improvement from additional model freedom, serves as a Bayesian analog of the more traditional \chisq test statistic  that compares the goodness-of-fit at the two models' maximum likelihood points. Further connections can be made to more traditional information criteria-based model comparison statistics, where for limiting cases the Suspiciousness can be interpreted as anlogous to $\Delta$DIC, but with a lesser penalty applied for additional model parameters. This is explored in greater detail in Ref.~\cite{fastismore}.

\subsection{MAP-based statistics}
In our large parameter space, MAP estimates from nested sampler chains are subject to significant sampling error and so cannot be accurately determined by simply selecting the sample with the highest posterior from a chain. To estimate the maximum posterior, we therefore perform additional maximization as follows. For each chain, we select the 15 samples with the highest reported posteriors. Starting at each of those 15 points,  we run two iterative optimization searches to maximize the posterior using the \cosmosis{} {\tt maxlike} sampler, which is an interface to the {\tt scipy.optimize} function~\cite{Jones:2001}, using the BFGS~\cite{Broyden:1970,Fletcher:1970,Goldfarb:1970,Shanno:1970} optimization algorithm. Of the resulting 15 MAP estimates, we select the one with the highest posterior.  
Based on limited studies for simulated analyses we find this produces reasonably accurate estimates of the maximum posterior probability,  (the error on $\chisq$ is probably less than about 0.5, though we have not quantified this rigorously), 
but still very noisy estimates of the associated parameter values. Given this, we  use the maximum posterior estimates for model comparison statistics, but we do not report MAP parameter values.

We use the MAP posterior estimates to compute the model comparison statistics  \dchisq,  $\Delta$AIC, and $\Delta$DIC.
The quantity \chisq here measures the goodness of fit at the best-fit point in parameter space,
\begin{equation}\label{eq:chisqdef}
    \chisq= -2\ln{\mathcal{L}^{\rm  max}}
\end{equation}
In practice we focus on the quantity $\dchisq/\Delta k$, where $\dchisq= \chisq_X - \chisq_0$ and  $\Delta k$ is the change in modeling degrees of freedom between models $\mathcal{M}_X$ and $\mathcal{M}_0$. The Akaike Information Criterion (AIC) is defined as \cite{Akaike:1974} 
\begin{equation}\label{eq:aicdef}
    {\rm AIC} = -2\ln{\mathcal{L}^{\rm max}}+ 2k
\end{equation}
where $k$ is the number of model parameters. The Deviance Information Criterion (DIC)
is defined by \cite{Spiegelhalter:2002}
\begin{align}
    {\rm DIC} &= -2 \ln{\mathcal{L}^{\rm max}} + 2p_{\rm DIC}.\label{eq:dicdef} \\
    \mbox{with} \;\; p_{\rm DIC} &= 2 \ln{\mathcal{L}^{\rm max}} - 2\langle\ln\mathcal{L}\rangle 
\end{align}
Note that here we follow Ref.~\cite{KiDS:2020ghu} in using a MAP-based calculation of the DIC statistic, rather than an alternative definition where $p_{\rm DIC}$ is instead equal to the Bayesian model dimensionality defined in \eq{eq:dbmd} (see e.g. Eq. (7.10) in Ref.~\cite{Gelman2013-ic}). This is motivated by the fact that the MAP-based calculation is not affected by the instabilities for small $d_{\rm BMD}$ described in the text below \eq{eq:dbmd_for_model}.

The numbers reported for MAP-based quantities in \sect{sec:res-modcomp} rely on a slightly different definitions than those given above. This is because  we would like to consider agreement with the Gaussian priors on DES nuisance parameters in addition to the \mpp and external likelihoods to assess goodness-of-fit. For example, if model $\mathcal{M}_X$ could get an excellent fit to the \mpp measurements, but as a consequence  required shear calibration parameters $m_i$ to take extreme values compared to the principled priors made as part of the \metacal analysis, we would not want to consider that model favored over $\mathcal{M}_0$. To account for this, when computing $\Delta\chisq$, $\Delta$AIC, and $\Delta$DIC, we treat the Gaussian priors as effectively part of the likelihood. 
In practice what this means is that we evaluate them using the maximum posterior 
$\mathcal P$ instead of the maximum likelihood $\mathcal{L}$ 
and introduce a correction to account for 
differences in the flat prior contributions between models $\mathcal{M}_X$ and $\mathcal{M}_0$.

To derive what the flat prior correction should be, 
we write the posterior  $\mathcal{P}$  for data $\mathbf d$, model $\mathcal{M}$, and parameters $\mathbf\Theta$ as
\begin{equation}
    \mathcal{P(\mathbf\Theta|\mathbf d,\mathcal{M})} = \frac{ \mathcal{L}(\mathbf d|\mathbf \Theta,\mathcal{M})\mathcal{G}(\mathbf\Theta|\mathcal{M})\pi^{\rm flat}(\mathcal{M})}{\mathcal{Z}(\mathbf d|\mathcal{M})},
\end{equation}
where $\mathcal{L}$ is the likelihood, $\mathcal{G}$ is the product of the various of Gaussian priors on nuisance parameters, and $\pi^{\rm flat}$ is the $\mathbf \Theta$-independent contribution to the prior from all flat priors. For the purpose of our model comparison statistics, we would like to define goodness-of-fit quantities like \eqs{eq:chisqdef},~(\ref{eq:aicdef}), and~(\ref{eq:dicdef}), but depending on the product $\mathcal{L}\mathcal{G}$ in place of just $\mathcal{L}$. The fact that $\pi^{\rm flat}$ does not depend on $\mathbf\Theta$ means that the best fit defined for $\mathcal{L}\pi^G$ will be the same vector $\mathbf\Theta^{\rm MAP}_{\mathcal{M}}$ which maximizes $\mathcal{P}$.
Suppressing arguments for conciseness, we  note that a \chisq-like posterior-based goodness-of-fit for model $\mathcal{M}_X$ is 
\begin{align}
    [\chisq_{\mathcal{P}}]_X &= -2\ln{(\mathcal{L}^{\rm MAP}_X\mathcal{G}^{\rm{MAP}}_X)} - 2\ln{\pi^{\rm flat}_X}\\
   & \equiv [\chisq_{\mathcal{LG}}]_X -  2\ln{\pi^{\rm flat}_X}.
\end{align}
Model comparison between $\mathcal{M}_X$ and $\mathcal{M}_0$ therefore involves the comparison
\begin{align}
    \dchisq_{\mathcal{LG}} &\equiv  [\chisq_{\mathcal{LG}}]_X - [\chisq_{\mathcal{LG}}]_0\\
    &= [\chisq_{\mathcal{P}}]_X - [\chisq_{\mathcal{P}}]_0 + 2\ln{\pi^{\rm flat}_X} - 2\ln{\pi^{\rm flat}_0}. 
\end{align}
Using this expression, the \dchisq model comparison statistic reported in the main body of this paper is computed  as:  
\begin{align}
    \dchisq \equiv &-2 (\ln{\mathcal{P}^{\rm MAP}_X} -\ln{\mathcal{P}^{\rm MAP}_0})\\&+ 2(\ln{\pi^{\rm flat}_X} - \ln{\pi^{\rm flat}_0}). \label{eq:dchisq_priorcorr}
\end{align}
Similar calculations for AIC and DIC  result in the same flat-prior correction term.

\section{Testing beyond-\texorpdfstring{\lcdm}{Lg} model response to \texorpdfstring{\xlens}{Lg}}\label{app:simXlens}

We can use a synthetic data study similar to those in \sect{sec:simdat_changeDV} to assess how beyond-\lcdm models respond to a synthetic \mpp data vector produced with the redshift-independent parameter $\xlens\neq 1$. As described in \sect{sec:changepipe}, the parameter \xlens describes a mismatch between the galaxy bias detected by galaxy clustering versus that from galaxy-galaxy lensing. This is of interest because such a mismatch  was found in the \mpp results for the alternative \redmagic lens sample studied in DES-Y3KP: when \xlens is included in the \redmagic \mpp analysis, its preferred value is significantly less than the fiducial value of one. The effect is thought to be caused by residual systematics related to the lens sample selection, and remains a topic of active investigation. This indication of residual systematics ultimately motivated the choice of \maglim over \redmagic as the fiducial lens sample for DES-Y3KP, as this \xlens effect is not present for the four-bin \maglim sample we are using for the analysis in this paper. Given this, we emphasize that this investigation is performed to understand the effects of such a systematic in extended model spaces, rather than as a validation test that our analysis must pass. 

The test proceeds as follows: we generate a simulated data vector with $\xlens=0.89$, which approximates the preferred value from the \redmagic analysis.
We then fit the \xlens-contaminated synthetic data with our beyond-\lcdm models  to see how they respond. Note that we use full MCMC chains for this as the \fastismore results indicated that the posterior shifts introduced by the \xlens contamination were too large to be accurately captured through importance sampling. 

\begin{figure}[t]
  \centering
\includegraphics[width=\linewidth]{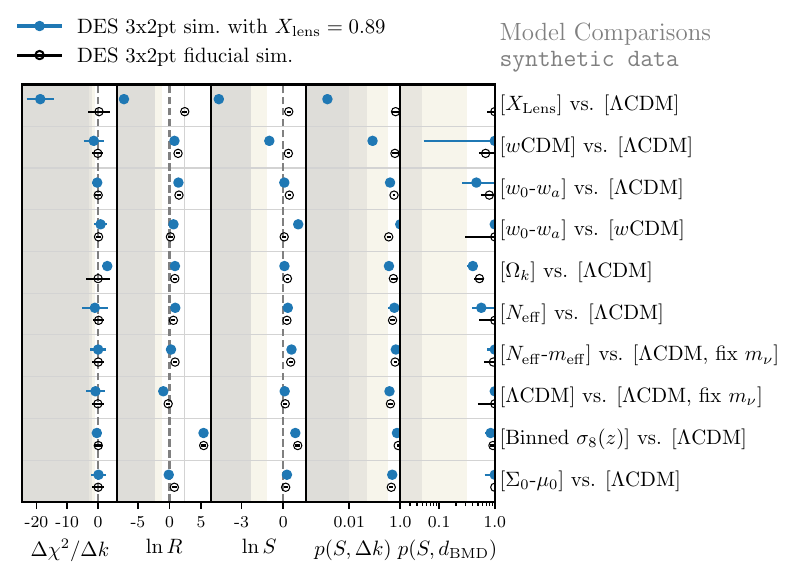}
  \caption{Model comparison metrics when run on a synthetic data vector produced with $\xlens=0.89$. Points mark the median of 200 sampling variance estimates output by {\sc Anesthetic}, and error bars (smaller than the points in most cases here) show the 16 and 84\% quantiles.}
  \label{fig:xlensmodcomp}
\end{figure} 
Estimated model comparison metrics between \lcdm and our extended models for such a contaminated data vector are shown in \fig{fig:xlensmodcomp}.
Compared to the baseline simulated analysis, the \xlens contamination causes model comparison metrics to shift slightly in favor of extended models relative to \lcdm, but none of these shifts are enough to cause an extended model to be strongly preferred to \lcdm. 
Compare this to the $\dchisq\sim 18$ found when fitting a $\lcdm+\xlens$ model (the correct model for the contamination). This is change in \chisq is comparable to, if a bit smaller than, the $\dchisq\sim 25$ improvement found when fitting the \redmagic \mpp data in DES-Y3KP. 
The \wcdm model shows perhaps the greatest sensitivity, and we do see $>1\sigma$ shifts upwards in the marginalized posterior on $w$ when contaminated with \xlens, which roughly agrees with the behavior seen in DES-Y3KP where the original \redmagic sample preferred $w>-1$ (c.f.\ Fig.~12 of DES-Y3KP), in contrast to the fiducial \maglim sample. 
These findings are not unexpected. As was discussed in DES-Y3KP,  the \xlens effect seen for the initial Y3 \redmagic analysis appears to be much more consistent with a lens sample systematic than with new physics. The study presented here supports this, finding that the
observed effect cannot be easily reproduced using the beyond-\lcdm models considered here.

\section{Comparison of Y1 and Y3 \texorpdfstring{\wowa}{Lg} constraints} \label{app:y1vy3}

\begin{figure}
  \centering
\includegraphics[width=\linewidth]{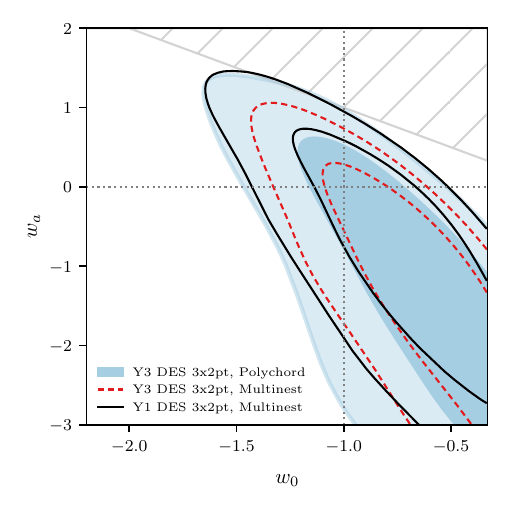}
  \caption{Comparison between the DES Y3 and Y1 \mpp constraints on dynamic dark energy parameters.  Dashed red contours show that analyzing the Y3 data with Multinest sampler settings matching what those used in Y1 artificially narrow the Y3 posterior estimate. }
  \label{fig:w0wa_y1comp}
\end{figure}

Figure~\ref{fig:w0wa_y1comp} shows a comparison between the DES Y3 \mpp constraints on the \wowa dynamic dark energy parameters reported above in the body of the text to those previously published  in DES-Y1Ext~\cite{Abbott:2018xao}. In that figure, the blue filled contour shows same result presented for DES Y3 \mpp constraints in \fig{fig:res_w0wa} of the main body of this paper, while the black contours show the 68 and 95\% confidence regions for the posterior estimate presented in DES-Y1Ext.
We see that there is little change between our Y3 \wowa results and those reported in the comparable DES Y1 analysis, although the modeling and analysis choices are similar in both cases. 

One analysis choice that has an impact on this comparison is the sampler used for parameter estimation: we use Polychord in the present analysis while Multinest was used in the DES-Y1Ext analysis.
To assess the impact of differences in this sampler choice on \wowa constraints we reanalyze the DES Y3 \mpp data using the same Multinest sampler \cite{Feroz:2007,Feroz:2008,Feroz:2013} settings used in DES-Y1Ext, but keeping all other analysis choices the same as in our fiducial Y3 analysis.  One of the main motivations for switching to using the Polychord sampler for DES Y3 analyses was that Multinest tends to undersample posterior tails~\cite{y3-samplers}, an effect which seems to be exacerbated in directions of parameter space where the posterior is more non-Gaussian. The impact of this undersampling is clearly visible for the \mpp \wowa constraints in \fig{fig:w0wa_y1comp}, with the dashed red Multinest contours suggesting tighter constraints than the (more correct) blue Polychord contours and than the DES-Y1Ext Multinest constraints. Quantitatively, Multinest underestimates the width of the DES Y3 \mpp marginalized  68 and 96\% confidence intervals for $w_p$ by 7 and 21\%, respectively, compared to the same quantities estimated using Polychord. This suggests that the Y1 \mpp \wowa constraints may be at least somewhat artificially tightened  due to the use of Multinest.

\end{document}

%% file: july14_DES-2021-0654_author_list.tex

\author{T.~M.~C.~Abbott}
\affiliation{Cerro Tololo Inter-American Observatory, NSF's National Optical-Infrared Astronomy Research Laboratory, Casilla 603, La Serena, Chile}
\author{M.~Aguena}
\affiliation{Laborat\'orio Interinstitucional de e-Astronomia - LIneA, Rua Gal. Jos\'e Cristino 77, Rio de Janeiro, RJ - 20921-400, Brazil}
\author{A.~Alarcon}
\affiliation{Argonne National Laboratory, 9700 South Cass Avenue, Lemont, IL 60439, USA}
\author{O.~Alves}
\affiliation{Department of Physics, University of Michigan, Ann Arbor, MI 48109, USA}
\affiliation{Laborat\'orio Interinstitucional de e-Astronomia - LIneA, Rua Gal. Jos\'e Cristino 77, Rio de Janeiro, RJ - 20921-400, Brazil}
\author{A.~Amon}
\affiliation{Institute of Astronomy, University of Cambridge, Madingley Road, Cambridge CB3 0HA, UK}
\affiliation{Kavli Institute for Cosmology, University of Cambridge, Madingley Road, Cambridge CB3 0HA, UK}
\author{F.~Andrade-Oliveira}
\affiliation{Department of Physics, University of Michigan, Ann Arbor, MI 48109, USA}
\author{J.~Annis}
\affiliation{Fermi National Accelerator Laboratory, P. O. Box 500, Batavia, IL 60510, USA}
\author{S.~Avila}
\affiliation{Instituto de Fisica Teorica UAM/CSIC, Universidad Autonoma de Madrid, 28049 Madrid, Spain}
\author{D.~Bacon}
\affiliation{Institute of Cosmology and Gravitation, University of Portsmouth, Portsmouth, PO1 3FX, UK}
\author{E.~Baxter}
\affiliation{Institute for Astronomy, University of Hawai'i, 2680 Woodlawn Drive, Honolulu, HI 96822, USA}
% ...

\collaboration{DES Collaboration}

%% file: figures/realy3dat.extparams_textable.tex
\centering
\renewcommand{\arraystretch}{1.5}
\begin{tabular}{|c|c|c|c|c|}
\hline
\multicolumn{2}{|c|}{  } & DES 3x2pt  & All External  & All data  \\
\hline
\hline
\multirow{1}{4em}{\centering $w$CDM} & $w_0$& $-0.94^{+0.31}_{-0.18} $& $-1.04^{+0.03}_{-0.03} $& $-1.03^{+0.03}_{-0.03} $\\
\cline{2-5}
\hline
\hline
\multirow{3}{4em}{\centering $w_0$-$w_a$} & $w_0$& $ \geq -1.40 $& $-0.94^{+0.08}_{-0.08} $& $-0.95^{+0.08}_{-0.08} $\\
\cline{2-5}
 & $w_a$& $-0.94^{+1.15}_{-1.15} $& $-0.45^{+0.36}_{-0.28} $& $-0.38^{+0.36}_{-0.28} $\\
\cline{2-5}
 & $w_{\rm p}$& $-0.99^{+0.28}_{-0.17} $& $-1.04^{+0.04}_{-0.03} $& $-1.03^{+0.04}_{-0.03} $\\
\cline{2-5}
\hline
\hline
\multirow{1}{4em}{\centering $\Omega_k$} & $10^2$$\Omega_k$& $ \geq -16 $& $0.08^{+0.18}_{-0.18} $& $0.09^{+0.17}_{-0.17} $\\
\cline{2-5}
\hline
\hline
\multirow{1}{4em}{\centering $N_{\rm eff}$} & $N_{\rm eff}$& $ \leq 7.84 $& $3.10^{+0.16}_{-0.17} $& $3.10^{+0.15}_{-0.16} $\\
\cline{2-5}
\hline
\hline
\multirow{2}{4em}{\centering $\Delta N_{\rm eff}>0.047$} & $\Delta N_{\rm eff}$& -& $ \leq 0.36 $& $ \leq 0.34 $\\
\cline{2-5}
 & $m_{\rm eff}$ [\rm{eV}]& -& $ \leq 0.18 $& $ \leq 0.14 $\\
\cline{2-5}
\hline
\hline
\multirow{2}{4em}{\centering $m_{\rm th}<10$eV} & $\Delta N_{\rm eff}$& -& $ \leq 0.23 $& $ \leq 0.28 $\\
\cline{2-5}
 & $m_{\rm eff}$ [\rm{eV}]& -& $ \leq 0.42 $& $ \leq 0.20 $\\
\cline{2-5}
\hline
\hline
\multirow{2}{4em}{\centering $\Sigma_0$-$\mu_0$} & $\Sigma_0$& $0.56^{+0.37}_{-0.48} $& $0.37^{+0.12}_{-0.09} $& $0.04^{+0.05}_{-0.05} $\\
\cline{2-5}
 & $\mu_0$&  - & $0.20^{+0.22}_{-0.22} $& $0.08^{+0.21}_{-0.19} $\\
\cline{2-5}
\hline
\hline
\multirow{9}{4em}{\centering Binned $\sigma_8(z)$} & $A_2^{P_{\rm lin}}$& $1.00^{+0.14}_{-0.21} $& $0.92^{+0.14}_{-0.23} $& $1.03^{+0.11}_{-0.14} $\\
\cline{2-5}
 & $A_3^{P_{\rm lin}}$& $0.88^{+0.14}_{-0.19} $& $0.95^{+0.17}_{-0.30} $& $0.98^{+0.11}_{-0.13} $\\
\cline{2-5}
 & $A_4^{P_{\rm lin}}$& $0.89^{+0.20}_{-0.26} $& $1.22^{+0.17}_{-0.33} $& $1.24^{+0.13}_{-0.16} $\\
\cline{2-5}
 & $A^{P_{\rm lin}}_{\rm CMB}$& -& $0.89^{+0.10}_{-0.22} $& $1.04^{+0.04}_{-0.06} $\\
\cline{2-5}
 & $\sigma_8^{[\rm{bin}\,1]}$& $0.75^{+0.05}_{-0.05} $& $0.83^{+0.09}_{-0.06} $& $0.78^{+0.02}_{-0.02} $\\
\cline{2-5}
 & $\sigma_8^{[\rm{bin}\,2]}$& $0.74^{+0.06}_{-0.07} $& $0.79^{+0.06}_{-0.06} $& $0.79^{+0.04}_{-0.04} $\\
\cline{2-5}
 & $\sigma_8^{[\rm{bin}\,3]}$& $0.70^{+0.06}_{-0.07} $& $0.80^{+0.07}_{-0.07} $& $0.76^{+0.04}_{-0.04} $\\
\cline{2-5}
 & $\sigma_8^{[\rm{bin}\,4]}$& $0.70^{+0.10}_{-0.09} $& $0.90^{+0.05}_{-0.04} $& $0.86^{+0.04}_{-0.05} $\\
\cline{2-5}
 & $\sigma_8^{[\rm{CMB}]}$& -& $0.78^{+0.02}_{-0.02} $& $0.79^{+0.01}_{-0.01} $\\
\cline{2-5}
\hline
\hline
\multirow{9}{4em}{\centering Binned $\sigma_8(z)$, hyperrank} & $A_2^{P_{\rm lin}}$& $1.16^{+0.16}_{-0.16} $& $0.92^{+0.14}_{-0.23} $& $1.28^{+0.07}_{-0.09} $\\
\cline{2-5}
 & $A_3^{P_{\rm lin}}$& $1.07^{+0.15}_{-0.17} $& $0.95^{+0.17}_{-0.30} $& $1.17^{+0.08}_{-0.10} $\\
\cline{2-5}
 & $A_4^{P_{\rm lin}}$& $0.85^{+0.13}_{-0.24} $& $1.22^{+0.17}_{-0.33} $& $1.51^{+0.12}_{-0.14} $\\
\cline{2-5}
 & $A^{P_{\rm lin}}_{\rm CMB}$& -& $0.89^{+0.10}_{-0.22} $& $1.26^{+0.03}_{-0.04} $\\
\cline{2-5}
 & $\sigma_8^{[\rm{bin}\,1]}$& $0.73^{+0.04}_{-0.04} $& $0.83^{+0.09}_{-0.06} $& $0.72^{+0.01}_{-0.01} $\\
\cline{2-5}
 & $\sigma_8^{[\rm{bin}\,2]}$& $0.78^{+0.06}_{-0.06} $& $0.79^{+0.06}_{-0.06} $& $0.81^{+0.02}_{-0.02} $\\
\cline{2-5}
 & $\sigma_8^{[\rm{bin}\,3]}$& $0.75^{+0.07}_{-0.06} $& $0.80^{+0.07}_{-0.07} $& $0.77^{+0.03}_{-0.03} $\\
\cline{2-5}
 & $\sigma_8^{[\rm{bin}\,4]}$& $0.67^{+0.07}_{-0.09} $& $0.90^{+0.05}_{-0.04} $& $0.88^{+0.04}_{-0.04} $\\
\cline{2-5}
 & $\sigma_8^{[\rm{CMB}]}$& -& $0.78^{+0.02}_{-0.02} $& $0.80^{+0.01}_{-0.01} $\\
\cline{2-5}
\hline
\end{tabular}

%% file: figures/realy3dat.mnu_textable_manual.tex
\begin{tabular}{lrr}
& 95\% upper bound on &$\sum m_{\rm \nu}$ [eV]  \\ \hline
Model & All External & All data \\
\hline
$\Lambda$CDM         &  0.14 &  0.14\\
 $w$CDM              &  0.17 &  0.19\\
 $w_0$--$w_a$        & 0.25  &  0.26\\
$\Omega_k$           & 0.16  &  0.15 \\
 $N_{\rm eff}$       & 0.14  & 0.16 \\
 $\Sigma_0$--$\mu_0$ & 0.21  & 0.14\\
Binned $\sigma_8(z)$ &  0.30 &  0.20\\
 $A_{\rm L}$         &  0.14 &  0.19\\
\hline
\end{tabular}